\documentclass[12pt]{article}
\usepackage{filecontents}
\usepackage{amsmath}
\usepackage{graphicx}
\usepackage{enumerate}
\usepackage{natbib}
\usepackage{url} % not crucial - just used below for the URL 

\usepackage{enumitem}
\usepackage{times}
\usepackage{bm}
\usepackage[plain,noend]{algorithm2e}
\usepackage{amssymb}
\usepackage{amsthm}
\usepackage{mathtools}
\usepackage{subcaption}
\usepackage{afterpage}
\usepackage{booktabs}
\usepackage{setspace}
\usepackage{xr-hyper}
\usepackage[hidelinks]{hyperref}
\usepackage{xr}
\usepackage[title]{appendix}
\usepackage{tikz}
\usetikzlibrary{arrows,shapes.arrows,shapes.geometric,shapes.multipart,backgrounds,decorations.pathmorphing,positioning,fit,automata}
\tikzset{
	>=stealth',
	true/.style={
		rectangle,
		draw=black, very thick,
		text width=6.5em,
		minimum height=2em,
		text centered,
		fill=gray, opacity = 0.5},
	punkt/.style={
		rectangle,
		rounded corners,
		draw=black, very thick,
		text width=6.5em,
		minimum height=2em,
		text centered},
	est/.style={
		circle,
		draw=black, very thick,
		text centered},
	shade/.style={
		circle,
		draw=black, very thick, fill=gray!50,
		text centered},
	weight/.style={
		circle,
		draw=black, very thick,
		text width=6.5em,
		minimum height=2em,
		text centered},
	pil/.style={
		->,
		thick,
		shorten <=2pt,
		shorten >=2pt,},
	double/.style={
		<->,
		thick,
		shorten <=2pt,
		shorten >=2pt,},
	dash/.style={
		dashed,
		thick,
		shorten <=2pt,
		shorten >=2pt,},
	dashdouble/.style={
		<->,
		dashed,
		thick,
		shorten <=2pt,
		shorten >=2pt,}
}

\newcommand{\blind}{1}

\def\T{{ \mathrm{\scriptscriptstyle T} }}

\newcommand{\bigCI}{\mathrel{\text{\scalebox{1.07}{$\perp\mkern-10mu\perp$}}}}

\newcommand{\pr}{\textrm{pr}}
\newcommand{\real}{\mathbb{R}}

\newcommand{\yone}{y^{(1)}}
\newcommand{\ytwo}{y^{(2)}}

\newcommand{\tw}{\textcolor{white}}
\newcommand{\tg}{\textcolor{gray!50}}

\newtheorem{theorem}{Theorem}
\newtheoremstyle{exampstyle}
{6pt} % Space above
{6pt} % Space below
{} % Body font
{} % Indent amount
{\itshape} % Theorem head font
{.} % Punctuation after theorem head
{.5em} % Space after theorem head
{} % Theorem head spec (can be left empty, meaning `normal')
\theoremstyle{exampstyle} 
\newtheorem{assumption}{Assumption}
\theoremstyle{exampstyle} 
\newtheorem{remark}{Remark}
\theoremstyle{exampstyle} 
\newtheorem{condition}{Condition}
\theoremstyle{exampstyle} 
\newtheorem{lemma}{Lemma}

\makeatletter
\newcommand*{\addFileDependency}[1]{% argument=file name and extension
  \typeout{(#1)}
  \@addtofilelist{#1}
  \IfFileExists{#1}{}{\typeout{No file #1.}}
}
\makeatother

\newcommand*{\myexternaldocument}[1]{%
    \externaldocument{#1}%
    \addFileDependency{#1.tex}%
    \addFileDependency{#1.aux}%
}

\myexternaldocument{supplement}

\begin{document}

\def\spacingset#1{\renewcommand{\baselinestretch}%
{#1}\small\normalsize} \spacingset{1}

%%%%%%%%%%%%%%%%%%%%%%%%%%%%%%%%%%%%%%%%%%%%%%%%%%%%%%%%%%%%%%%%%%%%%%%%%%%%%%

\if1\blind
{
  \title{\bf The Promises of Parallel Outcomes}
  \author{Ying Zhou,
%   \thanks{
%     The authors gratefully acknowledge \textit{please remember to list all relevant funding sources in the unblinded version}}\hspace{.2cm}\\
    Dingke Tang,
    Dehan Kong, 
    Linbo Wang \\
    {\small \itshape Department of Statistical Sciences, University of Toronto, 700 University Avenue, Toronto,}\\
    {\small \itshape Ontario M5G 1X6, Canada}\\
    {\small yingx.zhou@mail.utoronto.ca, dingke.tang@mail.utoronto.ca, dehan.kong@utoronto.ca,} \\
    {\small linbo.wang@utoronto.ca}}
  \maketitle
} \fi

\if0\blind
{
  \bigskip
  \bigskip
  \bigskip
  \begin{center}
    {\LARGE\bf The Promises of Parallel Outcomes}
\end{center}
  \medskip
} \fi

\bigskip
\begin{abstract}
A key challenge in causal inference from observational studies is the identification and estimation of causal effects in the presence of unmeasured confounding. In this paper, we introduce a novel approach for causal inference that leverages information in multiple outcomes to deal with unmeasured confounding.  The key assumption in our approach is conditional independence among multiple outcomes. In contrast to existing proposals in the literature, the roles of multiple outcomes in our key identification assumption are symmetric, hence the name parallel outcomes. We show nonparametric identifiability with at least three parallel outcomes and provide parametric estimation tools under a set of linear structural equation models. Our proposal is evaluated through a set of synthetic and real data analyses.
\end{abstract}

\noindent%
{\it Keywords:}  Causal inference; Latent confounding; Multivariate outcome; Non-parametric identification.
% \vfill

% \newpage

\spacingset{1.5} % DON'T change the spacing!

\section{Introduction}
\label{sec:intro}

Unmeasured confounding poses a major threat to the validity of causal conclusions drawn from observational studies. Over the past few  decades, there have been many frameworks developed to mitigate bias due to unmeasured confounding, such as the instrumental variable methods \citep[e.g.][]{wright1928tariff,hernan2006instruments,wang2018bounded}, causal structure learning \citep[e.g.][]{spirtes2000causation}, front-door adjustment \citep[e.g.][]{pearl2009causality},  invariant prediction \citep[e.g.][]{peters2016causal},  and negative controls \citep[e.g.][]{shi2020selective}.

In this paper, we contribute to this effort by introducing a novel approach for causal inference with unmeasured confounding. Our approach  leverages the information in multiple outcomes that are influenced by the same exposure. 
The key identification assumption in our approach is the independence among the multiple outcomes conditional on the common exposure and both measured and unmeasured confounders. Compared to existing proposals in the literature that use multiple outcomes for causal inference, our approach is unique in that the roles of multiple outcomes in our key identification assumption are symmetric. This is appealing as in many modern applications, a priori, there is no reason to discriminate one outcome over the others.  For example, in genetics applications studying associations between a risk factor and gene expression levels, gene expression data are often collected on multiple genes  \citep{leek2007capturing}. In medical applications, with the abundance of electronic health records, investigators may now study the effect of a risk factor, such as off-label drug usage, on many health outcomes at the same time \citep{eguale2016association}.
% In recommendation systems, the effect of recommendation can be studied on a wide range of products simultaneously \citep{sharma2018split}.
% Notably the widely-used click-through rate metric often overestimates the causal effect of a recommendation system due to latent confounding \citep{sharma2018split}.
In financial applications, it is often of interest to study the implications of a particular policy on the returns of multiple stocks \citep{menchero2010global}.

As a specific example, in Section \ref{sec:data}, we study the effect of smoking on serum vitamin C levels.
% with data from . 
To deal with unmeasured confounding between smoking and vitamin C, previous studies use genetic variants \citep{wehby_genetic_2011} or cigarette price \citep{leigh_instrumental_2004} as instrumental variables. As an alternative approach, we note that in the National Health and Nutrition Examination Survey 2005-2006,  participants took a series of lab tests that measured a variety of health indexes, including serum vitamin C. Many of these health indexes are influenced by smoking, which motivates our developments below that leverage information in multiple parallel outcomes for identifying and estimating causal effects.

The idea of causal inference with multiple outcomes has been explored previously in the literature. For example, \cite{mealli2013using} and \cite{mattei2013exploiting} used secondary outcomes to obtain tighter bounds for causal effects defined within certain subgroups known as principal strata. They assumed that a particular identification assumption, called the exclusion restriction, may be violated for the primary outcome of interest, but holds for a secondary outcome. Another line of research in this direction uses negative control outcomes, % The tradition of using negative controls dates back to the  Hill's criteria  for causation \citep{hill1965environment}, who advocated that a causal effect is more likely if the exposure has an effect on the primary outcome but not the auxiliary ones. 
% \cite{lipsitch_negative_2010} distinguished between two types of negative controls, namely positive control outcomes and negative control outcomes. 
with the key assumption that the exposure has no causal effect on the  negative control outcomes. \cite{rosenbaum1989role} showed that a negative control outcome can be used to test for hidden confounding. With an additional variable known as negative control exposure, \cite{miao_identifying_2018}  further showed that the ACE is nonparametrically identifiable, and \cite{shi2020multiply} developed a semiparametric inference procedure in the context of a categorical latent confounder and a binary exposure. \citet[\S 6]{miao2018confounding} extended
this framework by replacing the negative control outcome with   a positive control outcome.
They assumed that the exposure effect on the positive control outcome is non-zero, but known {a priori}. 
Our nonparametric identification strategies in Section \ref{sec:categorical case}  assume a similar causal structure to \citet[\S 6]{miao2018confounding} but do not require prior knowledge of the exposure effect. 

\section{A general framework for parallel outcomes}
\label{sec:framework}

Let $X \in \mathbb{R}$ be a scalar exposure, $U\in \mathbb{R}^r$ be  latent confounding variables and $V \in \mathbb{R}^q$ be baseline covariates. Under the potential outcome framework, $Y(x)=(Y^{(1)}(x),\ldots,\allowbreak Y^{(p)}(x))^{\T}  \in \mathbb{R}^p$ is the potential outcome had the subject  received exposure $x$. Following the stable unit treatment value assumption \citep{rubin1980comment}, the observed outcome $Y = Y(x)$ when $X=x$. We are interested in estimating the  mean potential outcome $E\{Y(x)\}$ with $n$ samples $\{(X_i, Y_i^{(1)},\ldots,Y_i^{(p)});\allowbreak i=1, \ldots, n\}$ independently drawn from the joint distribution of  $(X, Y^{(1)},\ldots,Y^{(p)})$.

% The following assumptions are commonly made in the causal inference literature.

% \begin{assumption}(Stable unit treatment value assumption): The potential outcomes for any unit do not vary with the treatments assigned to other units; for each unit, there are no different forms or versions of each treatment level, which leads to different potential outcomes.
% \label{assumption:sutva}
% \end{assumption} 
If we had measured covariates $U$ in addition to $V$, then under the following assumptions, $E\{Y(x)\}$ can be identified by 
$
    E\{Y(x)\} = E_{U,V}E(Y\mid X=x, U,V).
$

\begin{assumption}
\label{assumption:ignorability}
(Latent ignorability): For any $x$, $X \bigCI Y(x) \mid (U,V)$.
\end{assumption}

\begin{assumption}\label{assumption:positivity}
(Positivity): For any measurable set $\widetilde{\mathcal{X}}$ such that $\pr(X\in \widetilde{\mathcal{X}})>0$, we have $\pr(X\in \widetilde{\mathcal{X}} \mid U,V) > 0$ almost surely.
%(Positivity): Let $g_0(x\mid U,V) = f(X=x \mid U,V), x \in \mathcal{X}$ denote the conditional distribution of treatment $X$ given covariates. $\inf _{x\in\mathcal{X}}g_0(x \mid U,V)>0$ almost everywhere.
\end{assumption}

When $U$ is latent, however, $E\{Y(x)\}$ is not identifiable without further assumptions.  To address this problem, our approach leverages a conditional independence structure among the multiple outcomes $Y^{(1)}, \ldots,\allowbreak Y^{(p)}$ to identify $E\{Y(x)\}$. This structure is formally defined in 
Assumption \ref{assumption:shared-confounding}.

\begin{assumption}\label{assumption:shared-confounding}
(Parallel outcomes):  
% $Y^{(1)} \bigCI Y^{(2)} \bigCI \cdots \bigCI  Y^{(p)} \mid (U, V, X).$
$Y^{(1)}, Y^{(2)}, \ldots, Y^{(p)}$ are mutually independent conditional on $(U,V,X)$: $Y^{(1)} \bigCI Y^{(2)} \bigCI \cdots \bigCI  Y^{(p)} \mid (U, V, X).$
\end{assumption}

% \begin{remark}
% Assumption \ref{assumption:shared-confounding} appears similar to the shared confounding assumption among multiple causes \cite{wang2018blessings}
% suggested using multiple exposures to aid causal identification, assuming shared confounding among multiple exposures and independence of exposures conditional on the latent confounder. However,  as  the problem of unmeasured confounding involves both the exposures and outcome, 
% % ,ogburn2020counterexamples,grimmer2020ive
% the conditional independence structure that only involves exposures but not the outcome is not  sufficient for causal identification \citep{d2019multi,ogburn2019comment}. To achieve identification, one has to make additional assumptions involving the outcome  such as the parametric binary choice outcome model assumed in  \cite{kong2021multi}.
% In contrast, under a conditional independence structure among multiple outcomes conditional on the unmeasured confounder and exposure, it is possible to  nonparametrically identify the exposure effects with at least three parallel outcomes.
% \end{remark}

\begin{figure}[h]
% \begin{subfigure}{.4\textwidth}
\centering
\scalebox{0.8}{
\begin{tikzpicture}[->,>=stealth',shorten >=1pt,auto,node distance=2.3cm,
     semithick, scale=0.50]
     pre/.style={-,>=stealth,semithick,blue,ultra thick,line width = 1.5pt}]
     \tikzstyle{every state}=[fill=none,draw=black,text=black]
     \node[shade] (U)                    {$^{\tg{1}}U^{\tg{1}}$};
     \node[est] (Y2) [below right =1.6 cm of U] {$Y^{(2)}$};
     \node[est] (Y1) [left = 1.8 cm of Y2] {$Y^{(1)}$};
     \node[est] (Y3) [right = 1.8 cm of Y2] {$Y^{(3)}$};
     \node[est] (X) [above right =1.6 cm of Y2] {$^{\tw{1}}X^{\tw{1}}$};
     \path  
     (U) edge node {} (Y1)
     (U) edge node {} (Y2)
     (U) edge node {} (Y3)
     (X) edge node {} (Y1)
     (X) edge node {} (Y2)
     (X) edge node {} (Y3)
     (U) edge node {} (X);
 \end{tikzpicture}}
%  \caption{The simplest parallel-outcome model}
%  \label{fig:DAG2}
%  \end{subfigure}
% %  \bigskip
% %  \newline
% \begin{subfigure}{.6\textwidth}
%   \centering
%   \scalebox{0.8}{
% \begin{tikzpicture}[->,>=stealth',shorten >=1pt,auto,node distance=3.2cm,
%      semithick, scale=0.50]
%      pre/.style={-,>=stealth,semithick,blue,ultra thick,line width = 1.5pt}]
%      \tikzstyle{every state}=[fill=none,draw=black,text=black]
%      \node[shade] (U)                    {$^{\tg{1}}U^{\tg{1}}$};
%   \node[est] (Y2) [below right =1.6 cm of U] {$Y^{(2)}$};
%      \node[est] (Y1) [left = 1.8 cm of Y2] {$Y^{(1)}$};
%      \node[est] (Y3) [right = 1.8 cm of Y2] {$Y^{(3)}$};
%      \node[est] (X) [above right =1.6 cm of Y2] {$^{\tw{1}}X^{\tw{1}}$};
%           \node[shade] (W) [right  = 1.6cm of X] {$^{\tg{1}}W^{\tg{1}}$};
%      \path  
%      (U) edge node {} (Y1)
%      (U) edge node {} (Y2)
%      (U) edge node {} (Y3)
%      (X) edge node {} (Y1)
%      (X) edge node {} (Y2)
%      (X) edge node {} (Y3)
%      (U) edge node {} (X)
%       (W) edge node {} (X)
%          (W) edge node {} (Y3);
%  \end{tikzpicture}}
%  \caption{A parallel-outcome model with single-outcome confounding }
%  \label{fig:DAG-single-outcome-confounding}
% \end{subfigure}
 \caption{The simplest causal diagram associated with the parallel-outcome model when $p=3$. The baseline covariates $V$ are omitted for brevity. Variables $X, Y^{(1)}, Y^{(2)}, Y^{(3)}$ are observed; $U$ is unobserved.  }
 \label{fig:DAG2}
 \end{figure}
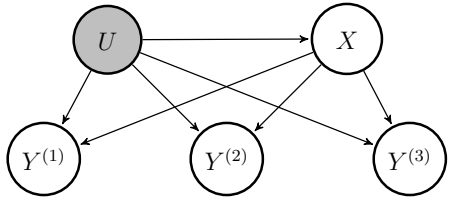

Figure \ref{fig:DAG2} gives the simplest causal diagram \citep{pearl2009causality} associated with the parallel-outcome model when $p=3$.  There is no  directed or bi-directed edges among $Y^{(j)}$'s, encoding the conditional independence in Assumption \ref{assumption:shared-confounding}.
This assumption may be partially tested and relaxed under a set of linear structural equation models; see Remark \ref{remark:fan} in \S\ref{sec:multi_identification}.

% \begin{remark}
%     In contrast to Figure \ref{fig:DAG-single-outcome-confounding}, 
% \end{remark}

% To provide more insights into Assumption \ref{assumption:shared-confounding}, consider the linear structural model assumed by \cite{wang2017confounder}:
% \begin{flalign*}
%     Y &= X \beta + U \gamma_Y + V \theta_Y + \varepsilon_Y,
% \end{flalign*}
% where $\varepsilon_Y \sim MVN(0, \Sigma).$ Assumption \ref{assumption:shared-confounding} holds under the independent-noise assumption that $\Sigma$ is a diagonal matrix \citep[e.g.][]{leek2008general,gagnon2013removing,sun2012multiple,wang2017confounder}.

 If one considers $Y^{(1)}$ as the primary outcome, then $Y^{(2)}$  and $Y^{(3)}$ in fact satisfy the conditions for negative control {exposures} as defined in \citet{miao2018confounding}.  Despite this connection, we coin the term parallel outcomes as in the literature, negative control exposures have commonly been used to refer to exposure variables; see \cite{shi2020selective} for examples. Moreover, unlike Assumption \ref{assumption:shared-confounding}, usually no assumptions are imposed among negative control exposures. Nevertheless, it is worth pointing out that from this perspective, our results in Section \ref{sec:categorical case} can also be interpreted as nonparametric identifiability of causal effects with a pair of negative control exposures that are conditional independent on $U$ and $X$. 

\section{Nonparametric identification}\label{sec:categorical case}

% \label{sec:categorical case}

% \subsection{Nonparametric identification under a binary model}

In this section, we study nonparametric identifiability of the mean potential outcome $E\{Y(x)\}.$  {We illustrate the main idea using the binary model in which  $X, U, Y^{(j)}, j=1,\ldots, p$ are all binary variables taking values 1 or 2, and then present  identification results for the general categorical case. Our results extend to continuous models, which require more technicalities and will be deferred to the Supplementary Material $\S$\ref{sec:identification_continuous}. }

In the following,  we suppress the dependence on baseline covariates $V$ for simplicity. 
For random variables $(W_1,\ldots,W_p) $, we denote their joint distribution as $\pr(w_1,\ldots,w_p)$, the conditional distribution of $ W_i$ given $ W_j=w_j$ as $\pr(w_i\mid w_j)$, and the marginal distribution of $ W_i $ as $\pr(w_i)$. 

In the case where $p=2$, it is easy to see that in the binary model, the observed data distribution $\pr(x, \yone, \ytwo)$ is  determined by 7 parameters, while there are 11 unknown parameters involving $\pr(u), \pr(x\mid u)$ and $\pr(y^{(j)}\mid u,x),j=1,2$. So nonparametric identification is generally not possible in this case; see the Supplementary Material $\S$\ref{sec:categorical_counterexample} for a counterexample.

We hence focus on the scenario where $p=3$. A quick calculation shows that in this case, both the observed data distribution and unknown parameters reside in 15-dimensional spaces. We shall show that under some additional conditions, there is indeed a one-to-one mapping between a subset of the parameter space and the observed data space. Consequently, these parameters are identifiable up to ordering, and the potential outcome distributions are identifiable. 
% Our results also apply if $X$ is a general categorical  variable.

Our identification approach is built on the matrix adjustment method used in \cite{rothman2008modern}, \cite{hu2008identification} and \cite{kuroki_measurement_2014}. Note that from the parallel-outcome assumption, we have 
\begin{align}
    \pr(y^{(2)},y^{(3)}\mid x) &= \sum\limits_{u} \pr(y^{(2)}\mid u,x)\pr(y^{(3)}\mid u,x)\pr(u\mid x), \label{eq:binary_ci1}\\
    \pr(y^{(1)},y^{(2)},y^{(3)}\mid x) &= \sum\limits_{u} \pr(y^{(1)}\mid u,x)\pr(y^{(2)}\mid u,x)\pr(y^{(3)}\mid u,x)\pr(u\mid x). \label{eq:binary_ci2}
\end{align}
Following the matrix adjustment method, we rewrite \eqref{eq:binary_ci1} and \eqref{eq:binary_ci2} in the form of matrix multiplication. 
Let $P(Y^{(2)},Y^{(3)}\mid x) = \left( \pr(Y^{(2)}=i, Y^{(3)}=j\mid X=x) \right)_{2\times 2}$ be a $2\times 2$ matrix whose $(i,j)$-th element is given by  $\pr(Y^{(2)}=i, Y^{(3)}=j\mid X=x)$. Similarly, we let 
$P(Y^{(2)}\mid U,x) = \left( \pr(Y^{(2)}=i\mid U=j, X=x) \right)_{2\times 2}, P(Y^{(3)}\mid U,x) = \left( \pr(Y^{(3)}=i\mid U=j, X=x) \right)_{2\times 2}.$ We also let $P_D(U\mid x) = \big( \pr(U=i\mid X=x) \big)_{2\times 2}$ be a diagonal matrix whose $(i,i)$-th element is given by $\pr(U=i\mid X=x)$. We define $P_D(y^{(1)}\mid U,x) = \left( \pr(Y^{(1)} = y^{(1)}\mid U=i, X=x) \right)_{2\times 2}$ similarly; here the subscript $D$ refers to a diagonal matrix.
Equations \eqref{eq:binary_ci1} and \eqref{eq:binary_ci2} can then be rewritten as
\begin{align}
    \label{eq:mp2}
    P(Y^{(2)},Y^{(3)}\mid x)&=P(Y^{(2)}\mid U,x) P_D(U\mid x)P(Y^{(3)}\mid U,x)^\T,\\
    \label{eq:mq2}
    P(y^{(1)},Y^{(2)},Y^{(3)}\mid x)&=P(Y^{(2)}\mid U,x) P_D(y^{(1)}\mid U,x) P_D(U\mid x)P(Y^{(3)}\mid U,x)^\T.
\end{align}

To eliminate common terms in \eqref{eq:mp2} and \eqref{eq:mq2}, we assume the following condition:
\begin{condition}[Full rank] \label{condition:full-rank}
For all $x$,     $P(Y^{(2)},Y^{(3)}\mid x)$ is of full rank.
 \end{condition}
 % \begin{condition}
% \label{condition:1}
%     For all $x$ and $j=1,2,3, U \nind Y^{(j)} \mid X=x.$
% \end{condition}
% \begin{lemma}
% \label{lemma:1}
% Under Assumptions \ref{assumption:ignorability}, \ref{assumption:shared-confounding} and Condition \ref{condition:1}, if both $Y^{(2)}$ and $Y^{(3)}$ are binary, then for all $x$, $P(Y^{(2)},Y^{(3)}\mid x)$ is of full rank.
% \end{lemma}
% It is an extension of the relevance condition \ref{condition:1} to the categorical case.  Similar to the binary case, it also implies the positivity assumption \ref{assumption:positivity}.
Under Condition \ref{condition:full-rank}, \eqref{eq:mp2} and \eqref{eq:mq2} imply that
\begin{equation}
\label{eqn:svd}
    P(y^{(1)},Y^{(2)},Y^{(3)}\mid x)P(Y^{(2)},Y^{(3)}\mid x)^{-1} = P(Y^{(2)}\mid U,x) P_D(y^{(1)}\mid U,x)P(Y^{(2)}\mid U,x)^{-1}.
\end{equation}
The left-hand side of equation \eqref{eqn:svd} can be identified from the observed data distribution, while the right-hand side is a canonical form of  eigendecomposition.  From the eigendecomposition of the left-hand side of \eqref{eqn:svd}, we have that  $P_D(y^{(1)}\mid U,x)$ is identifiable from data up to permutation of the eigenvalues.
%bCondition \ref{condition:full} ensures that the eigenvalues in $P_D(y^{(1)}\mid U,x)$ are not identical to each other. 
To recover the ordering of eigenvalues, we assume the following condition.
% \begin{condition}[No qualitative $U$--$X$ interaction]
% \label{condition:no-interaction}
%   {For at least one of  $j_1\in \{1,2,3\}$, we have that for any fixed level $y^{(j_1)}$ and all $x$,  $\pr(y^{(j_1)}\mid U=2,x)-\pr(y^{(j_1)}\mid U=1,x)$ have the same sign.}
% \end{condition}
\begin{condition}[No qualitative $U$--$X$ interaction]
 \label{condition:no-interaction2}
    % For at least one of  $j_1\in \{1,2,3\}$, we have that 
    For any fixed level $y^{(1)}$ of $Y^{(1)}$, $1 \leq u_1 < u_2 \leq k$,  $\pr(y^{(1)}\mid U=u_2,x)-\pr(y^{(1)}\mid U=u_1,x)$ are non-zero and have the same sign for all $x$.
 \end{condition}
% Note that Condition \ref{condition:1} ensures that the differences in Condition \ref{condition:no-interaction2} are non-zero. 

Under Condition \ref{condition:no-interaction2}, we can code $U$ such that $\pr(Y^{(1)}=1\mid u, x)$ is strictly increasing in $u$. We can hence identify $\pr(y^{(1)}\mid u,x)$ for each $u$ from $P_D(y^{(1)}\mid U,x)$. Furthermore, since by definition the columns of $P(Y^{(2)}\mid U,x)$ sum up to 1, we can also identify $P(Y^{(2)}\mid U,x)$ and hence $\pr(y^{(2)}\mid u,x)$  from equation \eqref{eqn:svd}.
By symmetry, $\pr(y^{(3)}\mid u, x)$ is also identifiable. It then follows from \eqref{eq:mp2} that $P_D(U\mid x)$ and hence $\pr(u)$ is identifiable. 

Finally,  the potential outcome distributions can be identified from the g-formula:
\begin{equation*}
    %\label{eq:potential_outcome_gformula}
    \pr\{y^{(j)}(x)\}=\sum\limits_{u} \pr(y^{(j)}\mid x,u)\pr(u), j=1,2,3.
\end{equation*}

These arguments also apply in a general categorical model, which gives us our main nonparametric identification result.

% \begin{theorem}
% \label{categorical_identification}
%  Under Assumptions \ref{assumption:ignorability}, \ref{assumption:shared-confounding} and Conditions \ref{condition:1}, \ref{condition:no-interaction}, for all $x$, the potential outcome distributions $\pr\{y^{(j)}(x)\}, j=1,2,3$ are identifiable in a binary parallel-outcome model in which  $X,U, Y^{(j)}, j=1,2,3$ are all binary variables.
% \end{theorem}

\begin{theorem} \label{thm:categorical}
 Suppose that the latent confounder $U$ and parallel outcomes $Y^{(2)}, Y^{(3)}$ are all categorical variables with cardinality $k$.  Suppose further that  Assumptions \ref{assumption:ignorability}, \ref{assumption:shared-confounding} and Conditions \ref{condition:full-rank} and \ref{condition:no-interaction2} hold.
%  \begin{condition}[Full rank] \label{condition:full-rank}
% For all $x$,     $P(Y^{(2)},Y^{(3)}\mid x)$ is of full rank.
%  \end{condition}
%  \begin{condition}[No qualitative $U$--$X$ interaction]
%  \label{condition:no-interaction2}
%     % For at least one of  $j_1\in \{1,2,3\}$, we have that 
%     For any fixed level $y^{(1)}$ of $Y^{(1)}$, $1 \leq u_1 < u_2 \leq k$,  $\pr(y^{(1)}\mid U=u_2,x)-\pr(y^{(1)}\mid U=u_1,x)$ are non-zero and have the same sign for all $x$.
%  \end{condition}
 Then for all $x$, the potential outcome distributions $\pr\{y^{(j)}(x)\}, j=1,2,3$ are identifiable.
\end{theorem}

\begin{remark}
\cite{kuroki_measurement_2014}'s procedure can be used to establish identifiability of causal effects assuming the absence of  arrow $X \rightarrow Y^{(3)}$ in Figure \ref{fig:DAG2}; see Figure S1 %\ref{fig:pearl} 
in the Supplementary Material for an illustration of their causal diagram. \cite{hu2008identification} used a similar approach to deal with a measurement error problem, in which their interest lies in the causal effect of the latent exposure on the outcome; see Figure S2 %\ref{fig:hu} 
in the Supplementary Material for an illustration of their causal diagram.
\end{remark}

% Note in Theorem \ref{categorical_identification}, no restrictions are placed on the cardinality of $X$.

% \subsection{Nonparametric identification in a general parallel-outcomes model}

% Theorem \ref{categorical_identification} can be generalized to a general parallel-outcome model where the latent confounders are categorical. Since the idea and procedure are similar to the binary model, we defer the discussion on this extension to $\S$?? in the Supplementary Material.  We also defer the discussion on continuous latent confounders to $\S$ \tblue{??} in the Supplementary Material. 

% where the latent confounder $U$ is categorical with $k$ levels, and at least two of the parallel outcomes have at least $k$ levels. Without loss of generality, we assume that  $\Ytwo$ and $\Ythree$ have $k$ levels, labeled as $1,\ldots,k.$  

In Theorem \ref{thm:categorical},  Condition \ref{condition:full-rank} can be checked against observed data. Moreover, in a binary model where { $U$,$Y^{(2)},Y^{(3)}$ are binary variables, Condition \ref{condition:full-rank} holds if the causal diagram in Figure \ref{fig:DAG2}  is faithful \citep{pearl2009causality}; see {Lemma \ref{lemma:1}} in the Supplementary Material $\S$\ref{sec:lemmas} for more details.} 
% Due to  eqn. \eqref{eq:mp2}, Condition \ref{condition:full-rank} implies that the unmeasured confounder $U$ has no more levels than $Y^{(2)}$ and $Y^{(3)}$.
It also implies the positivity assumption \ref{assumption:positivity} so the latter is not included in Theorem \ref{thm:categorical}.

% \begin{remark}
% We do not assume Assumption  in Theorem \ref{thm:categorical} as it is implied by Condition \ref{condition:full-rank}. In particular, Condition \ref{condition:full-rank} implies that for all $u,x$, $\pr(u\mid x)>0$, or equivalently, $\pr(x\mid u)>0,$ giving the positivity assumption. 
% \end{remark}

% If , then for all $x$ and $j=1,2,3$, 
% %$U \nind Y^{(j)} \mid X=x$, 
% $Y^{(j)}$ and $U$ are dependent conditional on $X=x$, which implies $P(Y^{(2)},Y^{(3)}\mid x)$ is of full rank in the binary model. For general categorical models,  Condition  can be verified empirically as it only involves observed data. 

Condition \ref{condition:no-interaction2}  requires that for at least one of the parallel outcomes, labeled as $Y^{(1)}$, the effect of latent confounder $U$ on the outcome  has the same direction across all exposure levels. In other words, there is no {qualitative} interaction between $U$ and $X$ on the additive scale in the model for $Y^{(1)}$. This is much weaker than and implied by the treatment effect homogeneity assumption that there is no interaction between $U$ and $X$ on the additive scale in the outcome model, an assumption that is commonly invoked in contexts such as instrumental variable models \citep[e.g.][A5.b]{wang2018bounded}. On the other hand, motivated by a reviewer's comment, we clarify that Condition \ref{condition:no-interaction2} is non-trivial either. Indeed, for each value of $x$, we can pick a permutation $\pi_{y,x}(\cdot)$ on $U$ such that $
        P(Y^{(1)}=y\mid U=\pi_{y,x}(i),x) < P(Y^{(1)}=y\mid U=\pi_{y,x}(j),x)
$ if $i<j.$
    % To emphasize that the choice of permutation may depend on the value of $x$, we add a subscript to $\pi_{y,x}(\cdot).$ 
    Condition \ref{condition:no-interaction2} assumes that there exists such a permutation $\pi_{y}(\cdot)$ that applies to all values of $x$. In other words, although one can label $U$ in an arbitrary order, Condition  \ref{condition:no-interaction2} requires that such an order must be independent of the value of $x$.

Arguing as in the proof of Theorem \ref{thm:categorical}, we may still arrive at local identifiability without Condition \ref{condition:no-interaction2}, that is, there is only a finite number of possible distributions for $\pr\{y^{(j)}(x)\}, j=1,2,3$ that are compatible with the observed data distribution. Condition \ref{condition:no-interaction2} is then needed to achieve global identifiability. We provide an illustration in the Supplementary Material $\S$\ref{example_condition4}.

% See Remark \ref{remark:condition_4} for a detailed discussion on Condition \ref{condition:no-interaction2}. 

In Theorem \ref{thm:categorical}, we do not place restrictions on the cardinality of the exposure $X$ and outcome $Y^{(1)},$ so it applies as long as there are two outcomes with no fewer levels than the unmeasured confounder $U$. 
% in other words, they may be categorical with  arbitrary cardinalities. 
When the cardinality of $Y^{(2)}$ or $Y^{(3)}$ is  larger than $k$, one can first combine different levels in $Y^{(2)}$ or $Y^{(3)}$ and then apply Theorem \ref{thm:categorical}. In particular, with at least two continuous outcomes, Theorem \ref{thm:categorical} implies that the causal effects are nonparametrically identifiable assuming a categorical confounder with arbitrarily many levels. Furthermore, Theorem \ref{thm:categorical} can be generalized to more than three outcomes by choosing different sets of three outcomes and applying Theorem \ref{thm:categorical} repeatedly.

{
Our nonparametric identification results suggest that it is possible to  consistently estimate causal effects under the parallel-outcome structure. In the Supplementary Material  $\S$\ref{section:categorical-estimation}-\ref{sec:simulation}, we describe simple nonparametric estimating procedures under the discrete model, and evaluate their finite sample performance through simulations.
}

In the Supplementary Material $\S$\ref{sec:identification_continuous}, we  extend  
Theorem \ref{thm:categorical} to a continuous model. We however, point out that our nonparametric results do not apply to the linear  structural equation models described later in \eqref{eq:X_multi} and \eqref{eq:Y_multi}, assuming that ${p}=3$
% with three parallel outcomes, 
% \begin{align*}
%     X &= \alpha_X U + \epsilon_X, \\
%     Y^{(j)} &= \alpha_j U + \beta_j X + \epsilon_j, j=1,2,3, 
% \end{align*}
and $U$, $\epsilon_X$, $\epsilon_j, j=1,2,3$ follow Gaussian distributions. In particular, Condition \ref{condition:continuous_indexing} is violated in this model. See  Supplementary Material $\S$\ref{sec:proof_gaussian_case}   for elaborations. This motivates our developments below.

% the causal effects cannot be identified using the nonparametric methods in . 

\section{Identification and estimation in linear structural equation models}
\label{sec:continuous_multiple}
\subsection{Identification}
\label{sec:multi_identification}

In this section, we consider causal effect identification under a set of linear structural equation models with multiple latent confounders. 
% In the following, we first study the identification problem in Section \ref{sec:multi_identification}. Then in Section \ref{section:multi_estimation}, we develop estimators in parallel to our identification procedures. 
We assume the following structural equation models:
\begin{align}
    X &= \alpha_X^\T U + \epsilon_X, \label{eq:X_multi}\\
    Y^{(j)} &= \alpha_j^\T U + \beta_j X + \epsilon_j,\quad j=1,\ldots,{p}, \label{eq:Y_multi}
\end{align}
where $U\in\real^r$ is an unobserved confounder and $\epsilon_X$, $\epsilon_1,\ldots, \epsilon_{p}$ are random errors so that $\allowbreak E(U\epsilon_X)=0\in\real^r$, $\textrm{Cov}(U)=I_r$, $(U,\epsilon_X)\bigCI(\epsilon_1,\ldots,\epsilon_{p})$,  $\epsilon_1,\ldots, \epsilon_{p}$ are mutually independent, and $\allowbreak \textrm{Cov}(X,Y^{(j)})\neq 0, j=1,\ldots, p$. All the random variables in these models are centered, so there are no intercepts. Let  $\sigma^2_X, \sigma^2_1, \ldots, \sigma^2_{p}$ be the variance of $\epsilon_X$, $\epsilon_1,\ldots, \epsilon_{p}$, respectively.
We are mainly interested in the parameters $\beta_j$, representing the causal effects of $X$ on $Y^{(j)},j=1,\ldots,p$.
% {\tred{$p$ and $\widetilde{p}$ might cause some confusions.(Dingke)}}

\begin{remark}
\label{remark:fan}
 The mutual independence among $\epsilon_1,\ldots, \epsilon_p$ implies Assumption \ref{assumption:shared-confounding}. This condition may be partially checked by examining the  error covariance matrix ${\textrm{Cov}}(\epsilon)$ \citep[e.g.][]{fan_large_2013}. See Section \ref{sec:data} for an illustration.
%  . In the data application, we estimate  ${\textrm{Cov}}(\epsilon)$ using a method that assumes the off-diagonal elements are sparse  
If  $\widehat{\textrm{Cov}(\epsilon)}$ is not diagonal, then Assumption \ref{assumption:shared-confounding} is violated.  In this case, one can find the maximal sub diagonal matrix of $\widehat{\textrm{Cov}(\epsilon)}$ and apply our proposed method to the corresponding outcomes. 
% where $\epsilon=(\epsilon_1,\ldots,\epsilon_p)$ may not be diagonal, which indicates violation of Assumption \ref{assumption:shared-confounding}. Instead, one may employ recent developments by \cite{fan_high_2011,fan_large_2013,bai_efficient_2016} to estimate the  
% If one is willing to assume the error $ \epsilon$ is Gaussian distributed, then uncorrelatedness is equivalent to independence. So even if the conditional independence assumptions only holds for a subset of outcomes, we may still keep these outcomes based on $\widehat{\textrm{Cov}(\epsilon)}$.} 
\end{remark}

% \kong{I do not think we need the condition 7. We can state here the first condition for identifiability. We only need to add a condition here: ``$\textrm{Cov}(X,Y^{(j)})= 0$ implies $\beta_j=0$, i.e., excluding the case that the effects are cancelled out.'' Suppose there are $p$ $Y^{(j)}$s that satisfy $\textrm{Cov}(X,Y^{(j)})\neq 0$. We can continue the writing here. We only need to change the original dimension from $ p $ to $ q$ with $ q\geq p$. For those  $Y^{(j)}$ whose causal effects are zero, the causal effects are identified, i.e., $ \beta_j=0$.}

\begin{remark}
The condition that $\textrm{Cov}(X,Y^{(j)})\neq 0$ can be checked from observed data. In the case that this condition is violated for some $Y^{(j)}$, one may apply the proposed method to the subset of outcomes with non-zero correlations with the exposure.
\end{remark}

Substituting \eqref{eq:X_multi} into \eqref{eq:Y_multi}, we get
\begin{align}\label{eq:Y_linear}
    Y^{(j)} = (\alpha_j^\T + \beta_j\alpha_X^\T)U + \beta_j\epsilon_X + \epsilon_j,
    \quad j=1,\ldots,p.
\end{align}
Let $Y = (Y^{(1)},\ldots,Y^{(p)})^\T$, $F=(\sigma_{X}^{-1}\epsilon_X,U_1,\ldots,U_r)^\T$, $\epsilon=(\epsilon_1,\ldots,\epsilon_p)^\T$. It follows that (i) $F\bigCI \epsilon $; (ii) $E(F)=0, \textrm{Cov}(F)=I_{r+1}$; (iii) $E(\epsilon)=0, \textrm{Cov}(\epsilon)=\textrm{diag}(\sigma_1^2,\ldots,\sigma_p^2)$. %\tred{Maybe we can rewrite these $\sigma$ as $\sigma_j$ and $\sigma_x$ instand of $\sigma_{\epsilon_j}$ and $\epsilon_{\sigma_X}$ to make notation simple?  (Dingke).} \tteal{done}
Hence \eqref{eq:Y_linear} is an orthogonal factor model with a matrix form % if: (1) $F$ and $\epsilon$ are independent; {\color{red} $E(F\epsilon^\T)=\textrm{Cov}(F,\epsilon)=0$  suffices} (2) $E(F)=0,\textrm{Cov}(F)=I_{r+1}$; (3) $E(\epsilon)=0,\textrm{Cov}(\epsilon)=\textrm{diag}(\sigma_{\epsilon_1}^2,\ldots,\sigma_{\epsilon_p}^2)$. 
%In our model, (1) and (3) are implied by Assumption \ref{assumption:shared-confounding} in Section \ref{sec:framework}, while (2) follows from Assumption \ref{asp:multiple_u}. 
% These conditions can be easily induced from \eqref{eq:multi_u_uncorrelated1}, \eqref{eq:multi_u_uncorrelated2} and \eqref{eq:multi_u_uncorrelated3}.
\begin{align}\label{eq:Y_matrix}
Y
=\Gamma^*F + \epsilon,
\end{align}
in which $\Gamma^*_{p\times (r+1)}$ is a loading matrix with the $j$-th row
$
    \Gamma^*_{j\cdot}=(
    \sigma_X\beta_j,  \alpha_{j 1}+\alpha_{X1}\beta_j,  \cdots,  \alpha_{j r}+\alpha_{Xr}\beta_j).
$
Under the following mild condition, $\Gamma^*$ can be identified up to rotation \citep[][Thm 5.1]{anderson_statistical_1956}. This condition  also implies that  $p\ge 2(r+1)+1$.
% \begin{align*}
%     \Gamma^*
%     =
%     \begin{pmatrix}
%         \sigma_X\beta_1 & \alpha_{11}+\alpha_{X1}\beta_1 & \cdots & \alpha_{1r}+\alpha_{Xr}\beta_1 \\
%         \sigma_X\beta_2 & \alpha_{21}+\alpha_{X1}\beta_2 & \cdots & \alpha_{2r}+\alpha_{Xr}\beta_2 \\
%         \vdots & \vdots & \ddots & \vdots \\
%         \sigma_X\beta_p & \alpha_{p1}+\alpha_{X1}\beta_p & \cdots & \alpha_{pr}+\alpha_{Xr}\beta_p
%     \end{pmatrix}.
% \end{align*}

%\tred{I think we can write a short paragraph here to summarize high-level idea that how sparsity helps to identify beta. Now we are merely present a result.(Dingke)}
\begin{condition}[submatrix rank]
\label{asp:identification_rotation}
    After removing any row,  there remain two disjoint submatrices of $\Gamma^*$ of rank $r+1$. 
\end{condition}

% Lemma \ref{lem:multi_confounder_identify0} says we are able to identify those outcomes with $\beta_j=0$ with condition \ref{asp:beta_sparsity} and \ref{asp:identification_rotation}. {\color{red} explanations of condition 10? what is implied by the rank requirement? } 
% In a matrix with full column rank, each column cannot be a linear combination of other columns. 

To further identify the rotation matrix, we  assume the following sparsity conditions on the the causal parameter $\beta=(\beta_1, \beta_2, \cdots, \beta_p)^\T$ and the loading matrix $\Gamma^*.$
% The following condition    assumes that the number of null $\beta_j$'s is greater than the dimension of unmeasured confounder.
% We want to further narrow down the possible rotations to help identify the causal effects. 
%  the first column of $\Gamma^*$, i.e. $\sigma_X\beta$, has sufficiently many zero entries. 
%  Intuitively, an arbitrary rotated loading matrix $\Gamma^*R$ have no more zero entries in its first column of $\Gamma^*$.
 \begin{condition}[sparsity]
\label{asp:beta_sparsity}
\begin{enumerate}[label=(\roman*)]
    \item $\|\beta\|_0 = s <p-r$;
    % \item Any submatrix consisting of $p-s+1$ rows of $\Gamma^*$ has full rank $r+1$; %\tred{Do we need to indicate that $p-s+1>r+1$ here?(Dingke)}\kong{This is OK because it is stated in Condition 5.}
    \item The submatrix of $\Gamma^*$ corresponding to the $p-s$ outcomes with $\beta_j=0$ has rank $r$; any other submatrix of $\Gamma^*$ consisting of $p-s$ rows  has the full rank $r+1$.
    % except for the combination of $p-s$ outcomes with $\beta_j=0$, which should have rank $r$.
\end{enumerate}
\end{condition}

Condition \ref{asp:beta_sparsity} assumes that the causal parameter is sparse, while ruling out some too sparse $\Gamma^*$. In (ii), the submatrix corresponding to the $p-s$ outcomes with $\beta_j=0$ is not of full rank because the entries in the first column are all zero. The other parts of Condition \ref{asp:beta_sparsity} are similar to the sparsity condition assumed in \cite{wang2017confounder}, and in parallel to conditions (i) and (iii) in \citet[][Theorem 3]{miao_identifying_2022}. We refer readers to \citet[][Remark 2.2]{wang2017confounder} and \cite{miao_identifying_2022} for discussions of this condition. 

Under Conditions \ref{asp:identification_rotation} and \ref{asp:beta_sparsity}, one can identify the rows of $\Gamma^*$ whose first element is zero. In other words, one can identify the index set of outcomes with zero causal effects: 
$S_0 = \{j: \beta_j= 0\}$.

% Condition \ref{asp:identification_rotation}(ii) says that any $p-s+1$ outcomes still make an $(r+1)$-factor model. Similar conditions to \ref{asp:identification_rotation}(ii) appear in \cite{miao_identifying_2022} and . In Condition \ref{asp:identification_rotation}(iii), 

%  Since multiplying a full column rank matrix on the right by a rotation matrix does not change the rank of the original matrix, Conditions \ref{asp:identification_rotation}(ii) and \ref{asp:identification_rotation}(iii) are compatible with the rotational indeterminacy of the loading matrix. 
 
%  In general, it is unlikely that  This inspires us to propose the identification strategy based on a sparsity assumption in this section.

% To identify $\beta_j,j=1,\ldots,p$, we need to impose the following conditions on the parameters.

% When $r=1$, Conditions \ref{asp:identification_rotation}(ii) and \ref{asp:identification_rotation}(iii) require $\alpha_1$ to be less sparse than $\beta$.

% Condition \ref{asp:identification_rotation}(iii) also implies there should be no more than $\min(r-1,p-s-r)$ rows of zeros. When $r=1$, no row of zeros is allowed. To guarantee this, we exclude outcomes with $\textrm{Cov}(Y^{(j)},X)=0$ in the analysis. This can be achieved by adding a pre-screening step to make sure $\textrm{Cov}(Y^{(j)},X)\ne0$ for $j=1,\ldots,p$.

\begin{lemma}\label{lem:multi_confounder_identify0}
Under  models \eqref{eq:X_multi}, \eqref{eq:Y_multi} and
Conditions \ref{asp:identification_rotation}, \ref{asp:beta_sparsity}, we have
\begin{align*}
    \|\Gamma^*_{\cdot 1}\|_0 \le \|(\Gamma^*R)_{\cdot 1} \|_0
\end{align*}
for any $(r+1)\times(r+1)$ rotation matrix $R$ with $R^\T R = I$; here $A_{\cdot j}$ denotes the $j$-th column of the matrix $A$. Furthermore,  
% When the minimum of $\|(\Gamma^*R)_{\cdot 1} \|_0$ is achieved, 0's are located at the rows of $R$ with the same row indices as the ones in the set $\{j:\beta_j=0\}$. \tblue{Rewrite this claim using mathematical expression.} %\tred{There exist multiple R such that the minimum is achieved? (Dingke)}\kong{Yes. See the following remark.}
{for any $R\in\arg\min_{R^\T R = I_{r+1}} \|(\Gamma^*R)_{\cdot 1} \|_0$, we have $\{j:(\Gamma^*R)_{j1}=0\}=S_0$.}

\end{lemma}
% \begin{remark}
% Here the equality holds when $ R=I$. However, there may be other $R$'s which can minimize $\|(\Gamma^*R)_{\cdot 1} \|_0$ as well. For all such $R$'s, the indices of zeros in $(\Gamma^*R)_{\cdot 1}$ are the same as the ones in $(\Gamma^*)_{\cdot 1}$. 
% \end{remark}

% Lemma \ref{lem:multi_confounder_identify0} 
% Let $k=r+1$. \tblue{Why change of notation here?}  
% Then the $\ell_0$ norm minimization problem is formulated as 
%  \begin{align}\label{eq:optimization_1}
%     \min_{w\in\real^{r+1}} \|\Gamma w\|_0 \textrm{\quad subject to }w^\T w=1.
%  \end{align}
% Lemma \ref{lem:multi_confounder_identify0} implies that
%  \begin{align}\label{eq:optimization_1}
%     \Gamma^* \in \arg\min_{w\in\real^{r+1}} \|\Gamma w\|_0 \textrm{\quad subject to }w^\T w=1.
%  \end{align}
% where $w\in\real^k$ can be considered as the first column of the rotation matrix $Q^\T R$.
% \begin{align*}
%     \min_{w\in\real^k} \|\Lambda w\|_0 \textrm{\quad subject to }w^\T w=1,
% \end{align*}
%More explicitly, we aim to solve
%\begin{align}\label{eq:optimization_1}
%    \min_{w_1,\ldots,w_k}\| w_1\Lambda_{\cdot1} + \cdots + w_k\Lambda_{\cdot k} \|_0\textrm{\quad subject to } w_1^2 + \cdots + w_k^2 = 1.
%\end{align}
% After identifying the $p-s$ zero elements in $\beta$ by solving \eqref{eq:optimization_2}, we now discuss how to 

 To identify the $s$ non-zero elements in $\beta$,  we extend the two stage least squares method in  \cite{miao2018confounding} to accommodate multiple confounders.  Following their terminology, we refer to 
%  In the following we will focus on $\beta_\ell$ where $\ell\in[p]\setminus S_0$.
% Recall that $Y^{(1)},\ldots,Y^{(p)}$ have the conditional independence \tblue{are mutual independent conditional on $(U,X)$ so that? }
% \begin{align}\label{eq:parallel_condition}
%     Y^{(j_1)}\bigCI Y^{(j_2)} \mid (U,X),\quad 1\le j_1<j_2\le p.
% \end{align}
% To this end, we assume the following condition. 
% To make progress, we observe that the outcomes with $\beta_j=0$ satisfy
% \begin{align}\label{eq:nco_condition}
%     Y^{(j)}\bigCI X\mid U,\quad Y^{(j)}\nbigCI U,
% \end{align}
% the latter is true since for $j$ in $S_0$, $\textrm{Cov}(X,Y^{(j)})=\alpha_j^\T\alpha_X\ne 0$ so that $\alpha_j \neq 0.$ These outcomes are 
the outcomes with indexes in $S_0$ as negative control outcomes, and the remaining outcomes as positive control outcomes.
%   In the following, we shall  
%   \tblue{Add more details here. And move the following sentences here: One key step in identifying $\beta^{(\ell)}$ is to regress $Y^{(\ell)}$ on $X$ and $E(W\mid X,Z^\ell)$. Condition \ref{condition:WZ} is used to eliminate the collinearity between $X$ and $E(W\mid X,Z^\ell)$. Otherwise, the coefficient of $X$ is undetermined. } 
{
% This extension is in a similar spirit to two stage least squares estimator with multiple endogenous regressors, but in our method, one should additionally include a set $Z^\ell$ defined in Condition \ref{condition:WZ} as regressors in both stages. Condition \ref{condition:WZ} is used to eliminate the collinearity between $X$ and $E(W\mid X,Z^\ell)$ when we regress $Y^{(\ell)}$ on $X$ and $E(W\mid X,Z^\ell)$ in the second stage. Otherwise, the coefficient of $X$ is undetermined.
% Outcomes satisfying \eqref{eq:nco_condition} are negative control outcomes,   All other outcomes are positive control outcomes with respect to $Y^{(\ell)}$ by the conditional independence assumption \ref{assumption:shared-confounding}. In the following, we shall  extend the two stage least squares method in  \cite{miao2018confounding} to multiple confounders.
In the first stage, we regress a set of negative control outcomes on the exposure and positive control outcomes, and obtain the fitted values for these negative control outcomes. In the second stage, we regress the outcome of interest   on the exposure and the fitted negative control outcomes. Condition \ref{condition:WZ} guarantees that  the exposure and other regressors in the second stage are not co-linear.
% Otherwise, the coefficient of the exposure is undetermined.
}

\begin{condition}[noncollinearity]
\label{condition:WZ}
% There exists a known vector $W=(Y^{(j_1)},\ldots,Y^{(j_r)})\in\real^r$,  where $\{j_1,\ldots,j_r\}\subset S_0$, such that for each $\ell\in [p]\setminus S_0$, $X$ is not a linear combination of $E(W\mid X,Z^\ell)$, where $Z^\ell = (Y^{(k_1)},\ldots,Y^{(k_{p-r-1})})\in\real^{p-r-1}$, with $\{k_1,\ldots,k_{p-r-1}\} = [p]\setminus\{j_1,\ldots,j_r, \ell\}$.
Denote $[p]=\{1,\ldots,p\}$. Let $W=(Y^{(j_1)},\ldots,Y^{(j_{p-s})})\in\real^{p-s}$, where $\{j_1,\ldots,j_{p-s}\}= S_0$. For each $\ell\in [p]\setminus S_0$,
$X$ is not a linear combination of $E(W\mid X,Z^\ell)$, where $Z^\ell = (Y^{(k_1)},\ldots,Y^{(k_{s-1})})\in\real^{s-1}$, with $\{k_1,\ldots,k_{s-1}\} = [p]\setminus\{j_1,\ldots,j_{p-s}, \ell\}$.
\end{condition}

Our identification results under the linear structural equation models can be summarized as follows.

\begin{theorem}
\label{thm:multi_identification}
Under models \eqref{eq:X_multi}, \eqref{eq:Y_multi} and
Conditions \ref{asp:identification_rotation}, \ref{asp:beta_sparsity}, and \ref{condition:WZ}, 
 the parameters $\beta_1,\ldots,\beta_p$ are identifiable. 
\end{theorem}

\subsection{Estimation}\label{section:multi_estimation}
Throughout this section, we use upper-case letters to denote random vectors/matrices or constant matrices, and calligraphy letters denote sample vectors/matrices. 
If $S$ is a set of indexes, then $A_{\cdot S}$ denotes the columns of $A$ with column indexes in $S$.
Suppose we observe independent and identically distributed data $\{(X_i, Y_i^{(1)},\ldots,Y_i^{(p)});\allowbreak i=1, \ldots, n\}$ following   models \eqref{eq:X_multi} and \eqref{eq:Y_multi}.
% In this section, we consider the estimation problem in \eqref{eq:X_multi} and \eqref{eq:Y_multi}.
Based on \eqref{eq:Y_matrix}, we have the factor model
\begin{equation}
\label{eqn:ss}
    \mathcal{Y}_{n\times p} =  \mathcal{F}_{n\times (r+1)}\Gamma^{*\T} + \mathcal{E}_{n\times p}, 
\end{equation}
% \kong{Do we need to use $\Gamma^{*}$ or $ \Gamma$ is already fine? The estimate is written as $ \widehat{\Gamma}$, not consistent.} \tteal{$\Gamma^*$ is the true loading matrix, $\Gamma=\Gamma^*Q$ is a rotated one from solving the factor model (here it's the solution from PCA).}
where $
    \mathcal{Y}_{n\times p} =
    (Y_i^{(j)})_{n\times p},$ 
    $\Gamma^*_{p\times(r+1)} = 
    (\sigma_{X}\beta,  \alpha_{\cdot1}+\alpha_{X1}\beta, \ldots, \alpha_{\cdot r}+\alpha_{Xr}\beta ),$ 
    $\mathcal{E}_{n\times p} = 
    (\epsilon_{i,j})_{n\times p},$ 
    $\mathcal{F}_{n\times(r+1)} =
    (\epsilon_{X}/\sigma_{X},U_1,\ldots, U_r).$

% \begin{align*}
%     \mathcal{Y}_{n\times p} &= 
%     \begin{pmatrix}
%     Y^{(1)}_1 & Y^{(2)}_1 & \cdots & Y^{(p)}_1 \\
%     Y^{(1)}_2 & Y^{(2)}_2 & \cdots & Y^{(2)}_n \\
%     \vdots & \vdots & \ddots & \vdots \\
%     Y^{(1)}_n & Y^{(2)}_n & \cdots & Y^{(p)}_n 
%     \end{pmatrix},
%     \Gamma^*_{p\times(r+1)} = 
%     \begin{pmatrix}
%     \sigma_{X}\beta_1 &  \alpha_{11}+\alpha_{X1}\beta_1 & \cdots & \alpha_{1r}+\alpha_{Xr}\beta_1 \\
%     \sigma_{X}\beta_2 &  \alpha_{21}+\alpha_{X1}\beta_2 & \cdots & \alpha_{2r}+\alpha_{Xr}\beta_2 \\
%     \vdots & \vdots & \ddots & \vdots \\
%     \sigma_{X}\beta_p &  \alpha_{p1}+\alpha_{X1}\beta_p & \cdots & \alpha_{pr}+\alpha_{Xr}\beta_p 
%     \end{pmatrix}, \\
%     \mathcal{F}_{n\times(r+1)} &= 
%     \begin{pmatrix}
%     \sigma_{X}^{-1}\epsilon_{X1} & U_{11} & \cdots & U_{r1} \\
%     \vdots & \vdots & \ddots & \vdots \\
%     \sigma_{X}^{-1}\epsilon_{Xn} 
%     & U_{1n} &
%     \cdots & U_{rn}
%     \end{pmatrix},
%     \mathcal{E}_{n\times p} = 
%     \begin{pmatrix}
%     \epsilon_{11} & \cdots & \epsilon_{p1} \\
%     \vdots & \ddots & \vdots \\
%     \epsilon_{1n} & \cdots & \epsilon_{pn} \\
%     \end{pmatrix}.
% \end{align*}

In \eqref{eqn:ss}, the number of factors $r+1$ can be estimated using standard tools such as the  Kaiser rule \citep{kaiser_application_1960}, while the loading matrix $\Gamma^*$ can be estimated up to rotation using the principal component method. Denote $\widehat{\Gamma}$ an estimate of $\Gamma^*$ up to rotation.

Lemma \ref{lem:multi_confounder_identify0} suggests that one can identify the set $S_0$ by solving an $\ell_0$-minimization problem
% Suppose   we identify $\Gamma^*$  up to rotation from a standard factor analysis \citep[][Thm 5.1]{anderson_statistical_1956}, or equivalently, we identify $\Gamma=\Gamma^*Q$ with $Q\in \mathbb{R}^{(r+1)\times (r+1)}$ being an unknown rotation matrix. 
% Let $w$ be the first column of any $(r+1)\times(r+1)$ rotation matrix, and $x=\Gamma w\in \real^p$.
% The $\ell_0$ norm minimization problem can then be formulated as
\begin{align}\label{eq:optimization_2}
    (x^*,w^*) = \arg\min_{x,w} \|x\|_0 \textrm{\quad subject to }w^\T w=1, x=\Gamma w,
\end{align}
where $\Gamma=\Gamma^*Q$ with $Q\in \mathbb{R}^{(r+1)\times (r+1)}$  an unknown rotation matrix. It follows that $S_0=\{j:x^*_j=0\}$. To account for the variability in $\widehat{\Gamma}$, we 
 introduce a threshold $\delta>0$ and solve the following problem instead:
% Intuitively, if $\widehat{\Lambda} = \Lambda+\Delta$ where $\|\Delta\|_{\max}$ is small enough, then the solution to \eqref{eq:optimization_2} also solves
\begin{align}\label{eq:optimization_5}
    (\widehat{y}^*,\widehat{w}^*)=\arg\min_{y,w} \sum_{i=1}^p \mathbb{I}(|y_i|>\delta) \textrm{\quad subject to }w^\T w=1, y=\widehat{\Gamma} w.  
\end{align}
 We let 
\begin{equation}\label{eq:S1}
\widehat{S}_0=\{j:|\widehat{y}^*_j|\le\delta\}
\end{equation}
be an estimate of the set $S_0$, $ \widehat\Sigma\in \mathbb{R}^{p\times p}$ be the sample covariance of $ (Y^{(1)}_i, \ldots, Y^{(p)}_i)^{\T} $, and $ \lambda_j(\widehat\Sigma) $ be the $j$th eigenvalue of $\widehat\Sigma $. The threshold in \eqref{eq:S1} is set as 
$
    \delta = \sqrt{2n^{-1}\log(p)\widehat{\sigma}^2}
$  \citep{donoho_ideal_1994}, 
where $\widehat{\sigma}^2=p^{-1}\sum_{\widehat{r}+2}^p \lambda_j (\widehat\Sigma) $. We include detailed derivation of $\widehat{\sigma}^2$ in the Supplementary Material $\S$\ref{threshold:derivation}. If there exists a known scalar $M>0$ such that $M\ge \| \widehat{y}^* \|_{\infty}$, then \eqref{eq:optimization_5} can be reformulated as a mixed integer programming problem \citep{feng_complementarity_2018}; see the Supplementary Material, $\S$\ref{sec:mip} for details. In the simulations and data application, we let $M=30$.

It remains to estimate $\beta_\ell$ for $\ell\in[p]\setminus \widehat{S}_0$.  A natural estimator of $\beta_\ell$ is the two stage least squares estimator  \citep{miao_identifying_2018}. Specifically,
let  $\mathcal{X}$ be the sample vector of $X$ and
% $W = \{Y^{(j)}:\zeta^*_j =0\}\in \mathbb{R}^{n\times |W|}$,
$\mathcal{W} = \mathcal{Y}_{\cdot \widehat{S}_0} \in \real^{n\times|\widehat{S}_0|} $, 
% where $S = \{j:\zeta^*_j =0\}$,
so that each column of $\mathcal{W}$ corresponds to an outcome $Y^{(j)}$ with $ \widehat{\beta}_j=0$.  For each $\ell\in[p]\setminus \widehat{S}_0$, we let
$\mathcal{Z}^\ell = \mathcal{Y}_{\cdot T} \in \real^{n\times |T|} $, where $T=[p]\setminus (\{\ell\}\cup \widehat{S}_0)$. To estimate $\beta_\ell,$ one   first regresses $\mathcal{W}$ on $(\mathcal{X},\mathcal{Z}^\ell)$, and obtains the fitted value $\widehat{\mathcal{W}}$. In the second step, one regresses $\mathcal{Y}_{\cdot \ell}$ on $(\mathcal{X},\widehat{\mathcal{W}})$, so that $\widehat{\beta}_\ell$ is the estimated coefficient of $\mathcal{X}$.

%For each $\ell\in\{j:\zeta^*_j=1\}$, we let $Z_\ell = \{Y^{(j)}:\zeta^*_j = 1, j\ne \ell\}$.
%Once we have decided on $W$ and $Z_\ell$, a natural estimator for $\beta_\ell$ following the proof of Theorem \ref{thm:multi_identification} is the two-stage least squares estimator which is popular in instrument variable literature. This is also the estimator used in \cite{miao_identifying_2018}. The two-stage least squares estimator first regresses $W$ on $(X,Z_\ell)$ and denotes the fitted value as $\widehat{W}$, then regresses $Y^{(\ell)}$ on $(X,\widehat{W})$. The estimated coefficient of $X$ is the estimate of $\beta_\ell$.

However, there may be collinearity among the regressors in the second step if $|\widehat{S}_0| > |T|$. This can happen, for example, if all the random variables in models \eqref{eq:X_multi} and \eqref{eq:Y_multi} follow Gaussian distributions. 
% However, if all random variables in models \eqref{eq:X_multi} and \eqref{eq:Y_multi} follow Gaussian distributions, there may be collinearity among the regressors in the second step if  $|\widehat{S}_0| > |T|.$
% the number of columns of $ \mathcal{W} $ is larger than that of $\mathcal{Z}^\ell$ \tblue{write this as $** >**$. }, which makes the two-stage least squares estimator infeasible. 
% This is formally stated in the following proposition.
% \begin{proposition}\label{lemma:gaussian_collinearity}
% If the variables $U$, $\epsilon_X$, $\epsilon_j,j=1,\ldots,p$ in \eqref{eq:X_multi} and \eqref{eq:Y_multi} are Gaussian and $|T|<|S|$, then Condition \ref{condition:WZ} does not hold. 
% \end{proposition}
%Hence in this case, the standard two-stage least squares estimator of $\beta_\ell$ is not applicable when $|Z_\ell|<|\widehat{W}|$.
%One possible approach is to randomly select a subset of $\widehat{W}$ to act as regressors in the second stage. However, this method has the disadvantages that it heavily depends on the result of the previous optimization step. If unfortunately an outcome is deemed as a negative control outcome by mistake, and it is chosen as one of the regressor in the second stage, the estimate will be severely biased. 
To address this, we suggest using ridge regression in the second stage when regressing $\mathcal{Y}_{\cdot \ell}$ on $(\mathcal{X},\widehat{\mathcal{W}})$. Denote $\mathcal{A} = (\mathcal{X},\mathcal{Z}^\ell)\in \mathbb{R}^{n\times (1+|T|)}$ the column-combined data matrix. We obtain $\widehat{\mathcal{W}} = \mathcal{A}(\mathcal{A}^\T \mathcal{A})^{-1}\mathcal{A}^\T \mathcal{W} $. Let $\widehat{\mathcal{B}} = (\mathcal{X},\widehat{\mathcal{W}})\in\real^{n\times (1+|\widehat{S}_0|)}$ be the column-combined data matrix. 
The proposed ridge estimator for $\beta_\ell$ is
$
    \widehat{\beta}_{\ell,ridge} = (\widehat{\mathcal{B}}^\T \widehat{\mathcal{B}} + \lambda \mathcal{D})^{-1}\widehat{\mathcal{B}}^\T \mathcal{Y}_{\cdot \ell},
$
where $\mathcal{D}=\textrm{diag}\{0,1,\ldots,1\}$ is a $(1+|\widehat{S}_0|)\times(1+|\widehat{S}_0|)$ diagonal matrix. The first diagonal element of $\mathcal{D}$ is $0$ because we do not penalize the regressor $\mathcal{X}$. In practice, the tuning parameter $\lambda$ can be chosen by 10-fold cross-validation. 

In summary, our estimator $\widehat{\beta}$ is defined as
\begin{align}\label{ridgeestimator}
    \widehat{\beta}_j =
    \begin{cases}
    0 & j\in \widehat{S}_0, \\
    (\widehat{\mathcal{B}}^\T \widehat{\mathcal{B}} + \lambda \mathcal{D})^{-1}\widehat{\mathcal{B}}^\T \mathcal{Y}_{\cdot \ell} & j\notin \widehat{S}_0.
    \end{cases}
\end{align}
where $\widehat{S}_0$ is defined in \eqref{eq:S1}.
 
% \tteal{In summary, the estimation procedure consists of a mixed integer programming problem, and a modified two stage least squares regression. The former is relatively more time-consuming than the latter. The whole procedure can be executed in \texttt{r}. We will illustrate the implementation details in Section \ref{sec:data}. }

%\tred{The first element of D is 0, which stands for non-standard ridge regression. Do we need to explain the reason? (Dingke)}
\section{Simulation}
In this section, we evaluate the performance of our proposed estimator \eqref{ridgeestimator}. The data are generated following models \eqref{eq:X_multi} and \eqref{eq:Y_multi}. We consider a two-dimensional unobserved confounder $U=(U_1,U_2)^\T $. We set $\sigma_{X} = 1$ and $\sigma_j = 1.5 + 0.25\{(j+2)\bmod{3}\}$ for $j=1, \ldots, p$, and generate $(U_1, U_2, \sigma_X^{-1}\epsilon_X,\sigma_1^{-1}\epsilon_1,\ldots,\sigma_p^{-1}\epsilon_p)^{\T}$ from $ \mathcal{N}(0,I_{p+3})$. Define $ \gamma\in \mathbb{R}^{7p}$ so that $\gamma_{(7k+1):(7k+7)} = (1.5, -1.8, 2.1, 2.4, -2.7, 3, -3.3), k=0,\ldots,p-1,$ where $\gamma_{i:j}$ denotes the subvector of $\gamma$ from the $i$th element to the $j$th element;  so $\gamma_{1:2p}$ is the first $ 2p$ entries of $ \gamma$.  We set $\alpha_X = (1,1)^\T $, $(\alpha_1^{\T}, \alpha_2^{\T})^{\T}=\gamma_{1:2p}$, $\beta_j = (-1)^{j}[1+\{(j+3)\bmod{4} \}]$ if $1\le j\le 0.4p$, and $\beta_j = 0$ if $0.4p< j\le p$. %For $ \alpha_1$ and $ \alpha_2 $
% \tteal{The numerical values of $\alpha$ and $\beta$ are set in a way to diversify the combinations of parameters.}
We consider $p=30,60,100$ and $n=500,1000,2000$ in our simulations. 

%dealing with a multi-dimensional unobserved confounder. We assume all variables are Gaussian distributed, and the data are generated from \eqref{eq:X_multi} and \eqref{eq:Y_multi} with $U=(U_1,U_2)^\T\in\real^2$. 

%Specifically, we let
%$(U_1, U_2, \sigma_{\epsilon_X}^{-1}\epsilon_X,\sigma_{\epsilon_1}^{-1}\epsilon_1,\ldots,\sigma_{\epsilon_p}^{-1}\epsilon_p)\sim \mathcal{N}(0,I_{p+3});
%\sigma_{\epsilon_X} = 1; 
%\sigma_{\epsilon_j} = 1.5 + 0.25\{(j+2)\bmod{3}\}; 
%\alpha_X^\T = (1,1); 
%$\alpha_{ji} = (-1)^{(m\bmod{3})+\mathbb{I}(m=6)} (1.5 + 0.3m), \textrm{ where }m=\{p(i-1)+j+6\}\bmod{7}, 1\le j \le p, 1\le i\le 2;$ 
%\beta_j = (-1)^{j}[1+\{(j+3)\bmod{4} \}] \text{ for }1\le j\le 0.4p, \beta_j = 0 \text{ for }0.4p< j\le p$. In this setting, $\sigma_{\epsilon_j}\in\{1.5, 1.75, 2\}$, $\alpha_{ji}\in\{1.5, -1.8, 2.1, 2.4, -2.7, 3, -3.3\}$, $\beta_j\in\{0,-1,2,-3,4\}$, for $j=1,\ldots,p, i=1,2$.

% The parameters are set in the following way. We let the absolute value of the $p$-vector $\alpha$ be repetitions of the sequence $(1.5,1.8,2.1,2.4,2.7,3)$ until all positions are filled. Similarly, the absolute value of the causal parameter $\beta$ has $0.6p$ 0's followed by repetitions of the sequence $(1,2,3,4)$. The signs of elements in $\alpha$ and $\beta$ are assigned to be positive or negative with probability 0.5. 
% The error terms $\epsilon_X\sim \mathcal{N}(0,1)$, $\epsilon_j\sim\mathcal{N}(0,\sigma_{\epsilon_j}),j=1,\ldots,p,$ where $\sigma_{\epsilon_j}$'s are repetitions of the sequence $(1.5,1.75,2)$.

{
% We use \texttt{r} to implement our estimator \eqref{ridgeestimator} in the simulations here and the data application in Section \ref{sec:data}. In particular, w
In our simulations and data application, we use the function \texttt{nScree} in the package \texttt{nFactors} to estimate the number of factors $r+1$ by the Kaiser rule, and the package \texttt{POET} \citep{fan_large_2013} to estimate the error covariance matrix $\textrm{Cov}(\epsilon)$ and check the conditional independence assumption \ref{assumption:shared-confounding}. We solve the optimization problem \eqref{eq:optimization_5} using the \texttt{gurobi} package.
}

In Table \ref{tab:simu_zero} we report the false positive rates and false negative rates over $100$ Monte Carlo runs, defined as $\bigr\rvert\{j:\widehat{\beta}_j\ne 0,\beta_j=0\}\bigr\rvert \bigr/ \bigr\rvert\{j:\widehat{\beta}_j\ne0\}\bigr\rvert$ and $\bigr\rvert\{j:\widehat{\beta}_j= 0,\beta_j\ne0\}\bigr\rvert \bigr/ \bigr\rvert\{j:\widehat{\beta}_j=0\}\bigr\rvert$, respectively. One can see that our selection procedure for negative control outcomes is conservative in that our false negative rates are zero for all $(n,p)$ combinations. In comparison, the false positive rate is non-zero, albeit very small, in the simulation settings considered here.  This is desirable as consistent estimation of causal parameters relies on valid negative control outcomes. If   $\beta_j \neq 0$ but $\widehat{\beta}_j = 0,$ then the estimates for all non-zero elements of $\beta$ may be biased as $Y^{(j)}$ is then mistaken as a negative control outcome in the ridge estimator \eqref{ridgeestimator}. In comparison, a false positive result does not introduce bias for  the ridge estimator \eqref{ridgeestimator}.

% If an outcome with $\beta_j\ne 0$ is mistakenly identified as an outcome with $\widehat{\beta}_j=0$, this outcome will be included in $W$ as a negative control outcome, which may introduce  On the other hand, if an outcome with $\beta_j= 0$ is mistakenly identified as an outcome with $\widehat{\beta}_j\ne 0$, it does not introduce additional bias and only slightly affects the efficiency. Therefore, we hope the false negative rate to be as close to zero as possible, while we can tolerate a nonzero false positive rate.  From the results, we can see that the false  and the false positive rates are also low. 

% In addition, we calculate the bias of the causal effect estimate given in \eqref{ridgeestimator}. Specifically,
In Table \ref{tab:simu_beta}, we report the average biases and the standard errors of the biases of $\widehat{\beta}_1,\widehat{\beta}_2,\widehat{\beta}_3,\widehat{\beta}_4$, corresponding to the causal effects of the first four outcomes.  {We can see that the biases are small relative to their standard deviations in all the scenarios considered here. }

\begin{table}[]
    \caption{{False positive rate  $\times 10000$ and false negative rate  $\times 10000$ of the proposed selection method for the negative control outcomes } }
    \centering
    \begin{tabular}{lrrrrrr}
    \toprule
    \multicolumn{1}{c}{}& \multicolumn{3}{c}{False positive rate $\times 10000$} & \multicolumn{3}{c}{False negative rate $\times 10000$}\\
    & $p=30$ & $p=60$ & $p=100$ & $p=30$ & $p=60$ & $p=100$ \\
    \midrule
    % type I error & $n=500$  &  0.44 & 0.11 & 0.12 \\
    % & $n=1000$ & 0.28 & 0.11 & 0.02 \\
    % & $n=2000$ & 0.28 & 0.08& 0.03 \\
    % type II error & $n=500$ & 0 & 0 & 0 \\
    %  & $n=1000$ & 0 & 0 & 0 \\
    %  & $n=2000$ & 0 & 0 & 0 \\
    $n=500$ & $58$ & $16$ & $10$ & $0$ & $0$ & $0$ \\
    $n=1000$ & $54$ & $16$ & $54$ & $0$ & $0$ & $0$\\
     $n=2000$ & $37$ & $12$ & $5$ & $0$ & $0$ & $0$\\
     \bottomrule
    \end{tabular}
    \label{tab:simu_zero}
\end{table}

% \begin{table}[]
%     \caption{Bias$\times100$ (Standard deviation $\times100$) for the proposed ridge estimator for estimating the causal parameters $(\beta_1,\beta_2,\beta_3,\beta_4)=(-1,2,-3,4)$. }
%     \centering
%     \begin{tabular}{c|c|c|c|c|c}
%       &  & $\widehat{\beta}_1$ & $\widehat{\beta}_2$ & $\widehat{\beta}_3$ & $\widehat{\beta}_4$ \\
%     $p = 30$ & $n=500$ &  1.09(7.31) & -0.95(10.93) & 0.16(10.65)& 0.75(7.72)  \\
%     & $n=1000$ &  $-0.19(6.62)$ & -0.11(6.68) & -0.28(7.63) & 0.08(5.99) \\
%     & $n=2000$ &  -0.64(4.34) & -0.18(4.93) & 0.03(4.77) & 0.19(3.40) \\
%     $p = 60$ & $n=500$ & -1.35(7.42) & 0.43(8.92) & -1.25(9.39) & 1.47(8.35) \\
%     & $n=1000$ & -0.58(5.53) & 0.71(5.49) & -1.81(8.35) & 0.14(6.04) \\
%     & $n=2000$ &  0.10(3.94) & -0.08(4.54) & -0.31(5.72) & 0.30(4.10) \\
%     $p=100$ & $n=500$ &  0.71(8.30) & 0.67(9.80) & -0.15(10.07) & 0.71(7.23) \\
%     & $n=1000$ &  -1.06(5.88) & 0.09(7.03) & 1.10(7.10) & 0.62(5.48) \\
%     & $n=2000$ &  0.47(3.38) & -0.30(3.84) & -1.13(5.37) & 0.45(4.13)
%     \end{tabular}
%     \label{tab:simu_beta}
% \end{table}

\begin{table}[]
    \caption{Bias$\times100$ (Standard error $\times100$) of the proposed ridge estimator for estimating the causal parameters $(\beta_1,\beta_2,\beta_3,\beta_4)=(-1,2,-3,4)$ }
    \centering
    \begin{tabular}{llrrrr}
    \toprule
       \multicolumn{1}{c}{} &\multicolumn{1}{c}{}  & \multicolumn{1}{c}{$\widehat{\beta}_1$} & \multicolumn{1}{c}{$\widehat{\beta}_2$} & \multicolumn{1}{c}{$\widehat{\beta}_3$} & \multicolumn{1}{c}{$\widehat{\beta}_4$} \\
       \midrule
    $p = 30$ & $n=500$ &  $0.75(0.75)$ & $1.26(0.98)$ & $-0.52(1.06)$ & $0.98(0.79)$  \\
    & $n=1000$ &  $0.03(0.60)$ & $0.61(0.67)$ & $-0.25(0.88)$ & $0.78(0.63)$ \\
    & $n=2000$ &  $-0.68(0.45)$ & $-0.12(0.49)$ & $0.03(0.49)$ & $0.10(0.36)$ \\\midrule
    $p = 60$ & $n=500$ & $-1.40(0.74)$ & $0.52(0.90)$ & $-1.27(0.94)$ & $1.43(0.83)$ \\
    & $n=1000$ & $-0.58(0.55)$ & $0.71(0.55)$ & $-1.81(0.84)$ & $0.14(0.60)$ \\
    & $n=2000$ &  $0.30(0.40)$ & $0.04(0.47)$ & $-0.70(0.58)$ & $0.15(0.40)$ \\\midrule
    $p=100$ & $n=500$ &  $-0.57(0.75)$ & $0.91(0.86)$ & $0.63(0.93)$ & $0.60(0.83)$ \\
    & $n=1000$ &  $-1.06(0.59)$ & $0.09(0.70)$ & $1.10(0.71)$ & $0.62(0.55)$ \\
    & $n=2000$ &  $0.47(0.34)$ & $-0.30(0.38)$ & $-1.13(0.54)$ & $0.45(0.41)$\\
    \bottomrule
    \end{tabular}
    \label{tab:simu_beta}
\end{table}

% \begin{table}[]
%     \centering
%     \begin{tabular}{c|c|c|c|c|c}
%       &  & $Y^{(1)}$ & $Y^{(2)}$ & $Y^{(3)}$ & $Y^{(4)}$ \\
%     $n=500$ &  ridge cv  & -0.0135(0.0742) & 0.0043(0.0892) & -0.0125(0.0939)& 0.0147(0.0835)  \\
%      & 2sls(selected) & -0.0121(0.0905) & 0.0167(0.0958) & -0.0240(0.0980) & 0.0278(0.0886)\\
%      $n=1000$ & ridge cv  & -0.0058(0.0553) & 0.0071(0.0549) & -0.0181(0.0835) & 0.0014(0.0604) \\
%      & 2sls(selected) & -0.0167(0.0672) & 0.0130(0.0657) & -0.0253(0.0836) & 0.0048(0.0680) \\
%     $n=2000$ & ridge cv & 0.0010(0.0394) & -0.0008(0.0454) & -0.0031(0.0572) & 0.0030(0.0410) \\
%     & 2sls(selected) & 0.0010(0.0458) & -0.0022(0.0478) & -0.0011(0.0593) & 0.0040(0.0439)
%     \end{tabular}
%     \caption{p=60}
%     \label{tab:simu_p60}
% \end{table}

% \begin{table}[]
%     \centering
%     \begin{tabular}{c|c|c|c|c|c}
%       &  & $Y^{(1)}$ & $Y^{(2)}$ & $Y^{(3)}$ & $Y^{(4)}$ \\
%     $n=500$ &  ridge cv  & 0.0071(0.0830) & 0.0067(0.0980) & -0.0015(0.1007) & 0.0071(0.0723) \\
%     & 2sls(selected) & 0.0265(0.0916) & 0.0052(0.1106) & -0.0128(0.1052) & 0.0422(0.0972) \\
%     $n=1000$ &  ridge cv & -0.0106(0.0588) & 0.0009(0.0703) & 0.0110(0.0710) & 0.0062(0.0548) \\
%     & 2sls(selected) & 0.0030(0.0624) & 0.0074(0.0756) & -0.0015(0.0756) & 0.0190(0.0673) \\
%     $n=2000$ & ridge cv & 0.0047(0.0338) & -0.0030(0.0384) & -0.0113(0.0537) & 0.0045(0.0413) \\
%     & 2sls(selected) & 0.0094(0.0341) & -0.0045(0.0456) & -0.0060(0.0565) & 0.0095(0.0548)
%     \end{tabular}
%     \caption{p=100}
%     \label{tab:simu_p100}
% \end{table}

\section{Real data Application}
\label{sec:data}

Studies have found that smoking, including secondhand smoking,
%have reduced amounts of vitamin C in their bodies.
are associated with a high risk of vitamin C inadequacy \citep{institute_of_medicine_dietary_2000}. In this section, we study how smoking affects serum vitamin C levels. As a byproduct, we also obtain the effects of smoking on a group of laboratory test results. 

The data we use come from the National Health and Nutrition Examination Survey 2005-2006. The exposure of interest is the serum cotinine level. Cotinine is a major metabolite of nicotine, and the measurement of serum cotinine is a marker for active and passive smoking \citep{benowitz_cotinine_1996}. 
{We consider 20 laboratory results, including serum vitamin C level, as outcomes in our analysis. We provide detailed descriptions of these outcomes  in the Supplementary Material, Table \ref{tab:description}.} These outcomes are selected from different categories in the laboratory data section using the criteria that they are representative of different health aspects, so that they are not likely to directly affect each other. % The description of these outcomes are in Table \ref{tab:description}  in the Supplementary Material. 
A total of 1852 samples with ages between $12$ and $85$ are included in the analysis, among which $925$ are females and $927$ are males. 

% We first perform a log transformation of the outcomes and the exposure. 
In our analysis, we adjust for observed confounders, including gender, age, and race, by regressing the log-transformed exposure and outcomes on these covariates. We take the residual corresponding to the exposure as $ X $ and the residuals corresponding to the outcomes as $Y^{(j)}$, $j=1,\ldots, 20$. We then  regress each $Y^{(j)}$ on $X$ and keep the outcomes with estimated regression coefficient greater than $
    \widehat{\sigma}_{(j)}\sqrt{2\log 20},
$  
where $ \widehat{\sigma}_{(j)} $ is the standard error of the estimated regression coefficient. 
This leaves us with 11 outcomes, including acrylamide, vitamin C, cadmium, urinary thiocyanate, retinyl palmitate, osmolality, vitamin D, vitamin B12, blood mercury, enterolactone, and C-reactive protein.

%Gender, age, and race are considered as observed covariates. 

%The analysis is done in the following manner. First, we regress the log transformed exposure and outcomes on gender, age and race, and let the (standardized?) residuals be the updated values of exposure and outcomes, denoted as $\tilde{X}$ and $\tilde{Y}^{(j)}, j=1,\ldots, 20$. 
%Then all variables are scaled to have variance 1. 
%Next, we screen out the outcomes which are uncorrelated with the exposure by hard thresholding. We keep the outcome if the magnitude of the coefficient from regressing the $\tilde{Y}^{(j)}$ on $\tilde{X}$ is greater than 
%$
%    \widehat{\sigma}_j\sqrt{2\log p},
%$
%where $\widehat{\sigma}_j$ is the standard error of the estimated coefficient, and $p=20$ in our case. 
After preprocessing, we apply our proposed methods in Section \ref{section:multi_estimation} to estimate the causal effects of serum cotinine level on the  outcomes. Using the Kaiser rule, we select three factors in the model \eqref{eqn:ss}, suggesting a two-dimensional latent confounder $U$. {
We then estimate the error covariance matrix $\textrm{Cov}(\epsilon)$  using the {\tt R} package \texttt{POET}.
All the off-diagonal entries in the error covariance matrix are estimated to be zero, with diagonal elements $\allowbreak \{0.31,0.51,0.33,0.38,0.70,0.57,$ $0.57,0.69,0.81,0.61,0.58\}$. This  provides empirical evidence for Assumption \ref{assumption:shared-confounding}. 
% To test the conditional independence assumption {\ref{assumption:shared-confounding}}, we use the packe \texttt{POET} \citep{fan_large_2013} to estimate  The estimated error covariance matrix is an 
% If we assume the error terms follow Gaussian distributions, this implies the errors are mutually independent. 
}
% \tblue{This comes before factor analysis?}
Our selection procedure described in \eqref{eq:S1} selects four positive control outcomes for which the causal effect of serum cotinine is potentially non-zero: acrylamide, vitamin C, cadmium, and urinary thiocyanate.
% \tblue{How does this compare with literature?}
% \tteal{We then use the modified two stage least squares estimator to estimate the causal effects of tobacco exposure on these four outcomes.} By \eqref{ridgeestimator}. The  causal effect estimates  are
% Our causal effect estimation procedure described in \eqref{ridgeestimator} shows that
% % 0.63, -0.31, 0.76, 0.40 (old using 1/(p-k))
% $0.64$, $-0.36$, $0.76$, $0.41$, \tteal{$0.63(0.025)$, $-0.31(0.074)$, $0.77(0.026)$, $0.41(0.030),$}
% respectively.
% \tblue{write these in English, not just numbers. Write it in sentences that scientists can understand, including the units. These numbers, as they are now, do not mean anything.}
{
Our causal effect estimation procedure described in \eqref{ridgeestimator} shows that 
% after log transformation and standardization of the data, the estimated causal effects of smoking on acrylamide, vitamin C, cadmium, and urinary thiocyanate are $0.63(0.025)$, $-0.31(0.074)$, $0.77(0.026)$, $0.41(0.030),$ respectively. 
increasing serum cotinine level by one fold causes acrylamide to increase by $8.6\% (\textrm{95\% CI}: [7.9\%, 9.3\%])$,   vitamin C  to decrease by $3.9\% (\textrm{95\% CI}: [2.1\%,5.7\%])$,
 cadmium  to increase by $11.7\%  (\textrm{95\% CI}: [10.9\%,12.5\%])$, and  urinary thiocyanate to increase by $9.2\%  (\textrm{95\% CI}: [7.8\%, 10.6\%])$. 
% multiply the serum cotinine level(ng/mL) by $e$ will multiply acrylamide(pmoL/G Hb) by $1.126 (\textrm{95\% CI}:[1.116,1.137])$, multiple vitamin C(mg/dL) by $0.944 (\textrm{95\% CI}:[0.919,0.969])$, multiple cadmium(ug/L) by $1.173 (\textrm{95\% CI}:[1.161,1.186])$, and multiple urinary thiocyanate(ng/mL) by $1.135 (\textrm{95\% CI}:[1.115,1.156])$
}
Note when reporting these confidence intervals, we do not take into account the randomness due to the selection of positive control outcomes. 
% \st{Hence, we conclude that tobacco exposure has negative effects on vitamin C} \citep{schectman_influence_1989,preston_influence_2003}, \st{positive effects on acrylamide, cadmium, urinary thiocyanate} \citep{mojska_acrylamide_2016,richter_cadmium_2017,ngogang_salivary_1983}, \tblue{Rewrite this sentence. Say our results agree with literature etc... You should not use references for our results... }  and no effect \tblue{Are we able to conclude no effect? or not significant? I am not sure. } on the remaining 7 outcomes. 
{
Our findings are consistent with the literature studying the effects of tobacco smoke on acrylamide \citep[e.g.][]{mojska_acrylamide_2016},  vitamic C \citep[e.g.][]{schectman_influence_1989,preston_influence_2003}, cadmium \citep[e.g.][]{richter_cadmium_2017}, and urinary thiocyanate \citep[e.g.][]{ngogang_salivary_1983}.
% For the remaining 7 outcomes, the causal effects are not evident from our results. 
}

%Following the procedure in Section \ref{section:multi_estimation}, 3 out of 4 methods adopted in \texttt{nScree} vote for a 3-factor model, which indicates a 2-dimensional latent confounder. 
%After that, we use Gurobi to solve the optimization problem, and the result shows there are 4 non-zero coefficients corresponding to acrylamide, vitamin c, cadmium, and urinary thiocyanate. Finally the 2-step mixed ridge regression gives estimates of 
% 0.63, -0.31, 0.76, 0.40 (old using 1/(p-k))
%0.64, -0.36, 0.76, 0.41
%respectively for the causal parameters of these 4 outcomes. 

\section{Discussion}

Our identification results in this paper show that 
the parallel-outcome framework provides a promising alternative to the multi-cause framework \citep{wang2018blessings}; the latter has attracted a lot of attention recently in the causal inference and machine learning community. It is also related
to, but conceptually different from existing methods based on secondary outcomes such as negative controls or proximal causal inference. In particular, the roles of $Y^{(1)}, \ldots, Y^{(p)}$ in the definition of parallel outcomes are symmetric, which
allows our method to be applied in a wide range of practical settings where multiple parallel outcomes are available.  In fact, our framework has been implicitly assumed in genetic applications for multiple hypothesis testing with latent confounding. As pointed out by \cite{leek2008general}, this problem can be represented by a linear model relating a multivariate response to an exposure and latent factors. They also assume that the errors in the linear model are normally distributed with independent noises, which implies the parallel outcome structure defined in Assumption \ref{assumption:shared-confounding}. 

In related work, \cite{wang2017confounder} further assume that the latent
confounding factors may be represented as a linear transformation of the exposure variable plus a Gaussian error. They develop an approach to identify the direct effect from an exposure to
multiple outcomes in the presence of unmeasured mediators.    Their model and ours can be seen as reparametrizations of each other. For example, when $r=1$, equation \eqref{eq:X_multi}
% $
%     X = \alpha_X U + \epsilon_X
% $
% where $\textrm{Var}(U)=1$, $\textrm{Cov}(U,\epsilon_X)=0$, 
  can be rewritten as
$
    U = a_X X + \epsilon_U,
$
where $a_X = \alpha_X(\alpha^2_X + \sigma^2_X)^{-1}$, $\epsilon_U = (\alpha^2_X + \sigma^2_X)^{-1}\sigma^2_X U - (\alpha^2_X + \sigma^2_X)^{-1}\alpha_X\epsilon_X$, and $\textrm{Cov}(X,\epsilon_U)=0$. Similarly, their model can be rewritten in the form of equation \eqref{eq:X_multi}.
% They assume conditions on sparsity and submatrix rank to help identify $\beta_j$. Their method originates from .... Our approach is an innovative one involving factor analysis, optimization, and negative control methods. Both approaches assume similar conditions on submatrix rank, while one advantage of our approach is that Condition \ref{asp:beta_sparsity} about sparsity is weaker than their sparsity assumption, as we allow the number of non-zero $\beta_j$'s to be approximately twice of that required by their assumption. 
The key difference between \cite{wang2017confounder}'s  and our parametric identification results in Section \ref{sec:multi_identification} is that we require a much weaker sparsity condition than \cite{wang2017confounder}. In particular, \cite{wang2017confounder} assume $\Vert \beta \Vert_0  \leq (p-r)/2$, while our Condition \ref{asp:beta_sparsity} requires that $\Vert \beta \Vert_0  < p-r$. This improvement in sparsity is possible due to a novel identification approach  we develop. 

The parallel-outcome assumption \ref{assumption:shared-confounding} might be violated if there are common latent mediators $M$ for the exposure effect on the outcomes. In this case, the causal effects are still identifiable if one assumes linear structural equation models on the relationships among $U,X,M,Y$; see the Supplementary Material,  $\S$\ref{sec:extension_mediators} for   detailed discussions.

\section*{Supplementary material}
\label{SM}
%Further material such as technical details, extended proofs, code, or additional  simulations, figures and examples may appear online, and should be briefly mentioned as Supplementary Material where appropriate.  
% Please submit any such content as a PDF file along with your paper, entitled `Supplementary material for Title-of-paper'.  After the acknowledgements, include a section `Supplementary material' in your paper, with the sentence `Supplementary material available at \Bka\ online includes $\ldots$', giving a brief indication of what is available.  
%However it should be possible to read and understand the paper without reading the supplementary material.

%Further instructions will be given when a paper is accepted.
  The Supplementary Material contains nonparametric identification results in the continuous model, and  methods and simulations for nonparametric estimation in the discrete model. It also contains additional examples and proofs of all the theorems and lemmas. 

\bibliographystyle{apalike}
\bibliography{references2}

\makeatletter\@input{yy.tex}\makeatother
\end{document}

\def\spacingset#1{\renewcommand{\baselinestretch}%
{#1}\small\normalsize} \spacingset{1}

%%%%%%%%%%%%%%%%%%%%%%%%%%%%%%%%%%%%%%%%%%%%%%%%%%%%%%%%%%%%%%%%%%%%%%%%%%%%%%

\if1\blind
{
  \title{\bf Supplementary Material for ``The Promises of Parallel Outcomes''}
  \author{Ying Zhou,
%   \thanks{
%     The authors gratefully acknowledge \textit{please remember to list all relevant funding sources in the unblinded version}}\hspace{.2cm}\\
    Dingke Tang,
    Dehan Kong, 
    Linbo Wang \\
    {\small \itshape Department of Statistical Sciences, University of Toronto, 700 University Avenue, Toronto,}\\
    {\small \itshape Ontario M5G 1X6, Canada}\\
    {\small yingx.zhou@mail.utoronto.ca, dingke.tang@mail.utoronto.ca, dehan.kong@utoronto.ca,}\\ {\small linbo.wang@utoronto.ca}}
  \maketitle
} \fi

\if0\blind
{
  \bigskip
  \bigskip
  \bigskip
  \begin{center}
    {\LARGE\bf The Promises of Parallel Outcomes}
\end{center}
  \medskip
} \fi

\bigskip

\begin{abstract}
    % The Supplementary Material contains methods for non-parametric estimation in the discrete model, non-parametric identification  in the continuous model, and simulations. It also contains an example that shows the non-identifiability with a categorical latent confounder and two parallel outcomes, an example that explains Condition \ref{condition:no-interaction2}, argumentation on Gaussian linear models, additional causal diagrams,  proofs of theorems and lemmas,  discussions on latent mediators, derivation of the threshold, description of outcomes in the real data application,  and discussions on  bounded completeness and injectivity. 
    The Supplementary Material is organized as follows. Section \ref{sec:identification_continuous} contains non-parametric identification results in the continuous model. Section \ref{sec:categorical_counterexample} contains an example that shows that the causal effects are not identifiable with a categorical latent confounder and two parallel outcomes. Section \ref{sec:additional_causal_diagrams} includes two causal diagrams for related works. In Section \ref{example_condition4}, we use an example to show local identifiability without Condition \ref{condition:no-interaction2} in the categorical model.
    Sections \ref{section:categorical-estimation} and \ref{sec:simulation} contain methods and simulations for non-parametric estimation in the discrete model. 
    In Section \ref{sec:proof_gaussian_case}, we explain why Condition \ref{condition:continuous_indexing} is violated in a linear Gaussian model.
    We derive the threshold in \eqref{eq:optimization_5} in Section \ref{threshold:derivation}, give the reformulation of \eqref{eq:optimization_5} to a mixed integer programming problem in Section \ref{sec:mip}, and provide a description of the outcomes in the real data example in Section \ref{sec:description_outcomes}. Sections \ref{sec:proof_thm1}-\ref{sec:proof_thm2} contain proofs of theorems and lemmas in the main text. Section \ref{sec:extension_mediators} 
   extends models \eqref{eq:X_multi} and \eqref{eq:Y_multi} to accomodate common latent mediators among outcomes. 
   %We discuss injectivity and bounded completeness, concepts that used in \tteal{non-parametric identification in the continuous model (Section \ref{sec:identification_continuous})} \tblue{???} in Section \ref{injectivity_bddcomplete}. 
   Sections \ref{sec:lemmas} and \ref{sec:lemma_proofs} contain additional lemmas and related proofs, and Sections \ref{sec:thm_s1_proof} and \ref{sec:thm_s2_proof} contain proofs of Theorem \ref{thm:continuous} and Theorem \ref{thm:categorical_m_est}.
\end{abstract}

%\vfill
%\clearpage

% \vfill

% \newpage

\spacingset{1.5} % DON'T change the spacing!

\section{Non-parametric identification with a continuous latent confounder}
\label{sec:identification_continuous}
\subsection{Main identification results}

In most empirical studies, the latent confounder $U$ is continuous. In parallel to the eigendecomposition of matrix  $P(y^{(1)},Y^{(2)},Y^{(3)}\mid x)P(Y^{(2)},Y^{(3)}\mid x)^{-1}$, we use spectral decomposition of the corresponding integral operator to identify the mean potential outcomes $E\{Y^{(j)}(X=x)\},j=1,2,3$. 
Specifically, let $\mathcal{Y}_1$, $\mathcal{Y}_2$, $\mathcal{Y}_3$, $\mathcal{U}$, $\mathcal{X}$ denote the support of  random variables $\Yone$, $\Ytwo$, $\Ythree$, $U$, and $X$, respectively. Similar to Theorem \ref{thm:categorical}, we do not place restrictions on the cardinality of the exposure $X$, but assume that outcomes $\Yone$, $\Ytwo$, $\Ythree$ are  continuous random variables. 
Throughout this section, we assume all the density functions are bounded, which is formally stated as below \citep[A1]{hu_instrumental_2008}.
% \st{The following conditions assume that the density functions are all bounded.}
% \tblue{This is a common condition? Any elaboration? Citation here?}

% \tblue{The discrete case satisfies the boundedness condition automatically?} \tred{Not for U and X?}

\begin{condition} \label{condition:bounded_density}
The joint distribution of $X$ and $U,\Yone,\Ytwo,\Ythree$ admits a bounded density with respect to the product measure of some dominating measure $\mu$ defined on $\mathcal{X}$, and the Lebesgue measure on $\mathcal{Y}^{(1)}\times\mathcal{Y}^{(2)}\times\mathcal{Y}^{(3)}\times\mathcal{U}$. % All marginal and conditional densities \tblue{used in this section?} are also bounded.
For any $A,B \subset \{U,\Yone, \Ytwo, \Ythree\}$ such that $A \cap B=\emptyset$ and $x\in\mathcal{X}$, the conditional density $f(A\mid B,X=x)$ is bounded. 
% The support $\mathcal{U}$ is bounded.  
% \tblue{joint density bounded does not imply marginal density bounded, nor conditional density bounded }\\
% \tblue{Revised condition: $\forall A,B \subset \{U,\Yone, \Ytwo, \Ythree\}$, $A \cap B=\emptyset$, . The support $\mathcal{U}$ is bounded.}
% The joint density of $(U,\Yone, \Ytwo,\Ythree)$ admits a bounded density with respect to the Lebesgue measure on $\mathcal{Y}^{(1)}\times\mathcal{Y}^{(2)}\times\mathcal{Y}^{(3)}\times\mathcal{U}$. All marginal and conditional densities are also bounded.
\end{condition}

% \begin{definition}
% Let $a$ and $b$ denote random variables with respective supports $\mathcal{A}$ and $\mathcal{B}$. Given two corresponding spaces $\mathcal{G}(\mathcal{A})$ and $\mathcal{G}(\mathcal{B})$ of functions with domains $\mathcal{A}$ and $\mathcal{B}$, respectively, let $T_{f(a,b)}$ denote the operator mapping $g$ to $T_{f(a,b)}g$ defined by
% \begin{equation*}
%     [T_{f(a,b)}g](b)=\int_{\mathcal{A}}f(a,b)g(a)\dif a,
% \end{equation*}
% where $f(a,b)$ is the kernel of the operator.
% \end{definition}

%\tblue{Organization of this part: Why separate Condition S1 from others? Why do you introduce these notation? The writing is too dry!!! You need to provide motivation for what you are doing, whether it be introduction of a notation as here, or conditions, or a procedure. This is writing a paper, not writing a homework!!! Why do you introduce all these notations here?} 
To facilitate the statement of our results, we first introduce the concept of an integral operator.
Let $\mathcal{L}^1_{\textrm{bnd}}(\mathcal{A})$ be the set of all bounded absolutely integrable function over the support $\mathcal{A}$. An integral operator denoted by $T_{k(a,b)}$ is an operator mapping $g\in\mathcal{L}^1_{\textrm{bnd}}(\mathcal{B})$ to $T_{k(a,b)}g\in\mathcal{L}^1_{\textrm{bnd}}(\mathcal{A})$, defined as $(T_{k(a,b)}g)(a)=\int_{\mathcal{B}}k(a,b)g(b)\dif b$; here $k(a,b)$ is called the kernel of the operator.  This generalizes matrix multiplication in the discrete case. Similar to matrix transpose, we let $(T_{k(a,b)^\T}g)(b)=\int_{\mathcal{A}}k(a,b)g(a)\dif a$.
In parallel to \eqref{eq:mp2} and \eqref{eq:mq2}, we have 
% \tred{note the notation change} %the relation among $(U,X,\Yone,\Ytwo,\Ythree)$ reflected in
\begin{align}
    T_{f(\Ytwo,\Ythree\mid x)} &= T_{f(\Ytwo\mid U,x)} \Delta_{f(U\mid x)}T_{f(\Ythree\mid U,x)^\T}, \label{eq:op1}\\
    T_{f(\yone,\Ytwo,\Ythree\mid x)} &= T_{f(\Ytwo\mid U,x)}\Delta_{f(\yone\mid U,x)} \Delta_{f(U\mid x)}T_{f(\Ythree\mid U,x)^\T}. \label{eq:op2}
\end{align}
The multiplication operators in \eqref{eq:op1} and \eqref{eq:op2}, defined as $
   \Delta_{m(U)}g(U)=m(U)g(U)
$, 
generalize the diagonal matrices in \eqref{eq:mp2} and \eqref{eq:mq2}. 
% If there is a unique integral operator $T$ such that
% \begin{equation*}
%     T_{\pr(\Yone,\Ytwo,\Ythree\mid X)(\ytwo,\ythree)} = T T_{\pr(\Ytwo,\Ythree\mid X)(\ytwo,\ythree)} ,
% \end{equation*}
% then we have $T=T_{\pr(\Ytwo\mid U,X)(\ytwo,u)}\Delta_{\pr(\Yone\mid U,X)(u)}T_{\pr(\Ytwo\mid U,X)(\ytwo,u)}^{-1}$. This is the spectral decomposition of the operator $T_{\pr(\Yone,\Ytwo,\Ythree\mid X)(\ytwo,\ythree)} T_{\pr(\Ytwo,\Ythree\mid X)(\ytwo,\ythree)} ^{-1}$. Theorem \ref{thm:continuous} summarizes the identification in this case. The proof is in the Supplemental Material.
%If the inverse of $T_{f(\Ytwo,\Ythree\mid X)(\ytwo,\ythree)}$ is properly defined over the range of $T_{f(\Ytwo,\Ythree\mid X)(\ytwo,\ythree)}$, 
Under Condition \ref{condition:injectivity} below, we have %can arrive at the spectral decomposition of the known operator on the left hand side, %which acts as matrix eigendecomposition in Section \ref{sec:categorical case},
\begin{align}
\label{eqn:spectral}
T_{f(\yone,\Ytwo,\Ythree\mid x)} T_{f(\Ytwo,\Ythree\mid x)} ^{-1} 
= T_{f(\Ytwo\mid U,x)}\Delta_{f(\yone\mid U,x)}T_{f(\Ytwo\mid U,x)}^{-1}.
\end{align}
%This is the spectral decomposition of the operator  $T$. Under certain conditions, 
By the spectral decomposition of the left-hand side of \eqref{condition:injectivity}, under Conditions {\ref{condition:no_degeneracy} and \ref{condition:continuous_indexing}} below,  we can identify $E\{Y^{(j)}(x)\}, j=1,2,3$  in a similar way as in Section  \ref{sec:categorical case}.

% Our identification result is presented in the following Theorem \ref{thm:continuous}, where the detailed proofs are deferred to the Supplementary Material. 

\begin{condition} \label{condition:injectivity}
    % The operators $T_{f(\Ytwo\mid U,x)}$, $T_{f(\Ytwo,\Ythree\mid x)}$ and $T_{f(\Ythree\mid \Ytwo,x)}$ are injective; here an operator $T_{k(a,b)}$ is said to be injective if its inverse is defined over the range of   $T_{k(a,b)}$.
    $f(\Ytwo\mid U,x)$ is bounded complete in $\Ytwo$, $f(\Ytwo,\Ythree\mid x)$ is bounded complete in $\Ytwo$, and $f(\Ythree\mid \Ytwo,x)$ is bounded complete in $\Ythree$; here a function $k(a,b)$ is said to be bounded complete in $a$ if for all $f(b)\in\mathcal{L}^1_{\textrm{bnd}}$, $\int k(a,b)f(b)\dif b=0$ implies $f(b)=0$. 
    
 \end{condition}
 \begin{condition} \label{condition:no_degeneracy}
 %For all $u_1,u_2\in \mathcal{U}$, the set $\{\yone: f(\Yone\mid U=u_1,X)\ne f(\Yone\mid U=u_2,X)$ has positive probability under the marginal of $\Yone$ whenever $u_1\ne u_2$.
If there exists $x$ such that $\Yone\mid u_1,x \stackrel{\mathclap{{d}}}{=} \Yone\mid u_2,x$, then $u_1=u_2$. 
% \tred{for one x?} %\footnotemark 
 \end{condition}
 \begin{condition} \label{condition:continuous_indexing}
    There exists a known continuous functional $M$ such that $M\{f(\Ytwo\mid U,X)(u,x)\}= g^{-1}\{h_1(u)+h_2(x)\}$, % or $M[f_{\Ythree\mid UX}(\cdot\mid u,x)]= h_1(u)h_2(x)$ for all $u\in\mathcal{U}$ and $x\in\mathcal{X}$, 
where $h_1$ is one-to-one, $h_1(u)$ is bounded from above or below for $u\in\mathcal{U}$, and $g$ is a known continuous link function. %e.g., identity, log, etc.
\end{condition}

\begin{theorem} \label{thm:continuous}
Suppose that the latent confounder $U$ and parallel outcomes $\Yone, Y^{(2)}, Y^{(3)}$ are all continuous variables.  Suppose further that  Assumptions \ref{assumption:ignorability}, \ref{assumption:shared-confounding}, and Conditions \ref{condition:bounded_density}-\ref{condition:continuous_indexing} hold.
% Then for all $x$, the potential outcome distributions $f\{y^{(j)}(x)\}, j=1,2,3$ are identifiable.
Then for all $x$, the mean potential outcomes $E\{Y^{(j)}(X=x)\}, j=1,2,3$ are identifiable.
\end{theorem}

%\footnotetext{Condition \ref{condition:no_degeneracy} is similar to assumption 4 in \cite{hu_instrumental_2008}. Condition \ref{condition:injectivity} and \ref{condition:continuous_indexing} are different from assumption 3 and 5 in \cite{hu_instrumental_2008}.}

% Theorem \ref{thm:continuous} summarizes the identification in this case. The proof is in the Supplemental Material.

% The corresponding relationship
% \begin{table}[h]
%     \centering
%     \begin{tabular}{c|c|c|c}
%       $X^*$  & $Y$ & $X$ & $Z$ \\
%         $U$ & $\Yone$ & $\Ytwo$ & $\Ythree$
%     \end{tabular}
%     \caption{Hu(2008) v.s. here}
%     \label{tab:my_label}
% \end{table}

The bounded completeness assumption in Condition \ref{condition:injectivity} generalizes  the full rank condition \ref{condition:full-rank}; this assumption was often used for nonparametric identification in the causal inference literature \citep[e.g.][]{miao_identifying_2018,yang2019causal}.  
% It implies the corresponding operators are injective;  see the Supplementary Material for more discussions. 
If the kernel $k(a,b)$ of an operator $T_{k(a,b)}$ is bounded complete, then $T_{k(a,b)}$ is injective. Hence it plays a similar role to the full rank condition in the discrete case; 
see $\S$\ref{injectivity_bddcomplete} for more discussions. 
%\tblue{???????? Please read this carefully!!!!!! Read the part one more time when you move this in the paper. } 

% \tred{ 
% Unlike the discrete case, when $U$ takes value in the real line, even if we can code $U$ such that for all $x$, $f(y^{(1)}\mid u,x)$ is strictly increasing in $u,$ it is in general not possible to identify $f(y^{(1)}\mid u,x)$ for a specific $u$ from the set $\{f(y^{(1)}\mid u,x): u \in \mathcal{U}\}.$ This is because there may be infinitely many increasing one-to-one mappings from $\mathcal{U}$ to $\mathcal{U}$. To identify $f(y^{(1)}\mid u,x)$ for a specific $u,$ we instead employ Conditions \ref{condition:no_degeneracy} and \ref{condition:continuous_indexing}.  Condition \ref{condition:no_degeneracy} is similar to the non-zero part in Condition \ref{condition:no-interaction2}. Condition \ref{condition:continuous_indexing} assumes that $U$ and $X$ do not interact in causing $Y^{(2)}$ on an arbitrary known scale defined by $g$ and $M$. For example, Condition \ref{condition:continuous_indexing} holds if 
% $
%       E(Y^{(2)}\mid U,X) = \alpha U + h_2(X),
%  $
%  or $\log\{\textrm{var}(Y^{(2)}\mid U,X)\} = \alpha U + h_2(X),$
%  where $\alpha\neq 0$ and $h_2$ is an arbitrary function of $X.$  
% }

Recall that in the discrete case, 
we can code $U$ such that $\pr(Y^{(1)}=1\mid u, x)$ is strictly increasing in $u$ for a fixed $x$. Condition \ref{condition:no-interaction2} is used to ensure that $\pr(Y^{(1)}=1\mid u, x)$ is strictly increasing in $u$ for all values of $x$,
%order the values of $u$ in a consistent way across different values of $x$. Condition \ref{condition:no-interaction2} is used to ensure the ascending order of  $\pr(Y^{(1)}=1\mid u, x), u=1,\ldots,k$ is the same regardless of the value of $x$, 
so that we can code $U$ in a consistant way across different values of $x$. 
% \kong{The previoius sentence seems vague. Use mathematical representation to rewrite it.} 
Otherwise we cannot apply the formula $\pr(u_i)=E_X\{\pr(u_i\mid x)\}$ to identify $\pr(u_i)$, which is needed in identifying the potential outcome distributions by the g-formula $\pr\{y^{(j)}(x)\}=\sum_{u}\pr\{y^{(j)}\mid x,u\}\pr(u),j=1,2,3.$
% We assume the descending order of $\pr(y^{(1)}\mid U,x), U=u_1,\ldots,u_k$ is the same for all values of $x$. In this way, the $j$th largest value of $\pr(y^{(1)}\mid U,x), U=u_1,\ldots,u_k$ always corresponds to a certain $u_s$, regardless of the value of $x$. So we can identify $\pr(u_s)=E_X(\pr(u_s\mid x))$, where $\pr(u_s\mid x)$ is solved from $P(Y^{(2)}\mid x) = P(Y^{(2)}\mid U,x)P(U\mid x)$
When $U$ takes value in the real line, the eigenvalues $f(y^{(1)}\mid U,x)$ are continuous functions of $U$. Even if we can code $U$ such that for all $x$, $f(y^{(1)}\mid u,x)$ is strictly increasing in $u,$ it is in general not possible to identify $f(y^{(1)}\mid u,x)$ for a specific $u$ from the set $\{f(y^{(1)}\mid u,x): u \in \mathcal{U}\}.$ This is because there may be infinitely many increasing one-to-one mappings from $\mathcal{U}$ to $\mathcal{U}$. To identify $f(y^{(1)}\mid u,x)$ for a specific $u,$ we instead employ Conditions \ref{condition:no_degeneracy} and \ref{condition:continuous_indexing}.

Similar to Condition \ref{condition:no-interaction2}, which is used to identify the distribution of $U$ up to relabelling in the discrete case, Conditions \ref{condition:no_degeneracy} and \ref{condition:continuous_indexing} are used to identify the distribution of a continuous variable $\widetilde{U}$, which is a one-to-one mapping of $U$. Condition S3 is similar to the non-zero part in Condition \ref{condition:no-interaction2}. Unlike Condition \ref{condition:no-interaction2} which is about the eigenvalues $\{f(y^{(1)}\mid u,x):u\in\mathcal{U}\}$, Condition \ref{condition:continuous_indexing} is about the eigenfunctions $\{e_u(y^{(2)}):u\in\mathcal{U}\}$, where $e_u: y^{(2)}\in\mathcal{Y}^{(2)}\mapsto f(y^{(2)}\mid u,x)\in \real_+$.  We aim to index each eigenfunction to a specific $u$ in a way that the index is not affected by $x$,
and Condition \ref{condition:continuous_indexing} serves this purpose. It assumes the effect of  $U$  on $Y^{(2)}$ is bounded (from above or below) on a known scale, and $U$ has no effect modification on the relationship between $X$ and $Y^{(2)}$ on the same scale.
%It assumes that $U$ has a limited influence on $Y^{(2)}$, and has no effect modification on the relationship between $X$ and $Y^{(2)}$, defined by $g$ and $M$. 
This is stronger than Condition \ref{condition:no-interaction2} for the discrete case, which only requires the direction of the confounding effect of $U$ on $Y^{(1)}$ is the same for all values of $x$. For example, Condition \ref{condition:continuous_indexing} holds if 
$
      E(Y^{(2)}\mid U,X) = h_1(U) + h_2(X),
 $
 or $\log\{\textrm{var}(Y^{(2)}\mid U,X)\} = h_1(U) + h_2(X),$
 where $h_1$ is one-to-one, $h_1(\mathcal{U})$ is bounded, and $h_2$ is an arbitrary function of $X.$

\subsection{Discussion on injectivity and bounded completeness}\label{injectivity_bddcomplete}

% \subsection{Completeness and injectivity}
% We say that a distribution of $(a,b)$ with density $k(a,b)$ is complete in $a$ if for all measurable function $f(b)$ with finite expectations, $\int_{\mathcal{B}}k(a,b)f(b)\dif b=0$ implies $f(b)=0$. 

% We say that a distribution of $(a,b)$ with density $k(a,b)$ is bounded complete in $a$ if for all bounded measurable function $f(b)$, $\int_{\mathcal{B}}k(a,b)f(b)\dif b=0$ implies $f(b)=0$.
Let $\mathcal{L}^1$ be the set of all absolutely integrable function over the support, and $\mathcal{L}^1_{\textrm{bnd}}$ be the set of bounded functions in $\mathcal{L}^1$.
%We say that a function $k(a,b)$ is complete in $a$ if for all $f(b)\in\mathcal{L}^1$, $\int_{\mathcal{B}}k(a,b)f(b)\dif b=0$ implies $f(b)=0$.
We say that a function $k(a,b)$ is bounded complete in $a$ if for all $f(b)\in\mathcal{L}^1_{\textrm{bnd}}$, $\int_{\mathcal{B}}k(a,b)f(b)\dif b=0$ implies $f(b)=0$.
%By definition, completeness implies bounded completeness. 
We say that a linear operator $T_{k(a,b)}$ is injective as a mapping from $\mathcal{G}(\mathcal{B})$ to $\mathcal{G}(\mathcal{A})$ if $(T_{k(a,b)}f)(a)=(T_{k(a,b)}g)(a)$ implies $f=g$.

\begin{proposition}
\label{prop:complete}
%When $\mathcal{G}=\mathcal{L}^1$, $T_{k(a,b)}$ is injective if and only its kernel $k(a,b)$ is complete in $a$. 
When $\mathcal{G}=\mathcal{L}^1_{\textrm{bnd}}$, $T_{k(a,b)}$ is injective if and only its kernel $k(a,b)$ is bounded complete in $a$. 
\end{proposition}

\begin{proof}[Proof of Proposition \ref{prop:complete}]
\begin{align*}
& T_{k(a,b)}\textrm{ is injective}\\
\Longleftrightarrow\quad & (T_{k(a,b)}f)(a)=(T_{k(a,b)}g)(a)\Rightarrow f=g \\
\Longleftrightarrow\quad &[T_{k(a,b)}(f-g)](a)=0 \Rightarrow f-g=0 \\
\Longleftrightarrow\quad & k(a,b)\textrm{ is bounded complete in }a.
\end{align*}
\end{proof}

%\subsection{Completeness and exponential family}\footnotemark
% \subsection[Completeness and exponential family]{Completeness and exponential family\footnotemark}

Another closely related concept is completeness. We say that a function $k(a,b)$ is complete in $a$ if for all $f(b)\in\mathcal{L}^1$, $\int_{\mathcal{B}}k(a,b)f(b)\dif b=0$ implies $f(b)=0$.
By definition, completeness implies bounded completeness. Completeness has been studied and used in literature. For example, exponential families are known to be complete, see Lemma \ref{lem:exponential_completeness}. We would like to mention that, usually exponentially family is defined as the distribution with the probability density function (or probability mass function) of the form $f_X(x\mid \theta)=\psi(\theta)h(x)\exp\{\eta(\theta)^\T\lambda(x)\}$, where $\theta$ is the parameter. In our context, the conditional distribution $f_{a\mid b}(a\mid b;\theta)$ is called exponential family in the sense that $b$ is considered as the parameter, while $\theta$ is taken as fixed.

% The integral operator $\int f_{b\mid a}(b\mid a)g(a)\dif a$
% \begin{align*}
%   & \int f_{b\mid a}(b\mid a)g(a)\dif a \\
%     = &\int \frac{f_{a,b}(a,b)}{f_a(a)}g(a)\dif a\\
%     = &f_b(b)\int \frac{f_{a,b}(a,b)}{f_b(b)}g(a)/f_a(a)\dif a\\
%     = &f_b(b)\int f_{a\mid b}(a\mid b) g(a)f^{-1}_a(a)\dif a
% \end{align*}

% Similarly, if $k(b,a)=f_{a\mid b}(a\mid b)=\psi(b)h(a)\exp\{\eta(b)\lambda(a)\}$ satisfies (i) $h(a)>0$, (ii) $\lambda(a)$ is one-to-one in $a$, (iii) the range of $\eta(b)$ contains an open set. Then $k(b,a)$ is complete in $b$, and the integral operator $[T_{k(b,a)}g](b):=\int f_{b\mid a}(b\mid a)g(a)\dif a$ is injective.

\begin{remark}
Suppose that 
\begin{align*}
f_{b\mid a}(b\mid a;\theta_1)=\psi_1(a;\theta_1)h_1(b;\theta_1)\exp\{\eta_1(a;\theta_1)^\T\lambda_1(b;\theta_1)\},
% f_a(a;\theta_2)&=\psi_2(\theta_2)h_2(a)\exp\{\eta_2(\theta_2)^\T\lambda_2(a)\}
\end{align*}
then the joint density of $(a,b)$ has the form 
% \begin{align*}
% f_{a,b}(a,b;\theta_1,\theta_2)&=f_{b\mid a}(b|a;\theta_1)f_a(a;\theta_2) \\
% &= h_2(a)\psi_1(a;\theta_1)h_1(b)\psi_2(\theta_2)\exp\{\eta_1(a;\theta_1)^\T\lambda_1(b)\}\exp\{\eta_2(\theta_2)^\T\lambda_2(a)\},
% \end{align*}
% If it holds that \\
% 1) $\psi_1(a;\theta_1)=s_1(a)t_1(\theta_1)$, $\eta_1(a;\theta_1)=s_2(a)t_2(\theta_1)$; \\
% 2) $\psi_1(a;\theta_1)>0$, $h_1(b)>0$, $\psi_2(\theta_2)>0$; \\
% 3) $\eta_1(a;\theta_1)$ is one-to-one in $a$, $\lambda_1(b)$ is one-to-one in $b$, $\eta_2(\theta_2)$ is one-to-one in $\theta_2$; \\
% 4) the range of $\lambda_1(b)$ and the range of $\lambda_2(a)$ both contain an open set. 
\begin{align*}
f_{a,b}(a,b) &= f_{b\mid a}(b\mid a;\theta_1)f_a(a;\theta_2) \\
&=f_a(a;\theta_2)\psi_1(a;\theta_1)h_1(b;\theta_1)\exp\{\eta_1(a;\theta_1)^\T\lambda_1(b;\theta_1)\}.
\end{align*}
By Lemma \ref{lem:exponential_completeness}, if it holds that\\
1) $\psi_1(a;\theta_1) >0$, $h_1(b;\theta_1) >0$; \\
2) $\eta_1(a;\theta_1)$ is one-to-one in $a$, $\lambda_1(b;\theta_1)$ is one-to-one in $b$; \\
3) $\eta_1(a;\theta_1)$ contains an open set, $\lambda_1(b;\theta_1)$ contains an open set.

% Then $T_{f_{a,b}(a,b)}$ and $T_{f_{b\mid a}(b,a)}$ both are injective. 
Then $f_{a,b}(a,b)$ is complete in $a$, and $f_{b\mid a}(b,a)$ is complete in $b$.
\end{remark}

\begin{remark}
There are not many families of distributions beyond the exponential family that are known to be complete. However, the weaker concept of bounded completeness has been studied and known to encompass a larger family of distributions. For example, the location family generated by an absolutely continuous distribution (with respect to Lebesgue measure) is bounded complete if and only if its characteristic function is zero free \citep{mattner_incomplete_1993}.
\end{remark}

\section{Counterexample for  non-parametric identification  with a categorical latent confounder and two outcomes considered in Section \ref{sec:categorical case}}
\label{sec:categorical_counterexample}

% In Section \ref{sec:categorical case}, we show that when treatment, unobserved confounder and three outcomes satisfy causal diagram in Fig. \ref{fig:DAG2}, one can identify the causal effects under some mild conditions. Here, we show that if one can only observe two outcomes instead of three, generally, one can not identify the causal effects. 

% Suppose $ (U, X, Y ^{(1)}, Y^{(2)}) $ are all binary. Since $U$ is unobserved, we can only observe the joint distribution of $(X, Y ^{(1)}, Y^{(2)})$. As all of them are binary, the joint distribution is determined by $2^3-1=7$ probabilities, i.e., $\pr(x=0,y^{(1)}=0,y^{(2)}=0)$, $\pr(x=0,y^{(1)}=0,y^{(2)}=1)$, $\pr(x=0,y^{(1)}=1,y^{(2)}=0)$, $\pr(x=0,y^{(1)}=1,y^{(2)}=1)$, $\pr(x=1,y^{(1)}=0,y^{(2)}=0)$, $\pr(x=1,y^{(1)}=0,y^{(2)}=1)$, $\pr(x=1,y^{(1)}=1,y^{(2)}=0)$. However, 
% %Suppose that we have two outcomes instead of three, and $(U,X,Y,Z)$ are all binary random variables. Since $U$ is unobserved, we have $2^3-1=7$ equations constructed from $\pr(x,y,z)$. 
% there are 11 unknown parameters: $\pr(u=0)$, $\pr(x=0\mid u=0)$, $\pr(x=0\mid u=1)$, $\pr(y^{(1)}=0\mid x=0,u=0)$, $\pr(y^{(1)}=0\mid x=0,u=1)$,
% $\pr(y^{(1)}=0\mid x=1,u=0)$, $\pr(y^{(1)}=0\mid x=1,u=1)$, $\pr(y^{(2)}=0\mid x=0,u=0)$, $\pr(y^{(2)}=0\mid x=0,u=1)$,
% $\pr(y^{(2)}=0\mid x=1,u=0)$, $\pr(y^{(2)}=0\mid x=1,u=1)$. Intuitively, we can not identify all these parameters. 

% Here, we give a counterexample. 

\setlength{\abovedisplayskip}{3pt}%
\setlength{\belowdisplayskip}{3pt}%
\setlength{\abovedisplayshortskip}{3pt}%
\setlength{\belowdisplayshortskip}{3pt}%
\setlength{\jot}{3pt}% Inter-equation spacing

% \tblue{Start with a description of what you try to show here.}
In this section, we show that  with a categorical latent confounder considered in Section \ref{sec:categorical case}, the potential outcome distribution cannot be identified with two outcomes. We will start with the binary model in which $U$, $\Yone$, $\Ytwo$ are binary variables, and then generalize to the case where $U$, $\Yone$, $\Ytwo$ have $k$ categories, where $k\ge 2, k\in \mathbb{Z}$. 

When there are two outcomes $Y^{(1)}$ and $Y^{(2)}$ in the binary model, the joint distribution of $(X,Y^{(1)},Y^{(2)}, U)$ is determined by 11 parameters, which we set as
\begin{align*}
    &\pr(U=1)=0.6,\quad\pr(X=1\mid U=1)=0.7,\quad\pr(X=1\mid U=2)=0.4, \\
    &\pr(Y^{(1)}=1\mid U=1,X=1)=0.1,\quad\pr(Y^{(1)}=1\mid U=2,X=1)=0.2,\\
    &\pr(Y^{(1)}=1\mid U=1,X=2)=0.3,\quad\pr(Y^{(1)}=1\mid U=2,X=2)=0.4,\\
    &\pr(Y^{(2)}=1\mid U=1,X=1)=0.1,\quad\pr(Y^{(2)}=1\mid U=2,X=1)=0.2,\\
    &\pr(Y^{(2)}=1\mid U=1,X=2)=0.3,\quad\pr(Y^{(2)}=1\mid U=2,X=2)=0.4.
\end{align*}
Then the joint distribution of $(X,Y^{(1)},Y^{(2)})$ is
\begin{align*}
    \pr(X=1,Y^{(1)}=1,Y^{(2)}=1)=0.0106,\quad\pr(X=1,Y^{(1)}=1,Y^{(2)}=2)=0.0634,\\
    \pr(X=1,Y^{(1)}=2,Y^{(2)}=1)=0.0634,\quad\pr(X=1,Y^{(1)}=2,Y^{(2)}=2)=0.4426,\\
    \pr(X=2,Y^{(1)}=1,Y^{(2)}=1)=0.0546,\quad\pr(X=2,Y^{(1)}=1,Y^{(2)}=2)=0.0954,\\
    \pr(X=2,Y^{(1)}=2,Y^{(2)}=1)=0.0954,\quad\pr(X=2,Y^{(1)}=2,Y^{(2)}=2)=0.1746.
\end{align*}
Given the observed joint distribution of $(X,Y^{(1)},Y^{(2)}) $, the values of $\pr(x)$, $\pr(\yone\mid x)$, $\pr(\ytwo\mid x)$, $\pr(\yone,\ytwo\mid x)$ are also determined. These values, also denoted by $ \pr(\cdot) $, will be used to construct another set of parameters which share the same observed distribution as the  case above.  

If we fix $X=1$, the conditional independence $\Yone \bigCI \Ytwo \mid (U,X=1)$ enables us to construct 3 equations by equating the observed distribution with expression of unknown parameters, i.e., 
% \tblue{this sentence does not make sense}
%$$
%\pr(\yone,\ytwo \mid X=1)=\sum_u\pr(\yone\mid u,X=1)\pr(\ytwo\mid u,X=1)\pr(u\mid X=1),
%$$
\begin{align*}
\pr(\Yone=1,\Ytwo=1 \mid X=1)=\sum_u\pr(\Yone=1\mid u,X=1)\pr(\Ytwo=1\mid u,X=1)\pr(u\mid X=1),\\
\pr(\Yone=1,\Ytwo=2 \mid X=1)=\sum_u\pr(\Yone=1\mid u,X=1)\pr(\Ytwo=2\mid u,X=1)\pr(u\mid X=1),\\
\pr(\Yone=2,\Ytwo=1 \mid X=1)=\sum_u\pr(\Yone=2\mid u,X=1)\pr(\Ytwo=1\mid u,X=1)\pr(u\mid X=1). 
\end{align*}
There are 5 unknown parameters: $\pr(U=1\mid X=1)$, $\pr(\Yone=1\mid U=1,X=1)$, $\pr(\Yone=1\mid U=2,X=1)$, $\pr(\Ytwo=1\mid U=1,X=1)$, and $\pr(\Ytwo=1\mid U=2,X=1)$. We will show that we can set $\pr(U=1\mid X=1)$ and $\pr(\Yone=1\mid U=1,X=1)$ arbitrarily, and solve the remaining 3 parameters from the 3 equations. This set of parameters is denoted by $\pr^*(\cdot)$ to distinguish them from the parameter values we set at the beginning. 

Let $\pr^*(U=1\mid X=1)=0.3$ and $\pr^*(Y^{(1)}=1\mid U=1,X=1)=0.4$, then
\begin{align*}
    &\pr^*(Y^{(1)}=1\mid U=2,X=1)\\
    =&\frac{\pr(Y^{(1)}=1\mid X=1)-\pr^*(Y^{(1)}=1\mid U=1,X=1)\pr^*(U=1\mid X=1)}{1-\pr^*(U=1\mid X=1)}\\
    %=&0.01083744,\\
    =&0.0108,\\
    &\pr^*(Y^{(2)}=1\mid U=2,X=1)\\
    =&\frac{\pr(Y^{(1)}=1,Y^{(2)}=1\mid X=1)-\pr(Y^{(2)}=1\mid X=1)\pr^*(Y^{(1)}=1\mid U=1.X=1)}{\{\pr^*(Y^{(1)}=1\mid U=2,X=1)-\pr^*(Y^{(1)}=1\mid U=1,X=1)\}\{1-\pr^*(U=1\mid X=1)\}}\\
    %=&0.1202532,\\
    =&0.1203,\\
    &\pr^*(Y^{(2)}=1\mid U=1,X=1)\\
    =&\frac{\pr(Y^{(2)}=1\mid X=1)-\pr^*(Y^{(2)}=1\mid U=2,X=1)\{1-\pr^*(U=1\mid X=1)\}}{\pr^*(U=1\mid X=1)}\\
    %=&0.1446966.
    =&0.1447.
\end{align*}
Now we repeat the process for $X=2$. Let $\pr^*(U=1\mid X=2)=0.2$ and $\pr^*(Y^{(1)}=1\mid U=1,X=2)=0.6$, then
\begin{align*}
    &\pr^*(Y^{(1)}=1\mid U=2,X=2)\\
    =&\frac{\pr(Y^{(1)}=1\mid X=2)-\pr^*(Y^{(1)}=1\mid U=1,X=2)\pr^*(U=1\mid X=2)}{1-\pr^*(U=1\mid X=2)}\\
    %=&0.2964286,\\
    =&0.2964,\\
    &\pr^*(Y^{(2)}=1\mid U=2,X=2)\\
    =&\frac{\pr(Y^{(1)}=1,Y^{(2)}=1\mid X=2)-\pr(Y^{(2)}=1\mid X=2)\pr^*(Y^{(1)}=1\mid U=1,X=2)}{\{\pr^*(Y^{(1)}=1\mid U=2,X=2)-\pr^*(Y^{(1)}=1\mid U=1,X=2)\}\{1-\pr^*(U=1\mid X=2)\}}\\
    %=&0.3470588,\\
    =&0.3471,\\
    &\pr^*(Y^{(2)}=1\mid U=1,X=2)\\
    =&\frac{\pr(Y^{(2)}=1\mid X=2)-\pr^*(Y^{(2)}=1\mid U=2,X=2)\{1-\pr^*(U=1\mid X=2)\}}{\pr^*(U=1\mid X=2)}\\
    %=&0.397479.
    =&0.3975.
\end{align*}
In addition, we have
\begin{align*}
&\pr^*(U=1)=\pr^*(U=1\mid X=1)\pr(X=1)+\pr^*(U=1\mid X=2)\pr(X=2)=0.258,\\
&\pr^*(X=1\mid U=1)=\frac{\pr^*(U=1\mid X=1)\pr(X=1)}{\pr^*(U=1)}%=0.6744186,\\
=0.6744,\\
&\pr^*(X=1\mid U=2)=\frac{\pr^*(U=2\mid X=1)\pr(X=1)}{1-\pr^*(U=1)}%=0.5471698.
=0.5472.
\end{align*}
To sum up, the parameter values
\begin{align*}
    &\pr^*(U=1)=0.258,\quad\pr^*(X=1\mid U=1)=0.6744, \quad\pr^*(X=1\mid U=2)=0.5472, \\
    &\pr^*(Y^{(1)}=1\mid U=1,X=1)=0.4,\quad\pr^*(Y^{(1)}=1\mid U=2,X=1)=0.0108,\\
    &\pr^*(Y^{(1)}=1\mid U=1,X=2)=0.6,\quad\pr^*(Y^{(1)}=1\mid U=2,X=2)=0.2964,\\
    &\pr^*(Y^{(2)}=1\mid U=1,X=1)=0.1447,\quad\pr^*(Y^{(2)}=1\mid U=2,X=1)=0.1203,\\
    &\pr^*(Y^{(2)}=1\mid U=1,X=2)=0.3975,\quad\pr^*(Y^{(2)}=1\mid U=2,X=2)=0.3471,
\end{align*}
give the same joint distribution of $(X, Y ^{(1)}, Y^{(2)})$ as the first set. However, these two sets of values indicate different causal effects. The first set gives $\pr\{Y^{(1)}(X=1)=1\}=0.14$ and $\pr\{Y^{(1)}(X=2)=1\}=0.34$, while the second set has $\pr^*\{Y^{(1)}(X=1)=1\}=0.111$ and $\pr^*\{Y^{(1)}(X=2)=1\}=0.375$.

The above construction of a counterexample can be generalized to the case where $Y^{(1)}$, $Y^{(2)}$, $U$ all have $k$ categories. 
% \tblue{Again, move sentence like this to the top. First state what you try to show!} 
In that case, we fix $X$ at a certain $x$, then we have $k^2-1$ equations by equating the observed $\pr(y^{(1)},y^{(2)}\mid x)$ with expression of parameters. The unknown parameters are $k-1$ of $\pr(u\mid x)$, $k(k-1)$ of $\pr(y^{(1)}\mid u,x)$, and $k(k-1)$ of $\pr(y^{(2)}\mid u,x)$. So there is a total of $(k-1)(2k+1)$ parameters. This leaves $(k-1)(2k+1)-(k^2-1)=k(k-1)$ degrees of freedom. Therefore, one can set $k(k-1)$ parameters freely. For example, we can first specify the values of $\pr^*(U=u\mid x),u=1,\ldots,k-1$, which takes $k-1$ degrees of freedom. Then, we assign the values of %all $\pr(y^{(1)}\mid u,x)$ except for those corresponding to a certain $u_0$, 
$\pr^*(Y^{(1)}=i\mid U=u,x),\ i=1,\ldots,k-1$; $u=1,\ldots,k-1$, which takes the remaining $(k-1)^2$ degrees of freedom.

After we specify the values of the $k(k-1)$ parameters, we can solve the remaining parameters. First, we solve $\pr^*(Y^{(1)}=i\mid U=k,x),i=1,\ldots,k-1$ from the following $k-1$ equations
$$
\pr(Y^{(1)}=i\mid x)=\sum_{u=1}^k\pr^*(Y^{(1)}=i\mid U=u,x)\pr^*(U=u\mid x),
$$
where $i=1, \ldots, k-1.$ We then solve $\pr^*(Y^{(2)}=j\mid U=u,x), j=1,\ldots,k-1;u=1\ldots,k$ from the following $k(k-1)$ equations
\begin{align*}
\pr(Y^{(1)}=i,Y^{(2)}=j \mid x)
= \sum_{u=1}^k\pr^*(Y^{(1)}=i\mid U=u,x)\pr^*(Y^{(2)}=j\mid U=u,x)\pr^*(U=u\mid x),
\end{align*}
where $i=1,\ldots,k;$ $j=1,\ldots,k-1$. This is a linear system with $k(k-1)$ equations and $k(k-1)$ parameters. It has a unique solution if the coefficient matrix is invertible. We can repeat the process for each $x$, and get
$
\pr^*(u) = \sum_x\pr^*(u\mid x)\pr(x)
$
as well as
$
\pr^*(x\mid u) = \pr^*(u\mid x)\pr(x)/\pr(u)^*.
$
In this way, we can find a set of parameters denoted by $\pr^*(\cdot)$ which is compatible with the observed distribution of $(X,Y^{(1)},Y^{(2)})$, but gives different potential outcome distributions $\pr\{y^{(1)}(x)\}$ and $\pr\{y^{(2)}(x)\}$ from the true values.

\section{Causal diagrams for related works}
\label{sec:additional_causal_diagrams}
\begin{figure}[H]
\centering
\scalebox{0.9}{
\begin{tikzpicture}[->,>=stealth',shorten >=1pt,auto,node distance=3.2cm,
     semithick, scale=0.50]
     pre/.style={-,>=stealth,semithick,blue,ultra thick,line width = 1.5pt}]
     \tikzstyle{every state}=[fill=none,draw=black,text=black]
     \node[est] (X)                    {$X$};
     \node[est] (Y) [right of =  X] {$Y$};
     \node[est] (Z) [above left =1cm of X] {$Z$};
     \node[shade] (U) [above right = 1cm of X] {$U$};
     \node[est] (W) [right of = U] {$W$};
     \path  
     (Z) edge node {} (X)
     (X) edge node {} (Y)
     (U) edge node {} (Z)
     (U) edge node {} (X)
     (U) edge node {} (W)
     (U) edge node {} (Y);
 \end{tikzpicture}}
 \caption{A causal diagram associated with the measurement error model of \cite{kuroki_measurement_2014}. They identified the effect of  exposure $X$ on outcome $Y$, with two  proxies of the latent confounder $U$, denoted as   $Z$ and   $W$. }\label{fig:pearl}
 \end{figure}
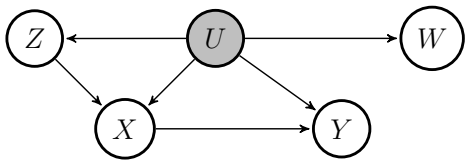

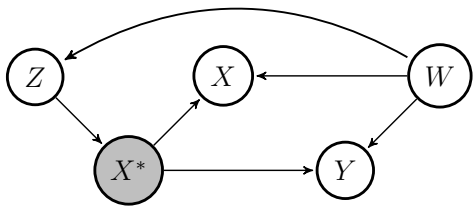
\begin{figure}[!h]
\centering
\scalebox{0.9}{
\begin{tikzpicture}[->,>=stealth',shorten >=1pt,auto,node distance=3.2cm,
     semithick, scale=0.50]
     pre/.style={-,>=stealth,semithick,blue,ultra thick,line width = 1.5pt}]
     \tikzstyle{every state}=[fill=none,draw=black,text=black]
     \node[shade] (U)                    {$X^{*}$};
     \node[est] (Y) [right of = U] {$Y$};
     \node[est] (Z) [above left =1cm of U] {$Z$};
     \node[est] (X) [above right = 1cm of U] {$X$};
     \node[est] (W) [right of = X] {$W$};
     \path  
     (U) edge node {} (Y)
     (U) edge node {} (X)
     (Z) edge node {} (U)
     (W) edge node {} (X)
     (W) edge node {} (Y);
     \path[pil, ->] (W) edge [bend right] node {} (Z);
 \end{tikzpicture}}
 \caption{A causal diagram associated with the measurement error model of \cite{hu2008identification}. They  identified the effect of a latent exposure $X^*$ on outcome $Y$, with an exposure proxy $X$, an instrument $Z$ and measured independent variables $W$. }\label{fig:hu}
 \end{figure}

Figure \ref{fig:pearl} is a causal diagram for the model in \cite{kuroki_measurement_2014}. Figure \ref{fig:hu} is a causal diagram for the model in \cite{hu2008identification}.
% \begin{figure}[!htbp]
% \centering

%  \caption{A causal diagram associated with a parallel-outcome model with single-outcome confounder $W$ when $p=3$. The baseline covariates $V$ is omitted for brevity. Variables $X, Y^{(1)}, Y^{(2)}, Y^{(3)}$ are observed; $U$ and $W$ are unobserved.  }\label{fig:DAG-single-outcome-confounding}
%  \end{figure}

%{\tred{why we need this figure? is was never refered by any paragraph (Dingke)} }

\section{Local identifiability without Condition \ref{condition:no-interaction2}}
\label{example_condition4}
% To see this, 
In this section, we use an example to show that in the categorical model in Section \ref{sec:categorical case}, local identifiability may be achieved without Condition \ref{condition:no-interaction2}, and Condition \ref{condition:no-interaction2} is then needed to achieve global identifiability. 

We consider the binary model as in Section \ref{sec:categorical case}.
Without Condition \ref{condition:no-interaction2}, it is possible that 
\begin{align*}
\text{Case I:}  \quad    \pr(y^{(1)}\mid U=1,X=1)<\pr(y^{(1)}\mid U=2,X=1) \\
    \pr(y^{(1)}\mid U=1,X=2)<\pr(y^{(1)}\mid U=2,X=2)
\end{align*}
or,
\begin{align*}
\text{Case II:}  \quad     \pr(y^{(1)}\mid U=1,X=1)>\pr(y^{(1)}\mid U=2,X=1) \\
    \pr(y^{(1)}\mid U=1,X=2)>\pr(y^{(1)}\mid U=2,X=2)
\end{align*}
or,
\begin{align*}
 \text{Case III:}  \quad   \pr(y^{(1)}\mid U=1,X=1)<\pr(y^{(1)}\mid U=2,X=1) \\
    \pr(y^{(1)}\mid U=1,X=2)>\pr(y^{(1)}\mid U=2,X=2)
\end{align*}
or,
\begin{align*}
\text{Case IV:}  \quad    \pr(y^{(1)}\mid U=1,X=1)>\pr(y^{(1)}\mid U=2,X=1) \\
    \pr(y^{(1)}\mid U=1,X=2)<\pr(y^{(1)}\mid U=2,X=2).
\end{align*}
Note that Case I and Case II are equivalent by permuting the labels of $U$, but they are not equivalent to Case III/IV.

More specifically, note that identification of causal effects is based on the following equations: \begin{align*}
    \pr(y^{(1)}\mid X=1) =& \pr(y^{(1)}\mid U=1, X=1)\pr(U=1\mid X=1) \\
    & \qquad + \pr(y^{(1)}\mid U=2, X=1)\{1-\pr(U=1\mid X=1)\} \\
    \pr(y^{(1)}\mid X=2) =& \pr(y^{(1)}\mid U=1, X=2)\pr(U=1\mid X=2) \\
    & \qquad + \pr(y^{(1)}\mid U=2, X=2)\{1-\pr(U=1\mid X=2)\}. \numberthis\label{eqn:iden}
\end{align*}

Suppose that $\{\pr(y^{(1)}\mid U=1,X=1),\pr(y^{(1)}\mid U=2,X=1)\} = \{0.1,0.6\},$ and $\{\pr(y^{(1)}\mid U=1,X=2),\pr(y^{(1)}\mid U=2,X=2)\} = \{0.4,0.8\},$ $\pr(y^{(1)}\mid X=1) =  \pr(y^{(1)}\mid X=2) = 0.5$ and $\pr(X=1)=0.5.$

Then we have under Case I, \eqref{eqn:iden} becomes
\begin{flalign*}
    0.5 &= 0.1 \pr(U=1\mid X=1) + 0.6 (1-\pr(U=1\mid X=1)) \\
    0.5 &= 0.4 \pr(U=1\mid X=2) + 0.8 (1-\pr(U=1\mid X=2)),
\end{flalign*}
so that $\pr(U=1\mid X=1) = 0.2$, $\pr(U=1\mid X=2) = 0.75$, $\pr(U=1) = 0.2 \times 0.5 + 0.75\times 0.5 = 0.475$, and 
$
    \pr\{y^{(1)}(1)\}  = \pr(y^{(1)}\mid X=1,U=1)\pr(U=1) + \pr(y^{(1)}\mid X=1,U=2)\pr(U=2) =  0.1 \times 0.475 + 0.6 \times 0.525 = 0.3625.
$

Under Case II, we have $\pr(U=1\mid X=1) = 0.8, \pr(U=1\mid X=2) = 0.25, \pr(U=1) = 0.8 \times 0.5 + 0.25\times 0.5 = 0.525,$ and 
$        \pr\{y^{(1)}(1)\}  =  0.6 \times 0.525 +  0.1 \times 0.475= 0.3625.
$

Under Case III, we can obtain similarly that $\pr(U=1\mid X=1) = 0.2, \pr(U=1\mid X=2) = 0.25, \pr(U=1) = 0.2 \times 0.5 + 0.25\times 0.5 = 0.225,$ and 
$        \pr\{y^{(1)}(1)\}  =  0.1 \times 0.225 + 0.6 \times 0.775 = 0.4875.
$

Under case IV, we have $\pr(U=1\mid X=1) = 0.8, \pr(U=1\mid X=2) = 0.75, \pr(U=1) = 0.8 \times 0.5 + 0.75\times 0.5 = 0.775,$ and 
$        \pr\{y^{(1)}(1)\}  = 0.6 \times 0.775 +  0.1 \times 0.225 = 0.4875.
$

One can now see clearly that there are two possible values for $\pr\{y^{(1)}(1)\}$, so $\pr\{y^{(1)}(1)\} $ is only locally identifiable without Condition \ref{condition:no-interaction2}. Since only Case I and Case II are allowed by Condition \ref{condition:no-interaction2}, and these two cases have the same value for $\pr\{y^{(1)}(1)\}$, the global identifiability is achieved under Condition \ref{condition:no-interaction2}.

\section{Non-parametric estimation under a categorical model} 
\label{section:categorical-estimation}

In this part, we consider the estimation problem assuming that both the exposure $X$ and parallel outcomes $\Yone, \Ytwo, \Ythree$ are categorical variables. 
Given identifiability, one may use the generalized least squares to estimate  the probabilities $\pr(y^{(j)}\mid u,x),j=1,2,3$, $\pr(x\mid u)$ and $\pr(u)$   from equations 
\begin{equation}
\label{eq:GLS}
\pr(y^{(1)},y^{(2)},y^{(3)},x) = \sum\limits_{u} \pr(y^{(1)}\mid u,x)\pr(y^{(2)}\mid u,x)\pr(y^{(3)}\mid u,x)\pr(x\mid u)\pr(u),
\end{equation}
 subject to the constraints that 
 \begin{equation}
 \label{eqn:constraint}
        \sum\limits_u \pr(u)=1, \sum\limits_x \pr(x\mid u)=1, \ \ \sum\limits_{y^{(j)}} \pr(y^{(j)}\mid u,x) = 1, \ \ \pr(y^{(1)}\mid u,x) \text{ is strictly increasing in } u.
 \end{equation}
The distributions $\pr\{y^{(j)}(x)\},j=1,2,3$ can then be estimated using plug-in estimators. 

Solving the generalized least squares problem in practice, however, can be challenging as the equations in \eqref{eq:GLS} are non-linear and involve a large number of parameters. As a result, the optimization problem may be very time-consuming and even worse, the results can be highly sensitive to the choice of starting values. To solve this problem, we propose to first follow the identification procedure in Section \ref{sec:categorical case}  and obtain  initial plug-in estimates of the probabilities $\pr(y^{(j)}\mid u,x),j=1,2,3$, $\pr(x\mid u)$ and $\pr(u)$. This step is computationally very efficient. We then use these initial estimates as warm starts  for the generalized least squares procedure. Our proposed estimating procedure is summarized in Algorithm \ref{algorithm:plug-in}.

%\afterpage{ 
%\clearpage
\begin{algorithm}[htbp]
\setstretch{1.1}
\caption{A generalized least squares procedure with warm starts for estimating the potential outcome distributions}
\label{algorithm:plug-in}
\begin{itemize}
    \item[1] For each $x$ and $y^{(1)}$:
    \begin{itemize}
        \item[1-1] Estimate $P(Y^{(2)},Y^{(3)}\mid x)$ and $P(y^{(1)},Y^{(2)},Y^{(3)}\mid x)$ by their empirical estimates, denoted by $\overline{P}(Y^{(2)},Y^{(3)}\mid x)$ and $\overline{P}(y^{(1)},Y^{(2)},Y^{(3)}\mid x)$;
        \item[1-2] Apply eigendecomposition on $\overline{P}(y^{(1)},Y^{(2)},Y^{(3)}\mid x)\overline{P}(Y^{(2)},Y^{(3)}\mid x)^{-1}$, and use its eigenvalues and associated normalized  eigenvectors (with $\ell_1$-norm 1) to construct $\widetilde{P}_D(y^{(1)}\mid U,x)$, $\widetilde{P}(Y^{(2)}\mid U,x)$ and $\widetilde{P}(Y^{(3)}\mid U,x)$ subject to the constraints in \eqref{eqn:constraint}; 
        \item[1-3] Estimate $P_D(U\mid x)$ by
                \begin{equation}
                \label{eq:M_est}
                \widetilde{P}_D(U\mid x)=\widetilde{P}(Y^{(2)}\mid U,x)^{-1}\overline{P}(Y^{(2)},Y^{(3)}\mid x)\{\widetilde{P}(Y^{(3)}\mid U,x)^\T\}^{-1};
                \end{equation}
    \end{itemize}
    % \item[2] For each $x$, find $\widehat{\pr}(u\mid x)$ and $\widehat{\pr}\{y^{(j)}\mid u,x\}$ that minimizes
    % $$
    %   \sum\limits_{y^{(1)},y^{(2)},y^{(3)}} \left\{\widetilde{\pr}(y^{(1)},y^{(2)},y^{(3)}\mid x) - \sum\limits_{u} \pr(y^{(1)}\mid u,x)\pr(y^{(2)}\mid u,x)\pr(y^{(3)}\mid u,x)\pr(u\mid x)\right\}^2
    % $$
    % subject to the constraint that 
    % $$
    %     \sum\limits_u \pr(u\mid x)=1, \ \ \sum\limits_{y^{(j)}} {\pr}\{y^{(j)}\mid u,x\} = 1, \ \ {\pr}\{y^{(1)}\mid u,x\} \text{ is strictly increasing in } u,
    % $$ 
    % using $\widetilde{\pr}(u\mid x)$ and $\widetilde{\pr}\{y^{(j)}\mid u,x\}$ as starting values;
    % \item[3] Estimate $\pr(u)$ by 
    % \begin{equation}
    %     \label{eq:pr_u_est}
    %     \widehat{\pr}(u)=\sum_x\widehat{\pr}(u\mid x)\widehat{\pr}(x),    
    % \end{equation}
    % where $\widehat{\pr}(x)$ is the empirical estimate of ${\pr}(x);$ 
    \item[2] Estimate $\pr(u)$ and $\pr(x \mid u)$ by 
    \begin{equation}
        \label{eq:pr_u_est}
        \widetilde{\pr}(u)=\sum_x\widetilde{\pr}(u\mid x)\overline{\pr}(x), \ \ 
        \widetilde{\pr}(x \mid u) = \widetilde{\pr}(u \mid x)\overline{\pr}(x)/\widetilde{\pr}(u),
    \end{equation}
    where $\overline{\pr}(x)$ is the empirical estimate of ${\pr}(x);$ 
    \item[3] Find $\widehat{\pr}(u)$, $\widehat{\pr}(x\mid u)$ and $\widehat{\pr}\{y^{(j)}\mid u,x\}$ that minimize
     \small
    $$
      \sum\limits_{x,y^{(1)},y^{(2)},y^{(3)}} \left\{\overline{\pr}(y^{(1)},y^{(2)},y^{(3)} \mid x)\overline{\pr}(x) - \sum\limits_{u} \pr(y^{(1)}\mid u,x)\pr(y^{(2)}\mid u,x)\pr(y^{(3)}\mid u,x)\pr(x\mid u)\pr(u)\right\}^2
    $$
    \normalsize
    subject to the constraints in \eqref{eqn:constraint},
    using $\widetilde{\pr}(u)$, $\widetilde{\pr}(x\mid u)$ and $\widetilde{\pr}(y^{(j)}\mid u,x)$ as starting values;
    
    \item[4] For each $x$, estimate the potential outcome distribution $\pr\{y^{(j)}(x)\},j=1,2,3$ by
    \begin{equation*}
    % \label{eq:categorical_causal_est}
        \widehat{\pr}\{y^{(j)}(x)\}=\sum\limits_u\widehat{\pr}(y^{(j)}\mid u,x)\widehat{\pr}(u).
    \end{equation*}
\end{itemize}
\end{algorithm}
%\thispagestyle{empty}
%\clearpage 
%} 

Theorem \ref{thm:categorical_m_est} summarizes the main theoretical properties of our proposed estimating procedure. In particular, it shows that the starting values $\widetilde{\pr}(u)$, $\widetilde{\pr}(x\mid u)$ and $\widetilde{\pr}(y^{(j)}\mid u,x)$ are root-n consistent for their population counterparts, and that the resulting estimator $\widehat{\pr}\{y^{(j)}(x)\}$ is consistent and asymptotically normal. A key step in the proof of the first statement is Lemma \ref{lm:perturbation}
%in the Supplementary Material
, which shows that  there exists an analytic continuation of eigendecomposition to its neighborhood. The second statement follows from standard M-estimation theory.

\begin{theorem}
\label{thm:categorical_m_est}
Under the conditions of Theorem \ref{thm:categorical} and standard regularity conditions, we have that if $X$ and $\Yone$ are also categorical variables, then \\
$1.\quad$ For each $u,x,y^{(j)},j=1,2,3$, $\widetilde{\pr}(u)-{\pr}(u)=O_p(n^{-1/2}),$ $\widetilde{\pr}(x\mid u)-{\pr}(x\mid u)=O_p(n^{-1/2}),$ $\widetilde{\pr}(y^{(j)}\mid u,x)-{\pr}(y^{(j)}\mid u,x)=O_p(n^{-1/2});$ \\
$2.\quad$ For each $x,y^{(j)},j=1,2,3$, $\sqrt{n}\left(\widehat{\pr}\{y^{(j)}(x)\}-\pr\{y^{(j)}(x)\}\right) \rightarrow_d N(0,\sigma_{jx}^2),$ where $\sigma_{jx}^2$ is the asymptotic variance of $\widehat{\pr}\{y^{(j)}(x)\}.$
\end{theorem}

%{\color{red} Remove all the equation numbers that are not referenced.} \tblue{if an equation is not referenced in main paper, but in supplement? e.g. 8 and 9}

\section{Simulations for binary models}
\label{sec:simulation}

%\subsection{Simulations under binary models}
In this part, the unmeasured confounder $U$ is simulated from a Bernoulli distribution taking value $1$ with probability $0.65.$ Conditional on $U,$ the exposure and outcomes are generated from the following models:
\begin{equation}
\label{eqn:simu}
\begin{aligned}
  (\pr(X=1\mid U=u): u=1,2) &= (0.4,0.7), \\
  (\pr(Y^{(1)}=1\mid U=u,X=x): u=1,2, x=1,2) &= \left(\begin{array}{cc}
       0.4& 0.3 \\
       0.7& 0.6 
  \end{array}\right),  \\
    (\pr(Y^{(2)}=1\mid U=u,X=x): u=1,2, x=1,2) &= \left(\begin{array}{cc}
       0.25& 0.45 \\
       0.45& 0.75 
  \end{array}\right), \\ 
    (\pr(Y^{(3)}=1\mid U=u,X=x): u=1,2, x=1,2) &= \left(\begin{array}{cc}
       0.15& 0.25 \\
       0.45& 0.65 
  \end{array}\right),
  \end{aligned}
\end{equation}
where for $j=1,2,3,$ $\pr(Y^{(j)}=1\mid U=u,X=x)$ is given by the $(u,x)$-th element of the corresponding matrix. 
In addition to  Algorithm \ref{algorithm:plug-in}, we also implement two comparison methods for estimating $\pr\{y^{(j)}(x)\}, j=1,2,3, x=1,2$: (1) Crude: empirical conditional probabilities $\widehat{\pr}(y^{(j)}\mid x)$; (2) Random start: instead of the warm starting values $\widetilde{\pr}(u)$, $\widetilde{\pr}(x\mid u)$ and $\widetilde{\pr}(y^{(j)}\mid u,x)$, we use random starting values from a uniform distribution on $[0,1]$ in Step 3 of Algorithm \ref{algorithm:plug-in}. All simulation results are based on 1000 Monte-Carlo runs, each with a sample size $n=1000$.  One can see from Table \ref{Table:sim_categorical_seed152} that the crude estimator is severely biased, while using the warm starts as proposed substantially improves the performance of the generalized least squares procedure.

\begin{table}[htp]
  \centering
    \caption{Bias$\times 100$ (Standard deviation $\times 100$)  of various estimators for estimating the  potential outcome distributions. The sample size is 1000
    %. For each parameter, the best estimator is highlighted in bold
    }
\begin{tabular}{crrr}
potential outcome & \multicolumn{1}{c}{crude} & \multicolumn{1}{c}{random start} & \multicolumn{1}{c}{warm start}\\\hline
$\pr\{Y^{(1)}(X=1)=1\}$ & $-$4.02 (2.21) & $-$2.71 (11.8) & $-$1.11 (10.4)\\
$\pr\{Y^{(1)}(X=2)=1\}$ & 4.06 (2.22) & 7.43 (12.0) & 2.32 (9.11)\\
$\pr\{Y^{(2)}(X=1)=1\}$ & $-$2.75 (1.97) & 19.2 (22.0) & 0.87 (9.56)\\
$\pr\{Y^{(2)}(X=2)=1\}$ & 4.08 (2.24) &  5.78 (9.63) & 1.84 (8.85)\\
$\pr\{Y^{(3)}(X=1)=1\}$ & $-$4.16 (1.89) & 24.5 (22.0) & 0.02 (10.5)\\
$\pr\{Y^{(3)}(X=2)=1\}$ & 5.47 (2.23) & 11.3 {(9.36)}& 3.19 (10.9)
\end{tabular}
\label{Table:sim_categorical_seed152}
\end{table}

% We also performed  simulations to evaluate the performance of our proposed estimator in the presence of single-outcome confounding. In particular, following Figure \ref{fig:DAG-single-outcome-confounding}, \kong{?? reference} we simulate an additional covariate $W$  from a Bernoulli distribution with $\pr(W=1)=0.3$. We simulate $U, Y^{(1)}, Y^{(2)}$ following the models in \eqref{eqn:simu}, and $X, Y^{(3)}$ from the following models:
% \begin{align*}
%     &\pr(X=1\mid U,W) = \textrm{expit}(a^X_0 + a^X_UU + a^X_WW), \\
%     &\pr(Y^{(3)}=1\mid U,W,X) = \textrm{expit}(a^{(3)}_0 + a^{(3)}_UU + a^{(3)}_XX + a^{(3)}_{UX}UX + a^{(3)}_WW), 
% \end{align*}
% where the model parameters are chosen so that when $\alpha_W^X = \alpha_W^{(3)} = 0,$ the resulting models coincide with the ones in \eqref{eqn:simu}.
% %Specifically, we let $a^X_W$ and $a^{(3)}_W$ be non-zero, while $a^{(1)}_W$, $a^{(2)}_W$ are null. 
% Table \ref{Table:sim_categorical_confounder_seed23} summarizes the estimation results for the causal effects $\pr\{Y^{(j)}(X=2)=1\}-\pr\{Y^{(j)}(X=1)=1\},j=1,2,3.$
% As our theory predicts, the  bias of causal effect estimate for $Y^{(3)}$ increases as the confounding introduced by $W$ becomes stronger. Nevertheless, the biases of causal effect estimates for $Y^{(1)}$ and $Y^{(2)}$ remain small compared to their standard deviations. 

\begin{table}[htp]
\centering
    \caption{ Bias$\times 100$ (Standard deviation $\times 100$) of the proposed estimator for estimating the average causal effect $\pr\{Y^{(j)}(X=2)=1\}-\pr\{Y^{(j)}(X=1)=1\},j=1,2,3$, when $X$ and $Y^{(3)}$ may be confounded by  variable $W$. The sample size is 1000 }{%
\begin{tabular}{crrr}
$(\alpha^X_W,\alpha^{(3)}_W)$  & \multicolumn{1}{c}{$Y^{(1)}$}&\multicolumn{1}{c}{$Y^{(2)}$}&\multicolumn{1}{c}{$Y^{(3)}$}\\
\midrule
$(0,0)$ & 4.8 (18) & 1.4 (15) &  3.6 (17)   \\
$(1,1)$ & 6.3 (20) & 1.6 (15) &  7.9 (16)   \\
$(2,2)$ & 6.6 (23) & $-$0.04 (17)  & 20.7 (17)  \\
$(3,3)$ & 5.1 (24) & 0.7 (19) & 36.1 (19)  \\
$(4,4)$ & 2.6 (25) & $-$1.4 (21) & 46.5 (20) 
\end{tabular}}
\label{Table:sim_categorical_confounder_seed23}
\end{table}

\section{Explanation on linear Gaussian structural equation models violating Condition \ref{condition:continuous_indexing}}
\label{sec:proof_gaussian_case}
% When the latent confounder is continuous, we point out in the paper that the assumptions needed for identification do not hold for the simplest linear structural model when all the variables follow Gaussian distributions. In particular, 
Consider the following linear Gaussian structural equation models with three parallel outcomes, 
\begin{align*}
    X &= \alpha_X U + \epsilon_X, \\
    Y^{(j)} &= \alpha_j U + \beta_j X + \epsilon_j, j=1,2,3, 
\end{align*}
where %$U\sim\mathcal{N}(0,1)$, $\epsilon_X,\epsilon_1,\epsilon_2,\epsilon_3$ are Gaussian variables with mean $0$ and finite variance $\sigma^2_X, \sigma^2_1, \sigma^2_2, \sigma^2_3$. 
$(U, \epsilon_X,\epsilon_1,\epsilon_2,\epsilon_3)^\T \sim \mathcal{N}(0,\diag(1, \sigma^2_X, \sigma^2_1, \sigma^2_2, \sigma^2_3)).$
% The only information from centered Gaussian variables are second-moment conditions. And there are $10$ second-moment equations but $11$ unknown parameters, so the causal parameters $\beta_j,j=1,2,3$ are not identifiable. \tteal{
We will show that in this case, Condition \ref{condition:continuous_indexing} is violated.
% In particular, \tteal{the conditional distribution $Y^{(2)}\mid (U=u,X=x)$ is a Gaussian distribution with mean $\alpha_2u+\beta_2x$ and variance $\sigma^2_{\epsilon_2}$. Since a Gaussian distribution is totally characterized by its mean and variance, we only need to consider $M$ to be expectation or variance. If $M$ is variance, then $h_1$ in Condition S4 has the form $h_1(u)=\sigma^2_{\epsilon_2}$, which is not a one-to-one mapping. If $M$ is expectation, $h_1$ in Condition S4 has the form $h_1(u)=\alpha_2u$, which is neither bounded from above nor below when $\alpha_2\ne 0$ and $U\sim \mathcal{N}(0,1)$.}

The conditional distribution of $Y^{(2)}$ given $(U,X)=(u,x)$ is a Gaussian distribution with mean $\alpha_2u+\beta_2x$ and variance $\sigma^2_2$. Since a Gaussian distribution is totally characterized by its mean and variance, for any continuous functional $M$, we have $M\{f(Y^{(2)}\mid U,X)(u,x)\}=\tilde{q}(\alpha_2u + \beta_2x,\sigma^2_2) = q(\alpha_2u + \beta_2x)$, where $\tilde{q}, q$ are continuous functions, and the second equality is by the fact $\sigma^2_2$ is a constant.  We want to show if $g\circ q(\alpha_2u + \beta_2x)=h_1(u)+h_2(x)$, then $h_1(u)$ is a linear function.

% \tteal{The conditional density of $Y^{(2)}$ given $(U,X)=(u,x)$ is $f_{\epsilon_2}(y^{(2)}-\alpha_2u - \beta_2x)$, where $f_{\epsilon_2}(\cdot)$ is the density function of $\epsilon_2$, and it does not involve $u$ or $x$. Therefore, for any continuous functional $M$, we have $M\{f(Y^{(2)}\mid U,X)(u,x)\}=M\{f_{\epsilon_2}(y^{(2)}-\alpha_2u - \beta_2x)\}=q(\alpha_2u + \beta_2x)$, where $q$ is a continuous function.  We want to show if $g\circ q(\alpha_2u + \beta_2x)=h_1(u)+h_2(x)$, then $h_1(u)$ is a linear function.}

Let $f=g\circ q$, then $f$ is also continuous by the continuity of $g$ and $q$. We have $f(\alpha_2u+\beta_2x)=h_1(u)+h_2(x)$. Let $x=0$, we have $f(\alpha_2u)=h_1(u) + h_2(0)$. Let $u=0$, we have $f(\beta_2x) = h_2(x) + h_1(0)$. Let $u=x=0$, we have $f(0) = h_1(0)+h_2(0)$. 

Therefore, $f(\alpha_2u)+f(\beta_2x) - f(0) = h_1(u) + h_2(x)$. 
Let $\tilde{f}(t) = f(t) - f(0)$, then $\tilde{f}(0)=0$, $\tilde{f}(\alpha_2u+\beta_2x) = f(\alpha_2u+\beta_2x) -f(0) = h_1(u) + h_2(x) - f(0) = f(\alpha_2u) + f(\beta_2x) - 2f(0) = \tilde{f}(\alpha_2u) + \tilde{f}(\beta_2x). $ By Lemma \ref{lem:linear_function}, $\tilde{f}(t)=ct$, where $c\in\real$ is a constant. So $f(t) = ct+ f(0)$ and $f(\alpha_2u+\beta_2x) = c(\alpha_2u+\beta_2x)+f(0) = c\alpha_2u + c\beta_2x + f(0)$. By comparing with $f(\alpha_2u+\beta_2x)=h_1(u)+h_2(x)$, we conclude $h_1$ is a linear function of the form $h_1(u)=au+b$. If $a=0$, then $h_1$ is not one-to-one; if $a\ne 0$, then $h_1(u)$ is neither bounded from above nor below when $U\sim \mathcal{N}(0,1)$. Both cases violate Condition \ref{condition:continuous_indexing}. 

% {\tred{Question: Do conditions in section 4 imply condition S2 holds? (Dingke)}} \tteal{For Gaussian case I think so.}

\section{Derivation of the threshold in \eqref{eq:optimization_5}}\label{threshold:derivation}
In \eqref{eq:optimization_5}, we need to set the threshold $ \delta $. Under the high dimensional framework, when distinguishing non-zero signals from noises, the threshold $\delta$ is often set %{\tred{Since we used the word ``often``, I think we need to provide reference to convince our audience.(Dingke)}} 
as $ \delta=\sqrt{2n^{-1}\log(p)\sigma^2}$, where $ \sigma^2$ is the error variance \citep{donoho_ideal_1994}. 

In our case, we consider the factor model 
\begin{align}\label{factor:model}
Y =\Gamma F + \varepsilon,
\end{align}
i.e., model \eqref{eq:Y_matrix} in the main paper. Recall in this model,  we have (i) $F\bigCI \varepsilon $; (ii) $E(F)=0,\textrm{Cov}(F)=I_{r+1}$; (iii) $E(\varepsilon)=0,\textrm{Cov}(\varepsilon)=\Lambda=\textrm{diag}(\sigma_{1}^2,\ldots,\sigma_{p}^2)$. %{\tred{Notation is not consistent here. Use $\sigma_i^2$? (Dingke)}}
Let $\Sigma = \textrm{Cov}(Y)$ and $\Lambda = \textrm{Cov}(\varepsilon)$, one can write $\Sigma = \Gamma \Gamma^{\T} + \Lambda$. 

The error term in model \eqref{factor:model} is a $p$-dimensional vector $\varepsilon$, and each component has a variance $\sigma_{j}^2 $ for $ j=1, \ldots, p$. Thus, we choose $ \sigma^2=p^{-1}\sum_{j=1}^p \sigma_{j}^2$. %By single algebra, we have $ \sigma^2=p^{-1}\sum_{j=1}^p \sigma_{\epsilon_j}^2=p^{-1}\textrm{tr}(\Lambda)=p^{-1}\textrm{tr}(\Sigma-\Gamma\Gamma^{\T})$. 

In practice, $\sigma^2$ is unknown, and needs to be estimated. This boils down to estimating $ \Sigma$, $\Gamma$, and $ \Lambda $. Denote $ \widehat\Sigma$ the sample convariance of the $(Y^{(1)}_i, \ldots, Y^{(p)}_i )$, $i=1,\ldots,n$. The eigen decomposition of $\widehat{\Sigma} $ is
$$
    \widehat{\Sigma} = \sum^p_{j=1}\widehat{\lambda}_j \widehat{v}_j\widehat{v}_j^\T,
$$ 
where $ \widehat{\lambda}_j $'s are the eigenvalues and $\widehat{v}_j$'s are orthonormal eigenvectors. For the loading matrix $\Gamma$, we use the principal component method to estimate it. Specifically, we have $ \widehat{\Gamma}\widehat{\Gamma}^{\T}=\sum^{\widehat{r}+1}_{j=1}\widehat{\lambda}_j \widehat{v}_j\widehat{v}_j^\T$, where $ \widehat{r}+1$ is the estimated number of factors. And the estimate of $ \Lambda $ is taken as $ \widehat{\Lambda}=\textrm{diag}(\widehat{\Sigma}-\widehat{\Gamma}\widehat{\Gamma}^{\T})$. 

Therefore, we can obtain $\widehat{\sigma}^2$ by
\begin{align*}
    &\widehat{\sigma}^2=p^{-1}\sum^p_{j=1}\widehat{\sigma}^2_{j} = p^{-1}\textrm{tr}(\widehat{\Lambda})=p^{-1}\textrm{tr}(\widehat{\Sigma}-\widehat{\Gamma}\widehat{\Gamma}^{\T})=p^{-1}\textrm{tr}(\sum^p_{j=\widehat{r}+2}\widehat{\lambda}_j \widehat{v}_j\widehat{v}_j^\T) \\
    =& p^{-1}\sum^p_{j=\widehat{r}+2}\textrm{tr}(\widehat{\lambda}_j \widehat{v}_j\widehat{v}_j^\T) =p^{-1}\sum^p_{j=\widehat{r}+2}\widehat{\lambda}_j\textrm{tr}(\widehat{v}_j^\T\widehat{v}_j) = p^{-1}\sum^p_{j=\widehat{r}+2}\widehat{\lambda}_j. 
\end{align*}

\section{Reformulation of \eqref{eq:optimization_5} to a mixed integer programming problem}
\label{sec:mip}
We assume there exists a known scalar $M>0$ such that $M\ge \| y^* \|_{\infty}$ for an optimal solution $y^*$ to \eqref{eq:optimization_5}.
By introducing binary variables $\zeta=(\zeta_1, \ldots, \zeta_p)^{\T} \in \{0,1\}^p$ with $\zeta_j=\mathbb{I}(y_j> \delta)$, the problem \eqref{eq:optimization_5} can be reformulated as the following mixed integer programming problem \citep{feng_complementarity_2018}
\begin{align*}
\begin{split}
    & \quad \min_{w,y,\zeta}\quad 1^\T_p \zeta = \sum_{j=1}^p \zeta_j \\
    \textrm{subject to } & \quad w^\T w=1,
    \quad y = \widehat{\Gamma} w,  \quad -M\zeta - \delta \le y\le M\zeta + \delta, \quad \zeta\in\{0,1\}^p.
\end{split}
\end{align*}

\section{Description of outcomes in the real data example}
\label{sec:description_outcomes}
In Table \ref{tab:description}, we include the detailed descriptions of each outcome used in Section \ref{sec:data}.

\begin{table}[htp]
    \centering
    \caption{Description of outcomes in the real data application}
    \label{tab:description}
    \makebox[\textwidth][c]{%
    \begin{tabular}{ll}
    \toprule
        \multicolumn{1}{p{3.5cm}}{Acrylamide} & \multicolumn{1}{p{12cm}}{a toxic and potentially cancer-causing chemical which is formed in high amounts in many types of food prepared/cooked at high temperatures.}\\
        \hline
        \multicolumn{1}{p{3.5cm}}{Vitamin C} & \multicolumn{1}{p{12cm}}{a water-soluble vitamin. It is an essential nutrient involved in the repair of tissue, the formation of collagen, and the enzymatic production of certain neurotransmitters.}\\
        \hline
        \multicolumn{1}{p{3.5cm}}{Cadmium} & \multicolumn{1}{p{12cm}}{a chemical element with the symbol Cd and atomic number 48.}\\
        \hline
        \multicolumn{1}{p{3.5cm}}{Urinary thiocyanate} & \multicolumn{1}{p{12cm}}{a biomarker of cyanide exposure from tobacco smoke or diet. It can disrupt thyroid function by competitively inhibiting iodide uptake.}\\
        \hline
        \multicolumn{1}{p{3.5cm}}{Retinyl palmitate} & \multicolumn{1}{p{12cm}}{the most abundant form of vitamin A storage in animals.}\\
        \hline
        \multicolumn{1}{p{3.5cm}}{Osmolality} & \multicolumn{1}{p{12cm}}{plasma osmolality measures the body's electrolyte–water balance.} \\
        \hline
        \multicolumn{1}{p{3.5cm}}{Vitamin D} & \multicolumn{1}{p{12cm}}{a group of fat-soluble secosteroids responsible for increasing intestinal absorption of calcium, magnesium, and phosphate, and many other biological effects.} \\
        \hline
        \multicolumn{1}{p{3.5cm}}{Vitamin B12} & \multicolumn{1}{p{12cm}}{a water-soluble vitamin involved in metabolism. It is one of eight B vitamins. It is required by animals, which use it as a cofactor in DNA synthesis, in both fatty acid and amino acid metabolism.}\\
        \hline
        \multicolumn{1}{p{3.5cm}}{Mercury, total} & \multicolumn{1}{p{12cm}}{a chemical element with the symbol Hg and atomic number 80}\\
        \hline
        \multicolumn{1}{p{3.5cm}}{Enterolactone} & \multicolumn{1}{p{12cm}}{a organic compound classified as an enterolignan. It is formed by the action of intestinal bacteria on plant lignan precursors present in the diet.}\\
        \hline
        \multicolumn{1}{p{3.5cm}}{C-reactive protein} & \multicolumn{1}{p{12cm}}{an annular pentameric protein found in blood plasma, whose circulating concentrations rise in response to inflammation.} \\
        \hline
        \multicolumn{1}{p{3.5cm}}{Total bilirubin} & \multicolumn{1}{p{12cm}}{a red-orange compound that occurs in the normal catabolic pathway that breaks down heme in vertebrates. Elevated levels may indicate liver damage or disease. }\\
        \hline
        \multicolumn{1}{p{3.5cm}}{Perchlorate, urine} & \multicolumn{1}{p{12cm}}{a polyatomic anion that can disrupt thyroid function by competitively inhibiting iodide uptake.}\\
        \hline
        \multicolumn{1}{p{3.5cm}}{Creatinine, urine} &\multicolumn{1}{p{12cm}}{a breakdown product of creatine phosphate from muscle and protein metabolism. High levels of creatinine can indicate that the kidney is not working well.} \\
        \hline
        \multicolumn{1}{p{3.5cm}}{Glucose, serum} &\multicolumn{1}{p{12cm}}{the amount of glucose in the fluid portion of the blood. It is the simplest and most direct single test available to test for diabetes.} \\
        \hline
        \multicolumn{1}{p{3.5cm}}{Iron, refigerated} & \multicolumn{1}{p{12cm}}{used in the diagnosis and treatment of diseases such as iron deficiency anemia, chronic renal disease, and hemochromatosis}\\
        \hline
        \multicolumn{1}{p{3.5cm}}{Serum total IgE antibody} & \multicolumn{1}{p{12cm}}{the amount of IgE antibodies in the blood and is the sum of all the forms of IgE. Total IgE testing is used to help diagnose some health conditions including certain types of infections and immune disorders.}\\
        \hline
        \multicolumn{1}{p{3.5cm}}{2,4,5-trichlorophenol} & \multicolumn{1}{p{12cm}}{metabolite of several organochlorine pesticides. }\\
        \hline
%        Mercury, inorganic (ug/L) & \\
        \multicolumn{1}{p{3.5cm}}{Mono-2-ethyl-5-carboxypentyl phthalate} & \multicolumn{1}{p{12cm}}{used extensively as plasticizers in a wide range of applications such as children’s toys, food packaging, personal care products, and medical supplies.}\\
        \hline
        \multicolumn{1}{p{3.5cm}}{Urinary bisphenol A} &\multicolumn{1}{p{12cm}}{used in the manufacture of polycarbonate plastics and epoxy resins, which have been used in baby bottles, as protective coatings on food containers, and as composites and sealants in dentistry.}\\
        \bottomrule
    \end{tabular}}
\end{table}

\section{Proof of Theorem \ref{thm:categorical}}
\label{sec:proof_thm1}
\begin{proof} 
 Condition \ref{condition:no-interaction2}  allows us to rearrange $\{u_1,\ldots,u_k\}$ in ascending order in terms of $\pr(y^{(1)}\mid u_i,x)$ for fixed $y^{(1)}$ and $x$, and relabel it as $\{u_{(1)},\ldots,u_{(k)}\}$. By Assumption \ref{assumption:shared-confounding}, we have
\begin{align}
\label{eq:klevel_ci1}
    \pr(y^{(2)},y^{(3)}\mid x)&=\sum^k_{i=1}\pr(y^{(2)}\mid u_{(i)},x)\pr(y^{(3)}\mid u_{(i)},x)\pr(u_{(i)}\mid x),\\
\label{eq:klevel_ci2}
    \pr(y^{(1)},y^{(2)},y^{(3)}\mid x)&=\sum^k_{i=1}\pr(y^{(1)}\mid u_{(i)},x)\pr(y^{(2)}\mid u_{(i)},x)\pr(y^{(3)}\mid u_{(i)},x)\pr(u_{(i)}\mid x).
\end{align}
Let $P(Y^{(2)},Y^{(3)}\mid x) = \left( \pr(Y^{(2)}=i, Y^{(3)}=j\mid X=x) \right)_{k\times k}$ be a $k$ by $k$ matrix whose $(i,j)$-th element is given by  $\pr(Y^{(2)}=i, Y^{(3)}=j\mid X=x)$. Similarly, we let $P(Y^{(2)}\mid U,x) = \left( \pr(Y^{(2)}=i\mid U=u_{(j)}, X=x) \right)_{k\times k}, P(Y^{(3)}\mid U,x) = \left( \pr(Y^{(3)}=i\mid U=u_{(j)}, X=x) \right)_{k\times k}.$ Additionally, $P_D(y^{(1)}\mid U,x)=\diag\{\pr(y^{(1)}\mid U=u_{(1)},x),\ldots,\pr(y^{(1)}\mid U=u_{(k)},x)\}$,  $P_D(U\mid x)=\diag\{\pr(U=u_{(1)}\mid x),\ldots,$ $\pr(U=u_{(k)}\mid x)\}$ are $k\times k$ diagonal matrices. Then in the form of matrix multiplication, \eqref{eq:klevel_ci1} and \eqref{eq:klevel_ci2} can be written as 
\begin{align}
    \label{eq:mp}
    P(Y^{(2)},Y^{(3)}\mid x)&=P(Y^{(2)}\mid U,x) P_D(U\mid x)P(Y^{(3)}\mid U,x)^\T,\\
    \label{eq:mq}
    P(y^{(1)},Y^{(2)},Y^{(3)}\mid x)&=P(Y^{(2)}\mid U,x) P_D(U\mid x)P_D(y^{(1)}\mid U,x)P(Y^{(3)}\mid U,x)^\T.
\end{align}

% where
% \begin{align*}
% P(Y^{(2)},Y^{(3)}\mid x)&={
% \left( \begin{array}{ccc}
% \pr(Y^{(2)}=1,Y^{(3)}=1\mid x) & \cdots & \pr(Y^{(2)}=1,Y^{(3)}=k\mid x)\\
% \vdots & \ddots &\vdots\\
% \pr(Y^{(2)}=k,Y^{(3)}=1\mid x) & \cdots & \pr(Y^{(2)}=k,Y^{(3)}=k\mid x)
% \end{array} 
% \right )},\\
% P(y^{(1)},Y^{(2)},Y^{(3)}\mid x)&={
% \left( \begin{array}{ccc}
% \pr(y^{(1)}, Y^{(2)}=1,Y^{(3)}=1\mid x) & \cdots & \pr(y^{(1)}, Y^{(2)}=1,Y^{(3)}=k\mid x)\\
% \vdots & \ddots &\vdots\\
% \pr(y^{(1)}, Y^{(2)}=k,Y^{(3)}=1\mid x) & \cdots & \pr(y^{(1)}, Y^{(2)}=k,Y^{(3)}=k\mid x)
% \end{array} 
% \right )},\\
% P(Y^{(3)}\mid U,x)&={
% \left( \begin{array}{ccc}
% \pr(Y^{(3)}=1\mid x,u_{(1)}) & \cdots & \pr(Y^{(3)}=k\mid x,u_{(1)})\\
% \vdots & \ddots &\vdots\\
% \pr(Y^{(3)}=1\mid x,u_{(k)}) & \cdots & \pr(Y^{(3)}=k\mid x,u_{(k)})
% \end{array} 
% \right )},\\
% P(Y^{(2)}\mid U,x)&={
% \left( \begin{array}{ccc}
% \pr(Y^{(2)}=1\mid x,u_{(1)}) & \cdots & \pr(Y^{(2)}=k\mid x,u_{(1)})\\
% \vdots & \ddots &\vdots\\
% \pr(Y^{(2)}=1\mid x,u_{(k)}) & \cdots & \pr(Y^{(2)}=k\mid x,u_{(k)})
% \end{array} 
% \right )},
% \end{align*}
%and $P_D(y^{(1)}\mid U,x)=\diag\{\pr(y\mid x,u_{(1)}),\ldots,\pr(y\mid x,u_{(k)})\}$,  $P_D(U\mid x)=\diag\{\pr(u_{(1)}\mid x),\ldots,$ $\pr(u_{(k)}\mid x)\}$ are $k\times k$ diagonal matrices. It is to be observed that each row of $R(z,u)$ and $S(w,u)$ sums to 1. We further assume that 
By Condition \ref{condition:full-rank}, $P(Y^{(2)},Y^{(3)}\mid x)$ is invertible. Then from \eqref{eq:mp} and \eqref{eq:mq} we have $P(y^{(1)},Y^{(2)},\allowbreak Y^{(3)}\mid x)P(Y^{(2)},Y^{(3)}\mid x)^{-1}=P(Y^{(2)}\mid U,x) P_D(y^{(1)}\mid U,x) P(Y^{(2)}\mid U,x)^{-1}$. Using the technique of eigendecomposition, $P_D(y^{(1)}\mid U,x)$ and $P(Y^{(2)}\mid U,x)$ can be recovered in the sense that the eigenvalues of $P(y^{(1)},Y^{(2)},Y^{(3)}\mid x)P(Y^{(2)},Y^{(3)}\mid x)^{-1}$ in descending order are the diagonal elements  of $P_D(y^{(1)}\mid U,x)$, i.e. $\pr(y^{(1)}\mid U=u_{(i)},x)$, and the corresponding eigenvectors scaled to $\ell_1$-norm one are columns of $P(Y^{(2)}\mid U,x)$. By symmetry, $P(Y^{(3)}\mid U,x)$ can be identified likewise. Therefore, {following     \eqref{eq:mp},} $P_D(U\mid x)=P(Y^{(2)}\mid U,x)^{-1}P(Y^{(2)},Y^{(3)}\mid x)\{P(Y^{(3)}\mid U,x)^\T\}^{-1}$ is identified. 
% Since Condition \ref{condition:no-interaction2} guarantees that the order of $\{u_{(1)},\ldots,u_{(k)}\}$ \tblue{what is this??} is the same for all $x$ when $y^{(1)}$ is fixed, we are able to 
We can then identify $\pr(u_{(i)})=E_X \{\pr(u_{(i)}\mid x)\}$. Finally, the potential outcome distributions can be identified from
\begin{equation*}
    \pr\{y^{(j)}(x)\} = \sum_{i=1}^k\pr(y^{(j)}\mid x,u_{(i)})\pr(u_{(i)}), j=1,2,3.
\end{equation*}

%Although $P_D(y^{(1)}\mid U,x)$, $P(Y^{(2)}\mid U,x)$, and $P(Y^{(3)}\mid U,x)$ are identified, we can not identify $\pr(y^{(j)}\mid u_i,x),j=1,2,3$ because we do not know the ordering of $\pr(y^{(j)}\mid u_{(i)},x)$. {\color{red} check whether my change is approporiate.}
%we do not know the correspondence between $(i)$ and $i$. 
%However, identifying $\pr\{y^{(j)}(x)\}$ is still possible because it only concerns the summation rather than individual $i$. \tblue{I don't think this is correct.}

\end{proof}

\section{Proof of Lemma \ref{lem:multi_confounder_identify0}}

\begin{proof}
Recall that
\begin{align*}
\Gamma^*_{p\times(r+1)} = 
    \begin{pmatrix}
        \sigma_{\epsilon_X}\beta_1 & \alpha_{11}+\alpha_{X1}\beta_1 & \cdots & \alpha_{1r}+\alpha_{Xr}\beta_1 \\
        \sigma_{\epsilon_X}\beta_2 & \alpha_{21}+\alpha_{X1}\beta_2 & \cdots & \alpha_{2r}+\alpha_{Xr}\beta_2 \\
        \vdots & \vdots & \ddots & \vdots \\
        \sigma_{\epsilon_X}\beta_p & \alpha_{p1}+\alpha_{X1}\beta_p & \cdots & \alpha_{pr}+\alpha_{Xr}\beta_p
    \end{pmatrix}.
\end{align*}
% Using results from factor analysis, we can identify $\Gamma^*$ up to a rotation under Condition \ref{asp:identification_rotation}. Let $\Gamma=\Gamma^*R$ be such an identified loading matrix, where $R$ is a $(r+1)\times(r+1)$ unknown rotation matrix with $R^\T R = I$.
% \tred{cross reference is problematic here(Dingke )}
By Condition \ref{asp:beta_sparsity}(i), $\|\Gamma^*_{\cdot 1}\|_0=s$. 
Condition \ref{asp:beta_sparsity}(ii)%{\tred{Condition 4? As there is no (ii) in Condition 3.(Dingke)}}
ensures that the minimum of $\|(\Gamma^*R)_{\cdot1}\|_0$ is $s$, where $R$ is any $(r+1)\times(r+1)$ matrix with $R^\T R=I$. We prove this by contradiction. 

If the minimum is $s-1$ or smaller and it is achieved for some $R^*$, then there are at least $p-s+1$ zeros in $(\Gamma^*R^*)_{\cdot1}$.  If $\|(\Gamma^*R^*)_{\cdot1}\|_0=s-1$, then the first column of $\Gamma^*R^*$ has $p-s+1$ zeros. We let $S=\{j:(\Gamma^*R^*)_{j1}=0\}$%\tred{$j$ or $j_1$? I think we have a typo here.(Dingke)}
%\tteal{it means the $(j,1)$ element of the matrix}
, hence $|S|=p-s+1>r+1$ by Condition \ref{asp:beta_sparsity}(i). Let $\Gamma^*_{S\cdot}$ be the $|S|\times (r+1)$ submatrix of $\Gamma^*$ consisting of rows indexed by $S$, then $\Gamma^*_{S\cdot}R^*$ is not full rank because its first column is all zero. Since the rotation matrix $R^*$ has rank $r+1$, we have $\textrm{rank}(\Gamma^*_{S\cdot}R^*)=\textrm{rank}(\Gamma^*_{S\cdot})$. We then conclude $\Gamma^*_{S\cdot}$ is not full rank. This violates Condition \ref{asp:beta_sparsity}(ii), which implies any submatrix of $\Gamma^*$ consisting of $p-s+1$ rows has full rank $r+1$. %{\tred{Condition 4(ii)? (Dingke)}}.
Contradition! So, the minimum of $\|(\Gamma^*R)_{\cdot1}\|_0$ is $s$.

% Since $p-s>r$ by Condition \ref{asp:beta_sparsity}, the matrix $\Gamma^* R$ has a column with $p-s+1$ zeros implies that the corresponding $(p-s+1)\times (r+1)$ submatrix of $\Gamma^*$ is not full rank. To see this, we denote this submatrix of $\Gamma^*$ by $A$. Since the rotation matrix $R$ has rank $k+1$, we have $\textrm{rank}(AR)=\textrm{rank}(A)$.
% We know that $AR$ is not full rank because it has a zero column, so $A$ must not be full rank. This violates Condition \ref{asp:identification_rotation}(ii). So, the minimum $\ell_0$-norm of $(\Gamma^*R)_{\cdot1}$ is $s$.

Finally, Condition \ref{asp:beta_sparsity}(ii) ensures when the minimum $\ell_0$-norm is achieved, it must be that the $p-s$ 0's appear at the rows with $\beta_j=0$. Any other combination of $p-s$ rows has full rank, thus the $(p-s)\times(r+1)$ submatrix of $\Gamma^*$ can never have a column of 0's after rotation.

\end{proof}
\section{Proof of Theorem \ref{thm:multi_identification}}
%\tred{Cross reference incorrect here(Dingke)}
\label{sec:proof_thm2}
\begin{proof}
In Lemma \ref{lem:multi_confounder_identify0}, we have identified the set $S_0=\{j:\beta_j=0\}$. Now we focus on $\{\ell:\beta_\ell\ne 0\}$.
Taking conditional expectation of $W$ on $(X,Z^\ell)$, we have
\begin{align*}
    E(W\mid X,Z^\ell) &= \Gamma^*_{S_0,-1} E(U\mid X,Z^\ell),
\end{align*}
% Here $\alpha_W$ is a $(p-s)\times r$ matrix.
where $\Gamma^*_{S_0,-1}$ is the $(p-s)\times r$ submatrix of $\Gamma^*$ with row indices in $S_0$ and column indices in $\{2,3,\ldots,r+1\}$. By Condition \ref{asp:beta_sparsity}(ii), $\Gamma^*_{S_0,-1}$ has rank $r$. Therefore, there exists $W^*=(W_{t_1},\ldots,W_{t_r})\in\real^r$ such that
\begin{align*}
    E(W^*\mid X,Z^\ell) &= \alpha_{W^*} E(U\mid X,Z^\ell),
\end{align*}
where $\alpha_{W^*}$ is a full rank $r\times r$ submatrix of $\Gamma^*_{S_0,-1}$.
Taking conditional expectation of $Y^{(\ell)}$ on $(X,Z^\ell)$, we have
\begin{align}
    \nonumber
    E(Y^{(\ell)}\mid X,Z^\ell) &= \beta_\ell X + \alpha_\ell^\T E(U\mid X,Z^\ell) \\
    &= \beta_\ell X + \alpha_\ell^\T\alpha_{W^*}^{-1}E(W^*\mid X,Z^\ell). \label{eq:beta_linear_identification}
\end{align}
Here $\alpha_\ell$, $E(U\mid X,Z^\ell)$, $E(W\mid X,Z^\ell)$ are $r$-vectors. 
%The $r\times r$ matrix $\alpha_W$ is invertible by Condition \ref{asp:beta_sparsity}(ii). To see this, we notice that $\alpha_W$ is an $r\times r$ submatrix of the $p\times r$ matrix $\Gamma^*_{S_0,-1}$ with row indices in $S_0=\{j:\beta_j=0\}$ and column indices in $\{2,3,\ldots,r+1\}$. By Condition \ref{asp:beta_sparsity}(ii), $\Gamma^*_{S_0,-1}$ has rank $r$, so $\alpha_W$ should also have rank $r$.
By Condition \ref{condition:WZ}, $X$ is not a linear combination of $E(W\mid X,Z^\ell)$. So $X$ is not a linear combination of $E(W^*\mid X,Z^\ell)$ as $W^*$ is a sub-vector of $W$. Therefore, $\beta_\ell$ can be identified through the regression in \eqref{eq:beta_linear_identification}. 

\end{proof}

% \section{Proof of Proposition \ref{lemma:gaussian_collinearity} \tblue{delete?}}
% \begin{proof}
% Under the Gaussian linear model assumption, $E(W\mid X,Z^\ell)$ is a linear combination of $X$ and $Z^\ell$:
% \begin{align}\label{eq:W_XZ}
%     E(W\mid X,Z^\ell) = 
%     \Theta_{d_W\times(d_Z+1)}
%     \begin{pmatrix}
%     X & Z^\ell
%     \end{pmatrix}^\T,
% \end{align}
% where $d_Z$ is the dimension of $Z^\ell$ and $d_W$ is the dimension of $W$. If $d_Z+1\le d_W$, i.e., $\dim(Z^\ell)<\dim(W)$, then $\Theta$ has rank at most $d_Z+1$. 

% If the first column of $\Theta$ is not a zero vector, i.e., at least one of $E(U\mid X,Z^\ell)$ is dependent on $X$, then we can let $v$ be a $d_W$-vector such that $v^\T \Theta=(
% 1,0,\cdots,0)\in \real^{d_Z+1}$. Such a $v$ exists because the number of parameters $d_W$ is no less than the number of equations $d_Z+1$. If the first column of $\Theta$ is a zero vector, then $E(W\mid X,Z^\ell) = E(W\mid Z^\ell)$ and $\beta_\ell$ is identified as the coefficient of $X$ in $E(Y^{(\ell)} \mid X,Z^\ell)$.

% Multiplying $v^\T$ from the left on \eqref{eq:W_XZ}, we get
% \begin{align*}
%     v^\T E(W\mid X,Z^\ell) &= X.
% \end{align*}
% So $X$ is a linear combination of $E(W\mid X,Z^\ell)$.
% % and the coefficient of $X$ in \ref{eq:beta_linear_identification} is not identified. Hence we require that $d+1>r$, i.e., $d\ge r$.
% \end{proof}

\section{Extension of \eqref{eq:X_multi}-\eqref{eq:Y_multi} to include common latent mediators among outcomes}\label{sec:extension_mediators}

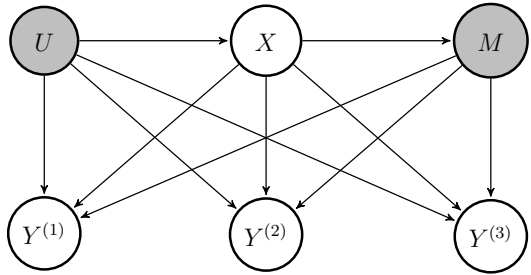
\begin{figure}
  \centering
\scalebox{0.8}{
\begin{tikzpicture}[->,>=stealth',shorten >=1pt,auto,node distance=2.3cm,
     semithick, scale=0.50]
     pre/.style={-,>=stealth,semithick,blue,ultra thick,line width = 1.5pt}]
     \tikzstyle{every state}=[fill=none,draw=black,text=black]
     \node[shade] (U)                    {$^{\tg{1}}U^{\tg{1}}$};
     \node[est] (X) [right = 2.5 cm of U] {$^{\tw{1}}X^{\tw{1}}$};
     \node[shade] (M) [right = 2.5 cm of X]                    {$^{\tg{1}}M^{\tg{1}}$};
     \node[est] (Y2) [below = 2 cm of X] {$Y^{(2)}$};
     \node[est] (Y1) [below = 2 cm of U] {$Y^{(1)}$};
     \node[est] (Y3) [below = 2 cm of M] {$Y^{(3)}$};
     \path  
     (U) edge node {} (Y1)
     (U) edge node {} (Y2)
     (U) edge node {} (Y3)
     (X) edge node {} (Y1)
     (X) edge node {} (Y2)
     (X) edge node {} (Y3)
     (U) edge node {} (X)
     (X) edge node {} (M)
     (M) edge node {} (Y1)
     (M) edge node {} (Y2)
     (M) edge node {} (Y3);
 \end{tikzpicture}}
 \caption{A causal diagram associated with the parallel-outcome framework when $p=3$ in the presence of unmeasured confounders and mediators.}
 \label{fig:DAG_UM}
 \end{figure}

We show in this section that our model in \eqref{eq:X_multi}-\eqref{eq:Y_multi} can be extended to accommodate common latent mediators among outcomes.
Figure \ref{fig:DAG_UM} provides a causal diagram in which there are common mediators for the effect of $X$ on $Y^{(j)}, j=1,2,3,$. In this case, the assumption that 
\begin{equation}
\label{eqn:cond_ind}
    Y^{(1)}\bigCI Y^{(2)} \bigCI Y^{(3)} \mid (U,X)
\end{equation}
fails to hold. However, we note that under the causal diagram in Figure \ref{fig:DAG_UM}, we have instead the assumption that 
\begin{equation}
\label{eqn:cond_ind2}
    Y^{(1)}\bigCI Y^{(2)} \bigCI Y^{(3)} \mid (U,X,M).
\end{equation}
Note that \eqref{eqn:cond_ind2} takes the same form as \eqref{eqn:cond_ind} except that $U$ is replaced with $U^* \equiv (U,M).$

Motivated by this, we shall now show that under the linear structural equation model described in Section \ref{sec:multi_identification}, we can  apply our estimation procedures, and still obtain the total effects of $X$ on $Y^{(j)}.$ Specifically,  suppose we have a latent confounder $U\in \mathbb{R}^r$ and a latent mediator $M\in \mathbb{R}^s$. Similar to  Section \ref{sec:multi_identification}, we assume the following linear structural equation models: 
\begin{align}
\label{eqn:X}
    X &= \alpha_X^\T U + \epsilon_X, \\ 
    \label{eqn:M}
    M &= \beta_M X + \epsilon_M, \\ 
    \label{eqn:Yj}
    Y^{(j)} &= \alpha_j^\T U + \beta_j X + \gamma_j^\T M + \epsilon_j. 
\end{align}
Define $ \beta_j^*=\beta_j+\gamma_j^\T \beta_M$, which characterizes the total effect of $ X $ on $Y^{(j)}$. By plugging \eqref{eqn:X} and \eqref{eqn:M} into \eqref{eqn:Yj}, we have 
\begin{align*}
Y^{(j)}=(\alpha_j+\alpha_X^\T\beta_j^*)U +  \gamma_j^\T \epsilon_M+\beta_j^*\epsilon_X  + \epsilon_j. 
\end{align*}
This has a similar form as equation \eqref{eq:Y_linear} in the paper, i.e., the above model is also an orthogonal factor model with latent confounders $U$ and $ \epsilon_M$ of dimension $ r+s$. By assuming similar conditions as Conditions \ref{asp:identification_rotation}-\ref{condition:WZ} in our main paper but replacing $\beta_j$ with $\beta_j^*$, we may still identify and estimate the causal parameters $ \beta_j^*$. 
%Note here we are only able to identify the total effect of $ X $ on $ Y^{(j)}$, but not the direct and indirect effects as the mediator $M$ is unobserved.   

% {\tred{I want to  clarify a question I have after I read this paragraph and comparison between our method with Wang's procedure (2017) in our response: If every random variables are Gaussian, we can write $X = \tilde{\alpha}_u^\T U + \tilde{\alpha}_w^\T W +\tilde{\epsilon}_x $ such that $(U,W) \perp  \tilde{\epsilon}_x $. Plug this equality into S20 we have $Y^{(j)} = (\alpha_j^\T + \tilde{\alpha}_u^\T\beta_j) U + (\gamma_j^\T + \tilde{\alpha}_w^\T\beta_j) M + \beta_j\tilde{\epsilon}_x+ \epsilon_j$ which is still a valid factor model? (W $\not\perp$ U, but we can transform (U,W) to make the transformed R.V. independent) So under proper condition it is still possible to identify $\beta_j$?  (Dingke)} } 
% \tteal{When $U$ and $M$ are scalar, I believe there are infinitely many reparameterizations.}

\section{Lemmas}\label{sec:lemmas}

\begin{lemma}
\label{lemma:1}
Under Assumptions \ref{assumption:ignorability}, \ref{assumption:shared-confounding}, if for all $x$ and $j=1,2,3, U \nind Y^{(j)} \mid X=x$, and $U$, $Y^{(2)}$, $Y^{(3)}$ are binary, then for all $x$, $P(Y^{(2)},Y^{(3)}\mid x)$ is of full rank.
\end{lemma}

\begin{lemma}
\label{lm:perturbation}
Let $\lambda_0$ be an algebraically simple eigenvalue of a matrix $M_0\in M_k(\real)$, where $M_k(\real)$ is the set of matrices of size $k\times k$ with entries in $\real$. Then there exists an open neighborhood $\mathcal{M}$ of $M_0$ in $M_k(\real)$, and two analytic functions
\begin{equation}
M\mapsto \Lambda(M),\quad M\mapsto X(M) 
\end{equation}
over $\mathcal{M}$, such that
\begin{itemize}[topsep=8pt,itemsep=0pt,partopsep=4pt, parsep=4pt]
    \item $\Lambda(M)$ is an eigenvalue of $M$.
    \item $X(M)$ is an eigenvector, associated with $\Lambda(M)$.
    \item $\Lambda(M_0)=\lambda_0$.
    \item $X(M_0)=X_0$, where $X_0$ is a normalized eigenvector pertaining to $\lambda_0$ with $1^\T_k X_0=1$; here $1_k$ denotes a $k$-vector whose elements are all $1$'s.
\end{itemize}
\end{lemma}

\begin{lemma}\label{lem:linear_function}
If $f$ is a continuous function, and $f(x+y) = f(x) + f(y)$, then $f(x) = cx$, where $c\in\real$ is a constant. 
\end{lemma}

\begin{lemma}
\label{lem:adjoint_def}
For $\mathcal{G}=\mathcal{L}^1, \mathcal{L}^1_{\textrm{bnd}}$, the adjoint of $T_{f_{a,b}(a,b)}$, denoted by $T_{f_{a,b}(a,b)}^*$, is an operator from $\mathcal{G}^*(\mathcal{A})$ to $\mathcal{G}^*(\mathcal{B})$ with the form
\begin{align}
\label{eq:adjoint}
    (T_{f_{a,b}(a,b)}^* g^*)(b)=\int f(a,b)g^*(a)\dif a.
\end{align}
\end{lemma}

\begin{lemma}
\label{lem:adjoint_injectivity}
% If $T_{f_{b\mid a}(b,a)}$ is injective, then $T_{f_{a,b}(a,b)}^{-1}$ exists and is densely defined over $\mathcal{G}(\mathcal{A})$ for $\mathcal{G}=\mathcal{L}^1,\mathcal{L}^1_{\textrm{bnd}}$.
If $T_{f_{b\mid a}(b,a)}$ and $T_{f_{a,b}(a,b)}$ are injective, then $T_{f_{a,b}(a,b)}^{-1}$ is densely defined over $\mathcal{G}(\mathcal{A})$ for $\mathcal{G}=\mathcal{L}^1,\mathcal{L}^1_{\textrm{bnd}}$.  %\footnotemark
\end{lemma}

%\footnotetext{Lemma \ref{lem:adjoint_def} is different from lemma 1 in \cite{hu_instrumental_2008}, as we have some doubt about their proof.}

\begin{lemma}
\label{lem:exponential_completeness}
If $k(a,b;\theta)=f_{a\mid b}(a\mid b;\theta)=\psi(b;\theta)h(a;\theta)\exp\{\eta(b;\theta)^\T\lambda(a;\theta)\}$ satisfies: (i) $\psi(b;\theta)>0$, (ii) $\eta(b;\theta)$ is one-to-one in $b$, (iii) the range of $\lambda(a;\theta)$ contains a k-dimensional open set, where $k$ is the dimension of $\eta(b;\theta)$. Then $k(a,b;\theta)$ is complete in $a$. 
\end{lemma}

\section{Proof of Lemmas}
\label{sec:lemma_proofs}

\subsection{Proof of Lemma \ref{lemma:1}}
\begin{proof}
    For $j=2,3,$  $P(Y^{(j)}\mid U, x)$ is full rank if and only if 
    $$
        \dfrac{\pr(Y^{(j)}=1\mid U=1, x)}{\pr(Y^{(j)}=2\mid U=1, x)} \neq \dfrac{\pr(Y^{(j)}=1\mid U=2, x)}{\pr(Y^{(j)}=2\mid U=2, x)},
    $$
    which holds if and only if $U\nind Y^{(j)}\mid X=x.$ By $U \nind Y^{(j)} \mid X=x$ for $j=1,2,3$, we also have that $U$ is not a constant variable so that $\pr(U=u)>0$ for $u=1,2$, implying that $P_D(U\mid x)$ is full rank. Equation \eqref{eq:mp2} then implies that $P(Y^{(2)},Y^{(3)}\mid x)$ is full rank.
\end{proof}

\subsection{Proof of Lemma \ref{lm:perturbation}}
\begin{proof}
This is a minor modification of the proof in   \citet[][p.90, Theorem 5.3]{serre_matrices_2010}. \\
Let $X_0$ be an eigenvector of $M_0$ associated with $\lambda_0$. We normalize $X_0$ such that $1^\T_k X_0=1$. It is easy to see that $\lambda_0$ is also a simple eigenvalue of $M_0^\T$. Due to Proposition 3.15 in \cite{serre_matrices_2010}, an eigenvector $Y_0$ of $M_0^\T$ associated with $\lambda_0$ satisfies $Y^\T_0X_0\ne0$. We also normalize $Y_0$ in such a way that $1_k^\T Y_0=1$, and denote $a=Y^\T_0X_0\neq 0$. %It must be true that $a\ne 0$.

Define a polynomial function $F$ over $M_k(\real)\times\real\times\real^k$, with values in $\real\times\real^k$, by
\begin{equation*}
    F(M,\lambda,x)=(1_k^\T x-1, Mx-\lambda x).
\end{equation*}
We have $F(M_0,\lambda_0,X_0)=(0,0)$.
The differential of $F$ with respect to $(\lambda,x)$, at the base point $(M_0,\lambda_0,X_0)$, is the linear map
\begin{equation*}
    (\mu,y)\overset{\delta}{\mapsto}(1_k^\T y,(M_0-\lambda_0)y-\mu X_0).
\end{equation*}
We first show that $\delta$ is one-to-one. Let $(\mu,y)$ be such that $\delta(\mu,y)=(0,0)$. Then $\mu = \frac{1}{a}\mu Y^\T_0X_0=\frac{1}{a}Y_0^\T(M_0-\lambda_0)y=\frac{1}{a}0^\T y=0$. Therefore, $(M_0-\lambda_0)y-\mu X_0=0$ implies that $(M_0-\lambda_0)y=0$. Inasmuch as $\lambda_0$ is simple, $y$ is colinear to $X_0$; now the fact that $1_k^\T y=0$ and $1_k^\T X_0=1$ yields $y=0$.

Because $\delta$ is a one-to-one endomorphism of $\real\times\real^k$, it is an isomorphism. We may then apply the implicit function theorem to $F$: there exists neighborhoods $\mathcal{M}$, $\mathcal{V}$, and $\mathcal{W}$ and analytic functions $(\Lambda,X):\mathcal{M}\mapsto\mathcal{V}$ such that
\begin{align*}
   {
\left( \begin{array}{c}
(M,\lambda,x)\in\mathcal{W}\\
F(M,\lambda,x)=(0,0)
\end{array} 
\right )}
\Longleftrightarrow
   {
\left( \begin{array}{c}
M\in\mathcal{M}\\
(\lambda,x)=(\Lambda(M),X(M))
\end{array} 
\right )}.
\end{align*}
Notice that $F=0$ implies that $(\lambda,x)$ is an eigenpair of $M$ and $1_k^{\T}x=1$. Therefore, there exists an open neighborhood $\mathcal{M}$ of $M_0$ in $M_k(\real)$, and two analytic functions $M\mapsto \Lambda(M)$, $M\mapsto X(M)$ over $\mathcal{M}$, such that
\begin{itemize}[topsep=8pt,itemsep=0pt,partopsep=4pt, parsep=4pt]
    \item $\Lambda(M)$ is an eigenvalue of $M$, $X(M)$ is an eigenvector associated with $\Lambda(M)$ normalized in a way such that $1_k^{\T}X(M)=1$.
    \item $\Lambda(M_0)=\lambda_0$, $X(M_0)=X_0$, $1_k^{\T}X_0=1$.
\end{itemize}
\end{proof}

\subsection{Proof of Lemma \ref{lem:linear_function}}

\begin{proof}
Let $y=0$, we have $f(x)=f(x) + f(0)$, therefore $f(0)=0$. 

Let $y=x$, we have $f(2x) = 2f(x)$. By induction, we know $f(nx)=nf(x)$ for any positive integer $n$. 

Let $y=-x$, we have $f(x) + f(-x) = f(x-x) = 0$, thus $f(-x) = -f(x)$. So $f(nx) = nf(x)$ for any integer $n$. 

Since $f(x) = f(\frac{x}{2} + \frac{x}{2}) = f(\frac{x}{2}) + f(\frac{x}{2}) = 2f(\frac{x}{2})$, we conclude $f(\frac{x}{2}) = \frac{1}{2}f(x)$. By induction, we have $f(\frac{x}{n})= \frac{1}{n}f(x)$ for any positive integer $n$. By the fact $f(-x)=-f(x)$, we have $f(\frac{x}{n})= \frac{1}{n}f(x)$ for any integer $n$.

Then for any rational number $\frac{m}{n}$, where $m,n$ are integers, $f(\frac{m}{n}) = f(m\frac{1}{n}) = mf(\frac{1}{n}) = \frac{m}{n}f(1).$ Let $f(1) = c$, then $f(x) = cx$ for $x\in\mathbb{Q}$. 

Let $g(x)=cx$ for $x\in\real$, then $f(x)=g(x)$ for $x\in\mathbb{Q}$. For any $x\in\real$, there exists a sequence of rational numbers $\{r_n\}$ such that $x = \lim_{n\to\infty} r_n $. Therefore, 
\begin{align*}
    f(x) = \lim_{n\to\infty}f(r_n) = \lim_{n\to\infty}g(r_n) = g(x),
\end{align*}
where the first and third equality is by the continuity of $f$ and $g$.

So $f(x)=g(x)=cx$ for $x\in\real$.

\end{proof}

\subsection{Proof of Lemma \ref{lem:adjoint_def}}
\begin{proof}
The adjoint $T^*$ of an operator $T\in \mathcal{L}(X,Y)$ is defined via
\begin{align*}
& \langle Tx,y^*\rangle=\langle x,T^*y^*\rangle  \quad
\textrm{or } \quad y^*T(x)=T^*y^*(x)
\end{align*}
with $T^*:Y^*\rightarrow X^*$ and $y^*\in Y^*$. The duality is given by
$
\langle r,s\rangle =\int r(x)s(x)\dif x
$
for $r\in \mathcal{G}(\mathcal{X})$ and $s\in \mathcal{G}^*(\mathcal{X})$ in our case. We can check that
\begin{align*}
&\langle T_{f_{a,b}(a,b)}g,h^*\rangle  \\
=& \int\int f(a,b)g(b)\dif b h^*(a)\dif a \\
=&\int g(b) \int f(a,b)h^*(a)\dif a\dif b \\
=&\langle g,T_{f_{a,b}(a,b)}^* h^*\rangle 
\end{align*}
where $g\in \mathcal{G}(\mathcal{B})$ and $h^*\in \mathcal{G}^*(\mathcal{A})$, so \eqref{eq:adjoint} is correct.
\end{proof}

\subsection{Proof of Lemma \ref{lem:adjoint_injectivity}}
\begin{proof}

The operator $T_{f_{b\mid a}(b,a)}$ is defined via
\begin{align*}
    (T_{f_{b\mid a}(b,a)}g)(b)=\int f(b\mid a)g(a)\dif a = \int \frac{f(a,b)}{f(a)}g(a)\dif a = \int f(a,b)\frac{g(a)}{f(a)}\dif a.
\end{align*}
By the injectivity assumption, if $(T_{f_{b\mid a}(b,a)}g)(a)=0$, then $g(a)=0$.

The operator $T_{f_{a,b}(a,b)}$ is defined via $(T_{f_{a,b}(a,b)}g)(a)=\int f(a,b)g(b)\dif b$.

By Lemma \ref{lem:adjoint_def}, the adjoint of $T_{f_{a,b}(a,b)}$, the operator $T_{f_{a,b}(a,b)}^*$ from $\mathcal{G}^*(\mathcal{A})$ to $\mathcal{G}^*(\mathcal{B})$, is given by
$
    (T_{f_{a,b}(a,b)}^* g^*)(b)=\int f(a,b)g^*(a)\dif a.
$
% To see this, we note that the adjoint $T^*$ of an operator $T\in \mathcal{L}(X,Y)$ is defined via
% \begin{align*}
% & \langle Tx,y^*\rangle=\langle x,T^*y^*\rangle  \quad
% \textrm{or } \quad y^*T(x)=T^*y^*(x)
% \end{align*}
% with $T^*:Y^*\rightarrow X^*$ and $y^*\in Y^*$. The duality is given by
% $
% \langle r,s\rangle =\int r(x)s(x)\dif x
% $
% for $r\in \mathcal{G}(\mathcal{X})$ and $s\in \mathcal{G}^*(\mathcal{X})$ in our case. We can check that
% \begin{align*}
% &\langle T_{f_{a,b}(a,b)}g,h^*\rangle  \\
% =& \int\int f(a,b)g(b)\dif b h^*(a)\dif a \\
% =&\int g(b) \int f(a,b)h^*(a)\dif a\dif b \\
% =&\langle g,T_{f_{a,b}(a,b)}^* h^*\rangle 
% \end{align*}
% where $g\in \mathcal{G}(\mathcal{B})$ and $h^*\in \mathcal{G}^*(\mathcal{A})$, so \eqref{eq:adjoint} is correct. 
%{\color{red} Is \eqref{eq:adjoint} definition or conclusion?}

For any $h^*\in\mathcal{G}^*(\mathcal{A})$, let $\ell(a)=h^*(a)f(a)$ so that $\ell \in \mathcal{G}^(\mathcal{A})$, then 
$$
(T_{f(b\mid a)(b,a)}\ell)(b)=\int f(b\mid a)\ell(a)\dif a  = \int f(a,b)\frac{\ell(a)}{f(a)}\dif a = \int f(a,b)h^*(a) = 0
$$
implies $\ell=0$. Since $f(a)\ne 0$, we must have $h^*=0$. So the injectivity of $T_{f_{a,b}(a,b)}^*$ is proved.

% Next, $T_{f_{a,b}(a,b)}$ can be shown to be injective when viewed as a mapping of $\overline{\mathcal{R}(T_{f_{a,b}(a,b)}^*)}$ into $\mathcal{G}(\mathcal{A})$, where $\overline{\mathcal{R}(T_{f_{a,b}(a,b)}^*)}$ denotes the closure in $\mathcal{G}(\mathcal{B})$ of the range of $T_{f_{a,b}(a,b)}^*$. Indeed, by Lemma VI.2.8 in \cite{dunford_linear_1988}, $\overline{\mathcal{R}(T_{f_{a,b}(a,b)}^*)}$ is the orthogonal complement of the null space of $T_{f_{a,b}(a,b)}$, denoted $\mathcal{N}(T_{f_{a,b}(a,b)})$. It follows that $T_{f_{a,b}(a,b)}^{-1}$ exists.

By Lemma VI.2.8 in \cite{dunford_linear_1988}, $\overline{\mathcal{R}(T_{f_{a,b}(a,b)})}$ is the orthogonal complement of $\mathcal{N}(T_{f_{a,b}(a,b)}^*)$, the null space of $T_{f_{a,b}(a,b)}^*$. Since $T_{f_{a,b}(a,b)}^*$ is injective, $\mathcal{N}(T_{f_{a,b}(a,b)}^*)=\{0\}$. Hence, $\overline{\mathcal{R}(T_{f_{a,b}(a,b)})}=\mathcal{G}(\mathcal{A})$. Since $T_{f_{a,b}(a,b)}$ is injective by assumption, we have that $T_{f_{a,b}(a,b)}^{-1}$ is densely defined over $\mathcal{G}(\mathcal{A})$ for $\mathcal{G}=\mathcal{L}^1,\mathcal{L}^1_{\textrm{bnd}}$.

% By Lemma \ref{lem:adjoint}, we have that the image of $T_{k(a,b)'}$ is dense.

\end{proof}

% \begin{lemma}
% \label{lem:adjoint}
% Let $X$, $Y$ be two normed spaces, and $T:X\rightarrow Y$ a bounded linear operator. The adjoint operator $T^*$ defined via $T^*f^*(x)=f^*(Tx)$ is injective if and only if the image of $T$ is dense.
% \end{lemma} 
% \begin{proof}
% % \href{https://math.stackexchange.com/questions/1076139/prove-that-t-is-injective-iff-imt-is-dense}{link}
% Let $f^*\in Y^*$ with $T^*(f^*)=0$. This means that
% $$
% f^*(Tx)=(T^*f^*)(x)=0
% $$
% for all $x\in X$. Let $y\in Y$. Since $T:X\rightarrow Y$ has dense range, there is a sequence $\{x_n\}$ in $X$ such that $y=\lim_{n\rightarrow\infty}T(x_n)$. Since $T$ and $f$ are both continuous ($T$ is a bounded linear operator and $f^*$ is a continuous linear functional), we have
% $$
% f^*(y)=\lim_{n\rightarrow\infty}f^*(Tx_n)=0.
% $$
% So $f^*=0$.
% \end{proof}

\subsection{Proof of Lemma \ref{lem:exponential_completeness}}
\begin{proof}
For brevity of notation, we suppress $\theta$ in the following proof.
Suppose that $\int g(b)k(a,b)\dif b=0$, which, in this setting, is
\begin{equation}
\label{eq:complete_exp1}
h(a)\int \tilde{g}(b)\exp\{\eta(b)^\T\lambda(a)\}\dif b=0,
\end{equation}
where $\tilde{g}(b)=g(b)\psi(b)$. Since the mapping $b\mapsto \eta(b)$ is one-to-one, let $t=\eta(b)$ or equivalently $b=\eta^{-1}(t)$. Then \eqref{eq:complete_exp1} becomes
\begin{equation}
    \label{eq:complete_exp2}
    h(a)\int\tilde{g}\{\eta^{-1}(t)\}[\dot{\eta}\{\eta^{-1}(t)\}]^{-1}\exp\{\lambda(a)^\T t\}\dif t=0,
\end{equation}
where $\dot{\eta}(b)=\partial\eta(b)/\partial b$ and $[\dot{\eta}\{\eta^{-1}(t)\}]^{-1}$ is the Jacobian matrix. 
%The left-hand side of \eqref{eq:complete_exp2} as a function of $\eta(b)$ is a Laplace transform of $\tilde{g}\{\lambda^{-1}(t)\}[\dot{\lambda}\{\lambda^{-1}(t)\}]^{-1}$, and it cannot be zero unless $\tilde{g}\{\lambda^{-1}(t)\}[\dot{\lambda}\{\lambda^{-1}(t)\}]^{-1}$ is zero almost everywhere.
By making a translation of the parameter space one can assume without loss of generality that the range of $\lambda(a)$ contains the rectangle $I=[-m,m]^k$. Let $\tilde{g}\{\eta^{-1}(t)\}[\dot{\eta}\{\eta^{-1}(t)\}]^{-1}=r(t)=r^+(t)-r^-(t)$ be such that
$$
\int r(t)\exp\{\lambda(a)^\T t\}\dif t=0\textrm{ for all } \lambda(a)\in I.
$$
Then
$$
\int r^+(t)\exp\{\lambda(a)^\T t\}\dif t=\int r^-(t)\exp\{\lambda(a)^\T t\}\dif t.
$$
Hence in particular let $\lambda(a)=0$,
$$
\int r^+(t)\dif t = \int r^-(t)\dif t.
$$
Dividing $r$ by a constant, one can take the common value of these two integrals to be 1, so that
$$
\dif P^+(t) = r^+(t)\dif t,\quad
\dif P^-(t) = r^-(t)\dif t 
$$
are probability measures, and
$$
\int \exp\{\lambda(a)^\T t\}\dif P^+(t)= \int \exp\{\lambda(a)^\T t\}\dif P^-(t)
$$
for all $\lambda(a)\in I$. From another point of view, these integrals can also be regarded as functions of the complex variables $\lambda_j(a)=\xi_j+i \delta_jz, j=1,\ldots,k$, with real parts strictly between $-m$ and $m$. By Theorem 2.7.1 in \cite{lehmann_testing_2005}, they are analytic functions of $\lambda_j(a)$ in the strip $R_j:-m< \xi_j < m,-\infty<\delta_j<\infty$ of the complex plane. For $\lambda_2(a),\ldots,\lambda_k(a)$ fixed, real, and between $-m$ and $m$, the equality of the integrals holds on the line segment $\{(\xi_1,\delta_1):-m<\xi_1<m,\delta_1=0\}$ and can therefore be extended to the strip $R_1$, in which the integrals are analytic. By induction the equality can be extended to the complex region $\{\lambda_1(a),\ldots,\lambda_k(a):(\xi_j,\delta_j)\in R_j \textrm{ for }j=1,\ldots,k\}$. It follows in particular that for all real $(\delta_1,\ldots,\delta_k)$,
$$
\int e^{i\sum\delta_j t_j}\dif P^+(t) = \int e^{i\sum\delta_j t_j}\dif P^-(t).
$$
These integrals are the characteristic functions of the distributions $P^+$ and $P^-$ respectively, and by the uniqueness theorem for characteristic functions, the two distributions $P^+$ and $P^-$ coincide. From the definition of these distributions it then follows $r^+(t)=r^-(t)$ almost everywhere, and hence that $r(t)=0$ almost everywhere. Since $[\dot{\lambda}\{\lambda^{-1}(t)\}]^{-1}$ is not zero, it can hold only if $\tilde{g}(b)$ is zero almost everywhere. Since $\psi(b)$ is not zero, $g(b)$ is zero almost everywhere. 
\end{proof}

\section{Proof of Theorem \ref{thm:continuous}}
\label{sec:thm_s1_proof}
\begin{proof}
% When $U$ is discrete, we applied eigendecomposition on matrices to identify eigenvalues and eigenvectors which are conditional probabilities of interest. Then we use these conditional probabilities to recover potential outcome distributions. The fact that matrices are linear operators sheds light on the case where $U$ is continuous, since we can similarly use integral operators and try to replicate the process. 

The proof can be decomposed into four main steps. \\

Step 1: We show that a known operator constructed from observed distributions has a form of eigenvalue-eigenfunction decomposition, where the eigenvalues and eigenfunctions are 
conditional densities that involve unobserved variables. \\

Step 2: We prove the eigenvalue-eigenfunction decomposition is unique, in the sense that the eigenvalues and corresponding eigenspaces are identified. \\

Step 3: We show that the eigenspace corresponding to each eigenvalue is one dimensional. And the eigenfunction which forms the eigenspace is uniquely determined. \\

Step 4: We uncover the correspondence between each eigen-pair and a value of the unobserved confounder. Then we restore the mean potential outcomes. \\

%We show the representation of the decomposition is unique, overcoming some potential issues similar to the matrix case when $U$ is discrete, e.g., the eigenfunctions may not be unique, or, the indexing of the eigenvalues and eigenfunctions among different $x$ may not be unified. 
%{\color{red} The sentence is not clear. Split this into several sentences.} \\

%Steps 1-3 are similar to the proof of Theorem 1 in \cite{hu_instrumental_2008}, while Step 4 is different. 

Before providing detailed proofs for each step, we introduce some relevant math definitions and theorem. 
\begin{itemize}
    \item An \textit{integral operator} denoted by $T_{k(a,b)}$ is an operator mapping $g\in\mathcal{G}(\mathcal{B})$ to $T_{k(a,b)}g\in\mathcal{G}(\mathcal{A})$, defined as $(T_{k(a,b)}g)(a)=\int_{\mathcal{B}}k(a,b)g(b)\dif b$, where $k(a,b)$ is called the \textit{kernel} of the operator.
    \item The \textit{spectrum} of a bounded linear operator $T$ is the set of all $ \lambda \in \mathbb{C}$ for which the operator $T-\lambda I$ does not have an inverse that is a bounded linear operator.
    \item A \textit{quasi-nilpotent} operator is an operator whose spectrum is $\{0\}$.
    \item An operator $T$ is said to \textit{commute} with another operator $A$ if $NA=AN$.
    \item A \textit{projection-valued measure} assigns a projection operator to each set in a field. Here the field is the Borel $\sigma$-field. A \textit{projection} $P$ is a linear operator such that $P^2=P$. Just like a real-valued measure, we can do integration with respect to a projection-valued measure to get a new operator. 
    \item (Theorem XV.4.5 in \cite{dunford_linear_1988}) If a bounded operator $T$ can be written as $T=A+N$, where $A$ is an operator of the form 
    $
    A = \int_{\sigma}\lambda P (\dif \lambda)
    $
    with $P$ a projection-valued measure supported on the spectrum $\sigma$, a subset of the complex plane, and $N$ is a quasi-nilpotent operator commuting with $A$, then there is a unique decomposition of $T$. 

\end{itemize}

%In this proof, first we show that a known operator constructed from observed distributions has a form of eigenvalue-eigenfunction decomposition, where the eigenvalues and eigenfunctions are unobserved conditional densities of interest. Then we prove this decomposition is unique, in the sense that the eigenvalues and corresponding eigenspaces are identified. Finally, we show the representation of the decomposition is unique, overcoming some potential issues similar to the matrix case when $U$ is discrete, e.g., the eigenfunctions may not be unique, or, the indexing of the eigenvalues and eigenfunctions among different $x$ may not be unified.

Step 1. %The form of an eigenvalue-eigenfunction decomposition.
% From the parallel-outcome assumption, for fixed $(\yone,x)$\footnotemark, we have
% \footnotetext{We assume $\Yone$ to be either continuous or discrete. When $\Yone$ is discrete, $f(\yone\mid u,x)$ is not the conditional probability density function, but the conditional probability mass function. Since $(X,\Yone)$ is fixed at $(x,\yone)$, $f(\yone\mid u,x)$ is a function of $u$.}
From the parallel-outcome assumption, for fixed $(\yone,x)$, we have
\begin{align*}
    f(y^{(2)},y^{(3)} \mid x) &=\int_{\mathcal{U}} f(y^{(2)}\mid u,x) f(y^{(3)}\mid u,x) f(u\mid x)\dif u,\\
    f(y^{(1)},y^{(2)},y^{(3)}\mid x) &=\int_{\mathcal{U}} f(y^{(1)}\mid u,x) f(y^{(2)}\mid u,x) f(y^{(3)}\mid u,x) f(u\mid x)\dif u.
\end{align*}
%Note that we assume $\Yone$ to be either continuous or discrete. When $\Yone$ is discrete, $f(\yone\mid u,x)$ is not the conditional probability density function, but the conditional probability mass function. Since $(X,\Yone)$ is fixed at $(x,\yone)$, $f(\yone\mid u,x)$ is a function of $u$. 
From the above equations, by Tonelli's theorem we have
\begin{align}
    % T_{f(\Ytwo,\Ythree\mid X)(\ytwo,\ythree)} &= T_{f(\Ytwo\mid U,X)(\ytwo,u)} \Delta_{f(U\mid X)(u)}T_{f(\Ythree\mid U,X)(u,\ythree)}, \label{eq:sm_op1}\\
    % T_{f(\Yone,\Ytwo,\Ythree\mid X)(\ytwo,\ythree)} &= T_{f(\Ytwo\mid U,X)(\ytwo,u)}\Delta_{f(\Yone\mid U,X)(u)} \Delta_{f(U\mid X)(u)}T_{f(\Ythree\mid U,X)(u,\ythree)}, \label{eq:sm_op2}
    T_{f(\Ytwo,\Ythree\mid x)} &= T_{f(\Ytwo\mid U,x)} \Delta_{f(U\mid x)}T_{f(\Ythree\mid U,x)^\T}, \label{eq:sm_op1}\\
    T_{f(\yone,\Ytwo,\Ythree\mid x)} &= T_{f(\Ytwo\mid U,x)}\Delta_{f(\yone\mid U,x)} \Delta_{f(U\mid x)}T_{f(\Ythree\mid U,x)^\T}, \label{eq:sm_op2}
\end{align}
which are exactly \eqref{eq:op1} and \eqref{eq:op2} in the main paper. % Recall that an integral operator denoted by $T_{k(a,b)}$ is an operator mapping $g\in\mathcal{G}(\mathcal{B})$ to $T_{k(a,b)}g\in\mathcal{G}(\mathcal{A})$, defined as $(T_{k(a,b)}g)(a)=\int_{\mathcal{A}}k(a,b)g(b)\dif b$, where $k(a,b)$ is called the kernel of the operator. 
The left-hand sides of \eqref{eq:sm_op1} and \eqref{eq:sm_op2} are known operators because the kernels are observed distributions. %{\color{red} Repeat equations \eqref{eq:op1} and \eqref{eq:op2} here. Also, remind the readers the notation, for example ``kernel''.}

Since $T_{f(\Ytwo\mid U,x)}$ is injective, by Condition \ref{condition:injectivity}, we can rewrite \eqref{eq:sm_op1} as
\begin{equation}
\label{eq:inverse_1}
T_{f(\Ytwo\mid U,x)}^{-1}T_{f(\Ytwo,\Ythree\mid x)} = \Delta_{f(U\mid x)}T_{f(\Ythree\mid U,x)^\T}.
\end{equation}
% The domain of the inverse is guaranteed to be dense in the range of $T_{f(\Ytwo,\Ythree\mid X)(\ytwo,\ythree)}$ as the right-hand side is a well-defined integral operator by absorbing the multiplication operator in the integral operator.
Plug \eqref{eq:inverse_1} into \eqref{eq:sm_op2}, we have
\begin{equation}
\label{eq:inverse_2}
    T_{f(\yone,\Ytwo,\Ythree\mid x)} = T_{f(\Ytwo\mid U,x)}\Delta_{f(\yone\mid U,x)} T_{f(\Ytwo\mid U,x)}^{-1}T_{f(\Ytwo,\Ythree\mid x)}.
\end{equation}
Since $T_{f(\Ytwo,\Ythree\mid x)}$ is injective, by Condition \ref{condition:injectivity}, we can
%The injectivity of $T_{f(\Ytwo,\Ythree\mid X)(\ytwo,\ythree)}$ by Condition \ref{condition:injectivity} 
apply $T_{f(\Ytwo,\Ythree\mid x)}^{-1}$ from the right on each side of \eqref{eq:inverse_2}, %which yields
\begin{equation}
\label{eq:continuous_eigendecomposition}
T_{f(\yone,\Ytwo,\Ythree\mid x)}T_{f(\Ytwo,\Ythree\mid x)}^{-1} = T_{f(\Ytwo\mid U,x)}\Delta_{f(\yone\mid U,x)} T_{f(\Ytwo\mid U,x)}^{-1}.
\end{equation}
% Since $T_{\pr(\Ytwo,\Ythree\mid X)(\ytwo,\ythree)}$ is injective by Condition \ref{condition:injectivity} and \eqref{eq:op1}, we have
% \begin{align*}
% T&\equiv T_{f(\Yone,\Ytwo,\Ythree\mid X)(\ytwo,\ythree)} T_{f(\Ytwo,\Ythree\mid X)(\ytwo,\ythree)} ^{-1}\\
% &=T_{f(\Ytwo\mid U,X)(\ytwo,u)}\Delta_{f(\Yone\mid U,X)(u)}T_{f(\Ytwo\mid U,X)(\ytwo,u)}^{-1}.
% \end{align*}
%This equivalence holds over a dense subset of $\mathcal{G}(\mathcal{Y}_2)$, and can be extended to the whole domain space $\mathcal{G}(\mathcal{Y}_2)$ by the standard extension procedure for linear operators.
% \tblue{The domain of the inverse operator? Should we require the injectivity of $T_{f(\Ytwo,\Ythree\mid X)(\ythree,\ytwo)}$ instead of $T_{f(\Ytwo,\Ythree\mid X)(\ytwo,\ythree)}$? \href{https://math.stackexchange.com/questions/3615979/when-is-the-range-of-a-bounded-linear-operator-dense}{(link})}
As $T_{f(\Ytwo,\Ythree\mid x)}^{-1}$ is densely defined over $\mathcal{L}^1_{\textrm{bnd}}(\mathcal{Y}_2)$ by Condition \ref{condition:injectivity} and Lemma \ref{lem:adjoint_injectivity}, the operator equivalence \eqref{eq:continuous_eigendecomposition} can be extended to the whole domain space $\mathcal{L}^1_{\textrm{bnd}}(\mathcal{Y}_2)$ using the standard extension procedure for linear operators.
The left-hand side of \eqref{eq:continuous_eigendecomposition} is a known operator since we can observe the joint distribution of $(X,\Yone,\Ytwo,\Ythree)$. The right-hand side is the spectral decomposition in the form of an eigenvalue-eigenfunction decomposition. 
%The above equation states that the known operator $T$ admits a spectral decomposition. 
The eigenvalues are elements from the multiplication operator, i.e., $f(\yone\mid U,x)$ for a given $(\yone,x)$ and for all $u$. The eigenfunctions are the kernels of the integral operators $T_{f(\Ytwo\mid U,x)}$, i.e., $f(\Ytwo\mid U,x)$ for a given $x$, and for all $(\ytwo,u)$.
To see this, we note that $\int_{\mathcal{U}}f(\ytwo \mid u,x)\delta(u-u_0)\dif u = f(\ytwo\mid u_0,x)$, where $\delta(\cdot)$ is the Dirac delta function. 
For any $u_0\in\mathcal{U}$, 
\begin{align}
\label{eq:eigenvalue_proof}
\begin{split}
    & T_{f(\Ytwo\mid U,x)}\Delta_{f(\yone\mid U,x)} T_{f(\Ytwo\mid U,x)}^{-1} f(\ytwo\mid u_0,x) \\
    = & T_{f(\Ytwo\mid U,x)}\Delta_{f(\yone\mid U,x)} \delta(u-u_0) \\
    = & T_{f(\Ytwo\mid U,x)} f(\yone\mid u,x)\delta(u-u_0) \\
    = & \int_{\mathcal{U}} f(\ytwo\mid u,x) f(\yone\mid u,x)\delta(u-u_0)\dif u \\
    = & f(\yone\mid u_0,x)f(\ytwo\mid u_0,x). 
\end{split}
\end{align}
Therefore $f(\Ytwo\mid u_0,x)$ is an eigenfunction corresponding to the eigenvalue $f(\yone\mid u_0,x)$. 
Although Dirac delta function is not a function, we make use of it to simplify the notation and to illustrate the idea. This is based on the fact that $f(\ytwo\mid u_0,x) = \lim_{n\rightarrow\infty}(T_{f(\Ytwo\mid U,x)}g_{n,u_0})(\ytwo)$, where $g_{n,u_0}=n1(|u-u_0|\le n^{-1})$, a sequence of absolutely integrable and bounded functions. We have shown that the operator equivalence \eqref{eq:continuous_eigendecomposition} can be extended to the whole domain space $\mathcal{L}^1_{\textrm{bnd}}(\mathcal{Y}_2)$. Thus $f(\ytwo\mid u_0,x)$ belongs to the extended domain of $T_{f(\Ytwo\mid U,x)}^{-1}$, and the derivation in \eqref{eq:eigenvalue_proof} can be seen as a limiting process.

Step 2. %Uniqueness of the decomposition. 
We will then show the uniqueness of the decomposition in \eqref{eq:continuous_eigendecomposition}. In this step, we follow similar proof technique used in \cite{hu_instrumental_2008},
and make use of Theorem XV.4.5 in \cite{dunford_linear_1988}.
% if a bounded operator $T$ can be written as $T=A+N$, where $A$ is an operator of the form 
% $
% A = \int_{\sigma}\lambda P (\dif \lambda)
% $
% with $P$ a projection-valued measure supported on the spectrum $\sigma$, a subset of the complex plane, and $N$ is a quasi-nilpotent operator commuting with $A$, then there is a unique decomposition of $T$. Here a quasi-nilpotent operator is defined as an operator whose spectrum is $\{0\}$.
In our case, this theorem applies with $N=0$ and $\sigma\subset\real$. % When $N=0$, we have $AN=NA=0$, so $N$ commutes with $A$. Besides, the spectrum of the bounded operator $N$ is $\{0\}$, so it is quasi-nilpotent.
It's easy to see $N$ is quasi-nilpotent and it commutes with any operator. The spectrum $\sigma$ of $A$ is the range of $f(\yone\mid U,x)$ as a function of $U$. Since $f(\yone\mid U,x)$ is bounded by Condition \ref{condition:bounded_density}, the operator $T$ is bounded as required by the theorem. As for $A$, the projection-valued measure $P$ assigned to any subset $\Lambda$ of $\real$ in our case is
% This theorem applies to our situation in the special case where $N=0$ and $\sigma\subset\mathbf{R}$. The spectrum $\sigma$ is the range of $f(\yone\mid U,x)$ as a function of $U$. Since the largest element of the spectrum is bounded by Condition \ref{condition:bounded_density}, the operator $T$ is bounded in the sense required by the theorem. 
%In our situation, the projection-valued measure $P$ assigned to any subset $\Lambda$ of $\mathbb{R}$ is
\begin{equation}
\label{eq:def_projection}
P(\Lambda)=T_{f(\Ytwo\mid U,x)}I_{\Lambda}T_{f(\Ytwo\mid U,x)}^{-1},
\end{equation}
where the operator $I_{\Lambda}$ is defined as
\begin{equation}\label{eq:def_i}
(I_{\Lambda}g)(u) = 1\{f(\Yone\mid U,X)\in\Lambda\}(u)g(u).
\end{equation}
Alternatively, we can define $P(\Lambda)$ by introducing the subspace
$$
\mathcal{S}(\Lambda)=\text{span}\{f(\Ytwo \mid u,x): u \text{ such that } f(\yone\mid u,x)\in \Lambda) \}
$$
for any subset $\Lambda$ of the spectrum $\sigma$. The projection $P(\Lambda)$ is then determined by specifying its range $\mathcal{S}(\Lambda)$ and its null space $\mathcal{S}(\sigma\backslash\Lambda)$.

Now we will show $\int_{\sigma}\lambda P(\dif\lambda)=T_{f(\Ytwo\mid U,x)}\Delta_{f(\yone\mid U,x)}T_{f(\Ytwo\mid U,x)}^{-1}$. We can write
\begin{align*}
\int_{\sigma}\lambda P(\dif\lambda)&=\int_{\sigma}\lambda\left(\frac{\dif}{\dif\lambda}P((-\infty,\lambda])\right)\dif\lambda \\
&=T_{f(\Ytwo\mid U,x)}\left(\int_{\sigma}\lambda\frac{\dif I_{(-\infty,\lambda]}}{\dif \lambda}\dif\lambda\right)T_{f(\Ytwo\mid U,x)}^{-1},
\end{align*}
where the second equality holds by \eqref{eq:def_projection}. To show that the middle term in parentheses is equivalent to $\Delta_{f(\yone\mid U,x)}$, we let it act on a function $g(u)$.
\begin{align*}
    \left(\int_{\sigma}\lambda\frac{\dif I_{[-\infty,\lambda]}}{\dif \lambda}\dif\lambda g\right)(u) &= \int_{\sigma}\lambda\frac{\dif}{\dif\lambda}1\{f(\Yone\mid U,X)(u)\in[-\infty,\lambda]\}g(u)\dif \lambda \\
    &= \int_{\sigma}\lambda\frac{\dif}{\dif\lambda}1\{f(\Yone\mid U,X)(u)-\lambda\le 0\}g(u)\dif \lambda \\
    &= \int_{\sigma}\lambda\delta\{f(\Yone\mid U,X)(u)-\lambda\}g(u)\dif \lambda \\
    &= \int_{\sigma}\lambda\delta\{\lambda-f(\Yone\mid U,X)(u)\}g(u)\dif \lambda \\
    &= f(\Yone\mid U,X)(u)g(u) \\
    &= (\Delta_{f(\yone\mid U,x)}g)(u).
\end{align*}
The first equality holds by \eqref{eq:def_i}. For the third and fourth equalities, we use the fact that the generalized differential of a step function $1(\lambda\le 0)$ is a Dirac delta $\delta(\lambda)$, and $\delta(\lambda)=\delta(-\lambda)$. The second last equation is by the fact that $\int\delta(\lambda-\lambda_0)h(\lambda)\dif \lambda=h(\lambda_0)$ for any function $h(\lambda)$ continuous at $\lambda=\lambda_0$ and here, for $h(\lambda)=\lambda$.

%where we have used the fact that the generalized differential of a step function $1(\lambda\le 0)$ is a Dirac delta $\delta(\lambda)$, as defined by the property that $\int\delta(\lambda-\lambda_0)h(\lambda)=h(\lambda_0)$ for any function $h(\lambda)$ continuous at $\lambda=\lambda_0$ and, in particular, for $h(\lambda)=\lambda$. Hence we can conclude that $\int_{\sigma}\lambda P(\dif\lambda)=T_{f(\Ytwo\mid U,X)(\ytwo,u)}\Delta_{f(\Yone\mid U,X)(u)}T_{f(\Ytwo\mid U,X)(\ytwo,u)}^{-1}$.

Step 3. Condition \ref{condition:no_degeneracy} ensures that the eigenspace corresponding to each eigenvalue is one dimensional. To show this, we utilize the fact that the operator $T_{f(\Ytwo\mid U,x)}$ that defines the eigenfunctions does not depend on $\Yone$, while the eigenvalues $f(\yone\mid U,x)$ do.
Let $D(\yone,u)=\{u':f(\yone\mid u',x)=f(\yone\mid u,x)\}$, the set of values of $U$ that index eigenfunctions sharing the same eigenvalue $f(\yone\mid u,x)$. Any linear combination of functions $f(\Ytwo\mid u',x)$ for $u'\in D(\yone,u)$ could be an eigenfunction of $T_{f(\yone,\Ytwo,\Ythree\mid x)} T_{f(\Ytwo,\Ythree\mid x)} ^{-1}$. If $v(u)\equiv\cap_{\yone\in\mathcal{Y}^{(1)}}\text{span}\{f(\Ytwo\mid u',x)_{u'\in D(\yone,u)}\}$ is not one dimensional, it must have at least two eigenfunctions, say $f(\Ytwo\mid u,x)$ and $f(\Ytwo\mid u',x)$. Thus, $\cap_{\yone\in\mathcal{Y}_1}D(\yone,u)$ must at least contain $u$ and $u'$. By the definition of $D(\yone,u)$, it must hold that $f(\yone\mid u,x)=f(\yone\mid u',x)$ for all $\yone\in\mathcal{Y}_1$. This violates Condition \ref{condition:no_degeneracy}. So we conclude $v(u)$ is one dimensional,
and it uniquely determines the eigenfunction $f(\Ytwo\mid u,x)$ after normalization to integrate to 1 by the requirement that $\int_{\mathcal{Y}_2}f(\ytwo\mid u,x)\dif\ytwo =1$.

Step 4. 
% Next, Condition \ref{condition:continuous_indexing} resolves the indexing ambiguity mentioned in 2. 
For $(X,\Yone)=(x,\yone)$, we have a set of eigenvalues $f(\yone\mid U,x)$ and corresponding eigenfunctions $f(\Ytwo\mid U,x)$ indexed by $U$. 
The ideal situation would be that we could connect each eigen-pair with a value of $U$. However this is not possible in our case, where only $x$ is known. For each pair $(u,x)$, we can act a known functional $M$, such as taking expectation, on the eigenfunction $f(\Ytwo\mid u,x)$ to derive some information on $(u,x)$, e.g., $h_1(u)+h_2(x)$ assuming $g$ is the identity function if Condition \ref{condition:continuous_indexing} holds. We shall not  assume that $h_2$ is known, since this is to some degree assuming the effect of $X$ on $\Ytwo$ is known. For example, if $M$ is the mean and $\Ytwo=\alpha U + \beta X + \epsilon$ where $E(\epsilon)=0$, then $h_2(x)=\beta x$ and $\beta$ is the causal effect of $X$ on $\Ytwo$. With $h_2$ unknown, we can not derive $h_1(u)$ for each $(u,x)$ pair. Nevertheless, if we know $h_1(u)+h_2(x)$ for a fixed $u$ and different $x_1,x_2$, then we can identify $h_2(x_2)-h_2(x_1)$. The following gives more details. 

%because if one considers another variable $\widetilde{U}$ related to $U$ through $U=R(\widetilde{U})$, we have
% \begin{align*}
%     M[\pr(\Ythree\mid \widetilde{U},X)(\widetilde{u},x)]=M[\pr(\Ythree\mid U,X)(R(\widetilde{u}),x)]= h_1(R(\widetilde{u}))+h_2(x),
% \end{align*}
By Condition \ref{condition:continuous_indexing}, assuming $h_1(u)$ is bounded from below, 
$$
\inf_{u\in\mathcal{U}} g[M\{f(\Ytwo\mid u,x)\}] = \inf_{u\in\mathcal{U}} h_1(u) + h_2(x).
$$
We take the supremum if $h_1(u)$ is not bounded from below but bounded from above. Then for $x_0$ and $x_1$, we can derive
$$
h_2(x_1) - h_2(x_0) = \inf_u g[M\{f(\Ytwo\mid u,x_1)\}] - \inf_u g[M\{f(\Ytwo\mid u,x_0)\}].
$$
Since the right-hand side is known, we can determine $h_2(x)$ up to a constant. Let $\widetilde{U}=h_1(U)+h_2(x_0)$, $\widetilde{h}_2(x)=h_2(x)-h_2(x_0)$, then $g[M\{f(\Ytwo\mid U,X)\}] = h_1(U)+h_2(X)=\widetilde{U} + \widetilde{h}_2(X)$. Now that $\widetilde{h}_2(X)$ is known, we can thus derive $\widetilde{U}$, which is a one-to-one mapping of $U$. That is to say, for $(X,\Yone)=(x,\yone)$, we have a set of eigenvalues $f(\yone\mid U,x)$ and corresponding eigenfunctions $f(\Ytwo\mid U,x)$ indexed by known $\widetilde{U}$ instead of unknown $U$. In other words, for a pair of eigenvalue and eigenfunction, $f(\yone\mid u,x)$ and $f(\Ytwo\mid u,x)$, although we do not know which $u$ they are associated to, we can relate them to $\widetilde{u}=h_1(u)+h_2(x_0)$ which is known. Hence we know $f(\Ytwo\mid \widetilde{U},x)$ and $f(\yone\mid\widetilde{U},x)$ rather than $f(\Ytwo\mid U,x)$ and $f(\yone\mid U,x)$. Since $f(\Ytwo\mid X)$ is observable, we are able to identify $f(\widetilde{U}\mid X)$ from $f(\Ytwo\mid X)=\int f(\Ytwo\mid \widetilde{U},X)f(\widetilde{U}\mid X)\dif\widetilde{U}$, or equivalently $f(\Ytwo\mid X)=T_{f(\Ytwo\mid \widetilde{U},x)}f(\widetilde{U}\mid X)$, if $T_{f(\Ytwo\mid \tilde{U},x)}$ is injective. 
%$T_{f(\Ytwo\mid X)(x,\ytwo)}=T_{f(\Ytwo\mid \widetilde{U},X)(\widetilde{u},\ytwo)}T_{f(\widetilde{U}\mid X)(x,\widetilde{u})}$ 
Actually, We can show injectivity of $T_{f(\Ytwo\mid U,x)}$ implies injectivity of $T_{f(\Ytwo\mid \tilde{U},x)}$. Let 
\begin{align*}
    0 &=\int f_{\Ytwo\mid \tilde{U},X=x}(\ytwo,\tilde{u})g(\tilde{u})\dif \tilde{u} \\
    &=\int f_{\Ytwo,U\mid X=x}(\ytwo,h^{-1}(\tilde{u}))\{f_{U\mid X=x}(h^{-1}(\tilde{u}))\}^{-1}g(\tilde{u})\dif\tilde{u} \\
    &=\int f_{\Ytwo,U\mid X=x}(\ytwo,u)\{f_{U\mid X=x}(u)\}^{-1}g(h(u))h'(u)\dif u \\ &= \int f_{\Ytwo\mid U,X=x}(\ytwo,u)g(h(u))h'(u)\dif u.
\end{align*}
The set $\{u:h'(u)=0\}$ has measure zero since $h(\cdot)$ is one-to-one. Hence $\int f_{\Ytwo\mid \tilde{U},X=x}(\ytwo,\tilde{u})g(\tilde{u})\dif \tilde{u}=0$ implies $g(\cdot)$ is zero.
The identification of $f(\widetilde{U}\mid X)$ leads to the identification of $f(\widetilde{U})=\int f(\widetilde{U}\mid X)f(X)\dif X$, 
and $f(\Ythree\mid \widetilde{U},X)$ is also identified by \eqref{eq:inverse_1} with $U$ replaced by $\widetilde{U}$.
Finally we can identify mean potential outcomes by g-formula $E\{Y^{(j)}(x)\}=E_{\widetilde{U}}E(Y^{(j)}\mid \widetilde{U},X=x)$, $j=1,2,3$. %Note that for $\Ytwo$ the result can be derived by exchanging the role of $\Ythree$ and $\Ytwo$.

\end{proof}

\section{Proof of Theorem \ref{thm:categorical_m_est}}
\label{sec:thm_s2_proof}
\begin{proof}
We only prove the first part of Theorem \ref{thm:categorical_m_est}. The second part follows from standard M-estimation theory.
We shall use Lemma \ref{lm:perturbation} in our proof.

For a square matrix $ A \in M_k(\real) $, define the Frobenius norm as $\lVert A\rVert_F=(\sum_{i,j=1}^k a_{ij}^2)^{1/2}$. Since entries of $P(Y^{(2)},Y^{(3)}\mid x)$ and $P(y^{(1)},Y^{(2)},Y^{(3)}\mid x)$ are estimated by sample means, each element of $\overline{P}(Y^{(2)},Y^{(3)}\mid x)$ and $\overline{P}(y^{(1)},Y^{(2)},Y^{(3)}\mid x)$ is $\sqrt{n}$-consistent. As the matrix dimension is fixed, entrywise convergence is equivalent to convergence in Frobenius norm, i.e., $\|\overline{P}(Y^{(2)},Y^{(3)}\mid x)-P(Y^{(2)},Y^{(3)}\mid x)\|_F=O_p(n^{-1/2})$ and  $\|\overline{P}(y^{(1)},Y^{(2)},Y^{(3)}\mid x)-P(y^{(1)},Y^{(2)},Y^{(3)}\mid x)\|_F=O_p(n^{-1/2})$. 

When $X$ is categorical, the determinant $\det\{P(Y^{(2)},Y^{(3)}\mid x)\}$ is bounded away from zero by Condition \ref{condition:full-rank}. Since matrix inversion is an analytic function, we have $\lVert\overline{P}(Y^{(2)},Y^{(3)}\mid x)^{-1}-P(Y^{(2)},Y^{(3)}\mid x)^{-1}\rVert_F = O_p(n^{-1/2})$. First we will show $\overline{P}(y^{(1)},Y^{(2)},Y^{(3)}\mid x)\overline{P}(Y^{(2)},Y^{(3)}\mid x)^{-1}$ is a $\sqrt{n}$-consistent estimator under the Frobenius norm. In particular, one has 
\begin{align*}
    &\lVert\overline{P}(y^{(1)},Y^{(2)},Y^{(3)}\mid x)\overline{P}(Y^{(2)},Y^{(3)}\mid x)^{-1} - P(y^{(1)},Y^{(2)},Y^{(3)}\mid x)P(Y^{(2)},Y^{(3)}\mid x)^{-1}\rVert_F \\
    =&\lVert\overline{P}(y^{(1)},Y^{(2)},Y^{(3)}\mid x)\overline{P}(Y^{(2)},Y^{(3)}\mid x)^{-1} - P(y^{(1)},Y^{(2)},Y^{(3)}\mid x)\overline{P}(Y^{(2)},Y^{(3)}\mid x)^{-1} \\
    &\quad + P(y^{(1)},Y^{(2)},Y^{(3)}\mid x)\overline{P}(Y^{(2)},Y^{(3)}\mid x)^{-1} - P(y^{(1)},Y^{(2)},Y^{(3)}\mid x)P(Y^{(2)},Y^{(3)}\mid x)^{-1}\rVert_F\\
    \le&\lVert\overline{P}(y^{(1)},Y^{(2)},Y^{(3)}\mid x)\overline{P}(Y^{(2)},Y^{(3)}\mid x)^{-1} - P(y^{(1)},Y^{(2)},Y^{(3)}\mid x)\overline{P}(Y^{(2)},Y^{(3)}\mid x)^{-1}\rVert_F \\
    &\quad+ \lVert P(y^{(1)},Y^{(2)},Y^{(3)}\mid x)\overline{P}(Y^{(2)},Y^{(3)}\mid x)^{-1} - P(y^{(1)},Y^{(2)},Y^{(3)}\mid x)P(Y^{(2)},Y^{(3)}\mid x)^{-1}\rVert_F\\
    \le&\lVert\overline{P}(y^{(1)},Y^{(2)},Y^{(3)}\mid x)-P(y^{(1)},Y^{(2)},Y^{(3)}\mid x)\rVert_F\lVert\overline{P}(Y^{(2)},Y^{(3)}\mid x)^{-1}\rVert_F \\
    &\quad+ \lVert P(y^{(1)},Y^{(2)},Y^{(3)}\mid x)\rVert_F\lVert\overline{P}(Y^{(2)},Y^{(3)}\mid x)^{-1}-P(Y^{(2)},Y^{(3)}\mid x)^{-1}\rVert_F\\
    =& O_p(n^{-1/2})\lVert\overline{P}(Y^{(2)},Y^{(3)}\mid x)^{-1}\rVert_F + \lVert P(y^{(1)},Y^{(2)},Y^{(3)}\mid x)\rVert_F O_p(n^{-1/2})\\
    =& O_p(n^{-1/2})O_p(1) + O(1)O_p(n^{-1/2})\\
    =& O_p(n^{-1/2}). 
\end{align*}

Let $M_0$ be such a matrix in Lemma \ref{lm:perturbation}, then one can find a closed neighborhood of $M_0$ over which $\Lambda(M)$ and $X(M)$ satisfy the Lipschitz condition. Therefore, the $n^{-1/2}$ convergence rate of $\widetilde{P}_D(y^{(1)} \mid U,x)$ and $\widetilde{P}(Y^{(2)}\mid U,x)$ can be inferred from the $n^{-1/2}$ convergence rate of $\widetilde{P}(y^{(1)},Y^{(2)},Y^{(3)}\mid x)\widetilde{P}(Y^{(2)},Y^{(3)}\mid x)^{-1}$. The same applies to $\widetilde{P}(Y^{(3)}\mid U,x)$ due to symmetry.  The $n^{-1/2}$ convergence rate of $\widetilde{P}_D(U\mid x)$ in \eqref{eq:M_est} can be verified using the technique in proving the $\sqrt{n}$-consistency of $\overline{P}(y^{(1)},Y^{(2)},Y^{(3)}\mid x)\overline{P}(Y^{(2)},Y^{(3)}\mid x)^{-1}$. According to \eqref{eq:pr_u_est}, one can obtain $\widetilde{\pr}(u)-\pr(u)=O_p(n^{-1/2})$ by noticing that $\widetilde{\pr}(u\mid x)$ and $\overline{\pr}(x)$ are both root-n consistent. %The estimated potential outcome distribution $\widehat{\pr}\{y^{(j)}(x)\}$ in \eqref{eq:categorical_causal_est}, as a linear combination of aforementioned estimators, is thus $\sqrt{n}$-consistent.

\end{proof}

\newpage
\bibliographystyle{apalike}
\bibliography{references2}

% \begin{thebibliography}{}

% \bibitem[Serre, 2010]{serre_matrices_2010}
% Serre, D. (2010).
% \newblock {\em Matrices: Theory and Applications}.
% \newblock Springer New York.

% \bibitem[van~der Vaart and Wellner, 2011]{van_der_vaart_local_2011}
% van~der Vaart, A. and Wellner, J.~A. (2011).
% \newblock A local maximal inequality under uniform entropy.
% \newblock {\em Electronic Journal of Statistics}, 5:192--203.

% \bibitem[van~der Vaart, 1998]{vaart_asymptotic_1998}
% van~der Vaart, A.~W. (1998).
% \newblock {\em Asymptotic Statistics}.
% \newblock Cambridge University Press.

% \end{thebibliography}

\makeatletter\@input{xx.tex}\makeatother